\documentclass[onecolumn,10pt]{IEEEtran}

\usepackage{graphicx,epsf,psfrag}
\usepackage{amsmath,amssymb}
\usepackage{amsfonts}
\usepackage{mathrsfs}
\usepackage{etoolbox}
\usepackage[noadjust]{cite}

\usepackage{latexsym}

\newcommand{\beq}{\begin{equation}}
\newcommand{\eeq}{\end{equation}}
\newcommand{\be}{\begin{equation}}
\newcommand{\ee}{\end{equation}}
\newcommand{\eps}{\epsilon}

\newcommand{\bi}{\begin{itemize}}
\newcommand{\ei}{\end{itemize}}

\newcommand{\calA}{\mathcal{A}}
\newcommand{\calB}{\mathcal{B}}
\newcommand{\calC}{\mathcal{C}}

\newcommand{\calE}{\mathcal{E}}

\newcommand{\calN}{\mathcal{N}}

\newcommand{\calP}{\mathcal{P}}

\newcommand{\calR}{\mathcal{R}}

\newcommand{\calU}{\mathcal{U}}

\newcommand{\calX}{\mathcal{X}}
\newcommand{\calY}{\mathcal{Y}}
\newcommand{\calZ}{\mathcal{Z}}

\newcommand{\bbE}{\mathbb{E}}

\newcommand{\bbP}{\mathbb{P}}

\newcommand{\bbR}{\mathbb{R}}

\newcommand{\scA}{\mathscr{A}}

\DeclareMathAlphabet{\mathbsf}{OT1}{cmss}{bx}{n}
\DeclareMathAlphabet{\mathssf}{OT1}{cmss}{m}{sl}

\newcommand{\rve}{\mathsf{e}}

\newcommand{\rvQ}{\mathsf{Q}}

\DeclareSymbolFont{bsfletters}{OT1}{cmss}{bx}{n}  
\DeclareSymbolFont{ssfletters}{OT1}{cmss}{m}{n}
\DeclareMathSymbol{\bsfGamma}{0}{bsfletters}{'000}
\DeclareMathSymbol{\ssfGamma}{0}{ssfletters}{'000}
\DeclareMathSymbol{\bsfDelta}{0}{bsfletters}{'001}
\DeclareMathSymbol{\ssfDelta}{0}{ssfletters}{'001}
\DeclareMathSymbol{\bsfTheta}{0}{bsfletters}{'002}
\DeclareMathSymbol{\ssfTheta}{0}{ssfletters}{'002}
\DeclareMathSymbol{\bsfLambda}{0}{bsfletters}{'003}
\DeclareMathSymbol{\ssfLambda}{0}{ssfletters}{'003}
\DeclareMathSymbol{\bsfXi}{0}{bsfletters}{'004}
\DeclareMathSymbol{\ssfXi}{0}{ssfletters}{'004}
\DeclareMathSymbol{\bsfPi}{0}{bsfletters}{'005}
\DeclareMathSymbol{\ssfPi}{0}{ssfletters}{'005}
\DeclareMathSymbol{\bsfSigma}{0}{bsfletters}{'006}
\DeclareMathSymbol{\ssfSigma}{0}{ssfletters}{'006}
\DeclareMathSymbol{\bsfUpsilon}{0}{bsfletters}{'007}
\DeclareMathSymbol{\ssfUpsilon}{0}{ssfletters}{'007}
\DeclareMathSymbol{\bsfPhi}{0}{bsfletters}{'010}
\DeclareMathSymbol{\ssfPhi}{0}{ssfletters}{'010}
\DeclareMathSymbol{\bsfPsi}{0}{bsfletters}{'011}
\DeclareMathSymbol{\ssfPsi}{0}{ssfletters}{'011}
\DeclareMathSymbol{\bsfOmega}{0}{bsfletters}{'012}
\DeclareMathSymbol{\ssfOmega}{0}{ssfletters}{'012}

\newcommand{\hati}{\hat{i}}
\newcommand{\hatI}{\hat{I}}

\newcommand{\tilX}{\tilde{X}}

\newcommand{\tilY}{\tilde{Y}}

\newcommand{\tilZ}{\tilde{Z}}

\newcommand{\barx}{\bar{x}}
\newcommand{\bary}{\bar{y}}

\newcommand{\barX}{\bar{X}}
\newcommand{\barY}{\bar{Y}}

\newcommand{\bbeta}{{\boldsymbol\beta}}

\newcommand{\ceil}[1]{\lceil{#1}\rceil}

\DeclareMathOperator{\var}{Var}

\ifcsmacro{theorem}{}{
\newtheorem{theorem}{Theorem}
\newtheorem{lemma}[theorem]{Lemma}
\newtheorem{proposition}[theorem]{Proposition}
\newtheorem{corollary}[theorem]{Corollary}
\newtheorem{definition}{Definition} 
\newtheorem{example}{Example}

}

\newcommand{\qednew}{\nobreak \ifvmode \relax \else
      \ifdim\lastskip<1.5em \hskip-\lastskip
      \hskip1.5em plus0em minus0.5em \fi \nobreak
      \vrule height0.75em width0.5em depth0.25em\fi}

\allowdisplaybreaks
\usepackage{diagbox}
\usepackage{hyperref}
\usepackage{xcolor}
\usepackage{comment}

\renewcommand{\hatI}{\widehat{I}}
\newcommand{\setA}{\calA}
\newcommand{\setB}{\calB}

\newcommand{\new}[1]{#1}

\title{A Second-Order Converse Bound for the Multiple-Access Channel via Wringing Dependence}

\author{Oliver Kosut
\thanks{This material is based upon work supported by the National Science Foundation under Grant No. CCF-1453718 and CCF-1908725.}
\thanks{O. Kosut is with the School of Electrical, Computer and Energy Engineering, Arizona State University, Tempe, AZ 85287 (e-mail: okosut@asu.edu).}}

\begin{document}

\maketitle

\begin{abstract}
A new converse bound is presented for the two-user multiple-access channel under the average probability of error constraint. This bound shows that for most channels of interest, the second-order coding rate---that is, the difference between the best achievable rates and the asymptotic capacity region as a function of blocklength $n$ with fixed probability of error---is $O(1/\sqrt{n})$ bits per channel use. The principal tool behind this converse proof is a new measure of dependence between two random variables called wringing dependence, as it is inspired by Ahlswede's wringing technique. The $O(1/\sqrt{n})$ gap is shown to hold for any channel satisfying certain regularity conditions, which includes all discrete-memoryless channels and the Gaussian multiple-access channel. Exact upper bounds as a function of the probability of error are proved for the coefficient in the $O(1/\sqrt{n})$ term, although for most channels they do not match existing achievable bounds.
\end{abstract}


\section{Introduction}

The multiple-access channel (MAC) is the fundamental information theory problem that addresses coordination among independent parties. In this problem, multiple transmitters\footnote{Throughout this paper, we will focus on the case with two transmitters.} independently send signals into a noisy channel, and a receiver attempts to recover a message from each transmitter. The MAC was alluded to by Shannon in \cite{Shannon1961}; the discrete-memoryless version was formally stated and its capacity region determined in \cite{Ahlswede1971,Liao1972,Slepian1973}. The capacity region for the Gaussian case was found in \cite{Wyner1974,Cover1975}.

These results were first-order asymptotic, meaning they considered the channel coding rates in the regime where the probability of error goes to zero and the blocklength goes to infinity. One may consider refinements to these results. For example, a strong converse states that, if the probability of error is fixed above zero and the blocklength goes to infinity, then the set of achievable rates is identical to the standard capacity region. The strong converse for the discrete-memoryless MAC was first proved by Dueck in \cite{Dueck1981}; this argument made use of the blowing-up lemma and a so-called wringing step. An alternative strong converse proof was presented by Ahlswede in \cite{Ahlswede1982}; this proof used Augustin's converse argument \cite{Augustin1966} in place of the blowing-up lemma, followed by a more refined wringing step. A strong converse for the Gaussian MAC was proved in \cite{Fong2016}, using an argument based on that of \cite{Ahlswede1982}.

One may refine the strong converse even further by fixing the probability of error, and asking how quickly the coding rates at blocklength $n$ approach the capacity region. This work dates back to Strassen \cite{Strassen}, who showed that for the point-to-point channel coding problem, the backoff from capacity at blocklength $n$ is $O(1/\sqrt{n})$, and also characterized the coefficient on this term. Recently, there has been renewed interest in this second-order (also known as \emph{dispersion}) regime following \new{\cite{Hayashi2009}, which refined Strassen's asymptotic analysis via the information spectrum, and \cite{Polyanskiy2010a}, which also focused} on non-asymptotic information theoretic bounds.

However, in the fixed-error second-order regime, the MAC has turned out to be significantly more difficult than the point-to-point channel. Achievable bounds are proved in \cite{Tan2014,Haim2012,Huang2012,MolavianJazi2012,Scarlett2015a,MolavianJazi2015}, each of which gives lower bounds  of order $O(1/\sqrt{n})$ on the back-off term in the coding rate. Second-order results for the related problem of the MAC with degraded message sets were presented in \cite{Scarlett2015,Scarlett2015b}, including matching second-order converse bounds. For the standard MAC under the maximal probability of error criterion, a second-order converse bound is presented in \cite{Moulin2013}. \new{Recently, a bound for the maximal probability of error version, based on the technique of the present paper, was presented in \cite{Wei2021}, which was published after the preprint of this paper. (See Sec.~\ref{sec:maximal} for a brief discussion of the maximal-error case.)} Herein we focus on the average probability of error case.  Second-order results for a random-access model, wherein an unknown number of transmitters send messages to a receiver, were derived in \cite{Effros2018}.

Despite this progress, the best converse bound for the second-order rate of the standard MAC with average probability of error has remained \cite{Ahlswede1982}. While \cite{Ahlswede1982} is primarily interested in proving a strong converse, rather than characterizing the asymptotic behavior of the coding rate, the converse bound presented there shows that
\be
\calR(n,\eps)\subseteq\calC+O\left(\frac{\log n}{\sqrt{n}}\right)
\ee
where $\calR(n,\eps)$ is the set of achievable rate pairs at blocklength $n$ and average probability of error $\eps$, and $\calC$ is the capacity region. In this paper, we improve upon the converse bound from \cite{Ahlswede1982} to show that for most MACs of interest---including discrete-memoryless MACs and the Gaussian MAC---the achievable rate region is bounded by
\be\label{eq:second_order}
\calR(n,\eps)\subseteq\calC+O\left(\frac{1}{\sqrt{n}}\right).
\ee
This result asserts that achievable second-order bounds of \cite{Tan2014,Haim2012,Huang2012,MolavianJazi2012,MolavianJazi2015,Scarlett2015a} are order-optimal; that is, the gap between the capacity region and the blocklength-$n$ achievable region, in either direction, is at most $O(1/\sqrt{n})$. We provide a specific upper bound on the coefficient in the $O(1/\sqrt{n})$ term, although for most channels it does not match the achievability bounds.

The main difficulty in proving a second-order converse for the MAC is to properly deal with the independence between the transmitters. The problem variant with degraded message sets, as studied in \cite{Scarlett2015,Scarlett2015b}, seems to be easier precisely because the transmitted signals are \emph{not} independent. The independence that is inherent to the standard MAC prohibits many of the methods to prove second-order converses for the point-to-point channel; for example, one cannot restrict the inputs to a fixed type (empirical distribution), which is one of the steps in the point-to-point converse in \cite{Polyanskiy2010a}, since imposing a fixed joint type on the two input signals creates dependence. An alternative approach adopted in \cite{Liu2019} to prove second-order converses uses the notion of \emph{reverse hypercontractivity}. This technique provides a strengthening of Fano's inequality, wherein the coding rate is upper bounded by the mutual information plus an $O(1/\sqrt{n})$ error term. However, this technique relies on the geometric average error criterion, which is stronger than the usual average error criterion (but weaker than the maximal error criterion). The method of \cite{Liu2019} can be applied to the average error criterion by first expurgating the code---i.e., removing some of the codewords with the largest probability of error. However, with the MAC, we cannot just expurgate codewords, we must expurgate codeword \emph{pairs}, which again introduces some dependence between inputs. For this reason, reverse hypercontractivity can be viewed as a replacement for the blowing-up lemma or Augustin's converse, but does not remove the need for wringing. Interestingly, the technique that we use here seems to be related to \emph{hypercontractivity}; see Sec.~\ref{sec:other_measures} for more details.

To handle the independence between transmitters, the strong converse of \cite{Ahlswede1982} adopted the following approach: given any MAC code, first expurgate it by restricting to those channel inputs with limited maximal probability of error. Of course, this expurgation introduces some dependence between the transmissions. Second, this dependence is ``wrung out'' by further restricting the channel inputs so as to restore some measure of independence between them. Our bound follows the same basic outline, but we use a different technique for wringing. Namely, we introduce a new dependence measure called \emph{wringing dependence}. In the wringing step, we restrict the channel inputs so that the wringing dependence between them is small. This method of wringing proves to be more efficient than that of \cite{Ahlswede1982}. In addition to being critical to our converse proof, the wringing dependence measure is interesting in its own right: it satisfies many natural properties of any dependence measure, including the data processing inequality, and all 7 of the axioms for dependence measures that R\'enyi proposed in \cite{Renyi1959}. Using this tool, we show that a bound of the form \eqref{eq:second_order} holds for any MAC that satisfies two regularity conditions. All discrete-memoryless MACs, and the Gaussian MAC, are shown to satisfy these conditions. 

The remainder of the paper is organized as follows. Sec.~\ref{sec:preliminaries} gives notational conventions and describes the setup for the MAC problem. Sec.~\ref{sec:wringing_dependence} is devoted to the wringing dependence: it is defined, some simple examples are presented, and its main properties are proved. Sec.~\ref{sec:fbl} gives a finite blocklength converse bound for the MAC; this bound includes the core steps of our converse argument based on the wringing dependence. In Sec.~\ref{sec:asymptotics}, second-order asymptotic bounds are proved, applying the finite blocklength bound from Sec.~\ref{sec:fbl} to prove \eqref{eq:second_order} under certain regularity conditions. Specifically, two second-order bounds are proved: one that applies to any channel that satisfies two regularity conditions, and a tighter bound that holds for discrete-memoryless channels. Sec.~\ref{sec:examples} illustrates the results with some specific example channels, including the Gaussian MAC. We conclude in Sec.~\ref{sec:conclusion}. Several of the more technical proofs are contained in appendices.

\section{Preliminaries}\label{sec:preliminaries}

\subsection{Notation}

Throughout, all logs and exponential have base $e$ unless otherwise specified; log base 2 is denoted $\log_2$. For a random variable, we use the corresponding calligraphic letter to indicate its alphabet; e.g. $X$ has alphabet $\calX$. While most results in the paper hold for arbitrary probability spaces, to simplify notation we do not typically specify the event space. For an alphabet $\calX$, the set of all distributions on that alphabet is denoted $\calP(\calX)$. Given two alphabets $\calX,\calY$, the channel $W$ from $\calX$ to $\calY$ is a collection $(W_x)_{x\in\calX}$ where $W_x\in\calP(\calY)$ for each $x\in\calX$. The set of all channels from $\calX$ to $\calY$ is denoted $\calP(\calX\to\calY)$. We will also sometimes use the notation $P_{Y|X}$ for a channel from $\calX$ to $\calY$ where $P_{Y|X=x}\in\calP(\calY)$ is the conditional distribution given $X=x$. We use $\bbE [X]$ for expectation of a real-valued random variable $X$; usually the underlying distribution will be clear from context, but if not we write $\bbE_P [X]$ to mean $\int X dP$. For variance, $\var (X)$ or $\var_P (X)$ are used in the same way. The probability of an event is denoted with $\bbP$ in a similar manner. For a set $\setA\subset\calX$, we write the indicator function for $\setA$ as $1(x\in \setA)$. For an integer $n$, we denote $[n]=\{1,\ldots,n\}$. A sequence $x^n\in\calX^n$ means $x^n=(x_1,\ldots,x_n)$. We adopt the standard $O(\cdot)$ and $o(\cdot)$ notations. Specifically, for functions $f(n),g(n)$, we write $g(n)=O(f(n))$ to indicate
\be
\limsup_{n\to\infty}\left|\frac{g(n)}{f(n)}\right|<\infty.
\ee
Similarly, $g(n)=o(f(n))$ means $\lim_{n\to\infty} g(n)/f(n)=0$. We also use this notation when the limit goes to $0$ instead of infinity; for example $g(\delta)=O(f(\delta))$ means $\limsup_{\delta\to 0} |g(\delta)/f(\delta)|<\infty$. We write $|x|^+=\max\{0,x\}$ for positive part.

We also adopt the following standard definitions. Given two distributions $P,Q\in\calP(\calX)$, the Kullback-Leibler divergence is denoted
\be
D(P\|Q)=\bbE_P \left[\log \frac{dP}{dQ}\right]
\ee
where $\frac{dP}{dQ}$ is the Radon-Nikodym derivative. We will also need the R\'enyi divergence of order $\infty$, given by
\be
D_\infty(P\|Q)=\sup_{\setA\subset\calX} \log \frac{P(\setA)}{Q(\setA)}
\ee
where the supremum is over all events $\setA$ in the probability space. The total variational distance is
\be
d_{TV}(P,Q)=\sup_{\setA\subset\calX} |P(\setA)-Q(\setA)|.
\ee
The hypothesis testing fundamental limit is given by
\be
\beta_\alpha(P,Q)=\inf_{\substack{T:\calX\to[0,1],\\
\bbE_P [T(X)]\ge \alpha}} \bbE_Q [T(X)].
\ee
Here, $T(x)$ represents the probability that a hypothesis test outputs hypothesis $1$ when $X=x$. The divergence variance is denoted
\be
V(P\|Q)=\var_P \left(\log \frac{dP}{dQ}\right).
\ee
\new{The third absolute moment of the log-likelihood ratio is given by
\be
T(P\|Q)=\bbE_P\left[\left|\log\frac{dP}{dQ}-D(P\|Q)\right|^3\right].
\ee}%
For distributions $P_X\in\calP(\calX),Q_Y\in\calP(\calY)$ and a channel $W\in\calP(\calX\to\calY)$, the conditional divergence and conditional divergence variance are denoted
\begin{align}
D(W\|Q_Y|P_X)&=\int dP_X(x) D(W_x\|Q_Y),\\
V(W\|Q_Y|P_X)&=\int dP_X(x) V(W_x\|Q_Y).
\end{align}
Given joint distribution $P_{XY}\in\calP(\calX\times\calY)$, the mutual information is given by
\be
I(X;Y)=D(P_{Y|X}\|P_Y|P_X)
\ee
where $P_{X},P_{Y},P_{Y|X}$ are the induced marginal and conditional distributions. The conditional mutual information is given by
\be
I(X;Y|Z)=D(P_{Y|XZ}\|P_{Y|Z}|P_{\new{X}Z}).
\ee
For a discrete distribution $P_X$, the entropy is
\be
H(X)=\sum_{x\in\calX} -P_X(x)\log P_X(x).
\ee
We also use $H_b(p)$ to denote the binary entropy; i.e. $H_b(p)=H(X)$ where $X\sim\text{Ber}(p)$.

\subsection{Multiple-Access Channel Problem Setup}

A one-shot multiple-access channel (MAC) with two users is given by a channel $W\in\calP(\calX\times\calY\to \calZ)$ where $\calX$ and $\calY$ are the input alphabets, and $\calZ$ is the output alphabet. A (stochastic) code is given by
\begin{enumerate}
\item a user 1 encoder $P_{X|I_1}\in\calP([M_1]\to\calX)$,
\item a user 2 encoder $P_{Y|I_2}\in\calP([M_2]\to\calY)$,
\item a decoder $P_{\hatI_1,\hatI_2|Z}\in\calP(\calZ\to [M_1]\times [M_2])$.
\end{enumerate}
The average probability of error is given by $\bbP((\hatI_1,\hatI_2)\ne (I_1,I_2))$
where $(I_1,I_2)$ represent the messages, which are uniformly distribution over $[M_1]\times[M_2]$, and 
\be
(X,Y,Z,\hatI_1,\hatI_2)\new{|(I_1,I_2)=(i_1,i_2)\sim P_{X|I_1=i_1}(x)P_{Y|I_2=i_2}(y)W_{xy}(z) P_{\hatI_1,\hatI_2|Z=z}(\hati_1,\hati_2)}.
\ee
\new{Here, recall that $W$ is the channel distribution from $(X,Y)$ to $Z$.} A code with message counts $M_1,M_2$ and average probability of error at most $\eps$ is called an $(M_1,M_2,\eps)$ code.

Given a one-shot channel $W$, the $n$-length product channel is given by
\be
W_{x^ny^n}=\prod_{t=1}^n W_{x_ty_t}.
\ee
For $n$-length channels, we also impose cost-constraints on the channel inputs. Specifically, there are functions $b_1:\calX\to\bbR$, $b_2:\calY\to\bbR$, and constants $B_1,B_2\in\bbR$; we assume that the encoders $P_{X^n|I_1},P_{Y^n|I_2}$ are such that the channel inputs $X^n,Y^n$ satisfy \new{the following almost surely:}
\be\label{eq:cost_constraints}
\frac{1}{n}\sum_{t=1}^n b_1(X_t)\le B_1,\qquad
\frac{1}{n}\sum_{t=1}^n b_2(Y_t)\le B_2.
\ee
Of course, a lack of cost constraint is included in this model simply by taking $b_1(x)=b_2(y)=0$ for all $x,y$. We consider $(W,b_1,b_2,B_1,B_2)$ to constitute the channel specification. We say an $(n,M_1,M_2,\eps)$ code is a code for $n$-length channel with average probability of error $\eps$. For any blocklength $n$ and probability of error $\eps\in(0,1)$, the set of achievable rates are
\be\label{eq:Rstar_def}
\calR(n,\eps)=\left\{\left(\frac{\log M_1}{n},\frac{\log M_2}{n}\right): \exists \text{ an }(n,M_1,M_2,\eps)\text{ code}\right\}.
\ee
The operational definition for the capacity region is given by\footnote{Recall that the lim-inf of a sequence of sets $\setA_n$ is $\bigcup_{n\ge 1} \bigcap_{k\ge n} \setA_k$.}
\be
\calC=\bigcap_{\eps>0}\, \liminf_{n\to\infty}\, \calR(n,\eps).
\ee
The first-order asymptotic result, proved in \cite{Ahlswede1971,Liao1972,Slepian1973,Wyner1974,Cover1975}, is that the capacity region is
\be\label{eq:capacity_region}
\calC=\bigcup_{\substack{P_{UXY}:X\perp Y|U,\\ \bbE [b_1(X)]\le B_1,\\ \bbE [b_2(Y)]\le B_2}}\left\{(R_1,R_2): R_1+R_2\le I(X,Y;Z|U),\ R_1\le I(X;Z|Y,U),\ R_2\le I(Y;Z|X,U)\right\}
\ee
where $X\perp Y|U$ indicates that $X$ and $Y$ are independent given $U$. Here, $U$ is the time-sharing random variable.\footnote{We have chosen to use $U$ rather than the more standard $Q$, since the letter $Q$ is primarily used for other concepts in this paper.} Using Carath\'eodory's theorem, we can restrict the alphabet cardinality of $U$ in the union to $|\calU|\le 6$.

Because of the multi-dimensional nature of achievable rate regions for network information theory problems such as the MAC, articulating second-order results can be a bit complicated. There are at least three equivalent methods for describing these results: (i) characterize the region of second-order coding rate pairs around a specific point on the boundary of the capacity region, (ii) fix an angle of approach to a point on the capacity region boundary, or (iii) bound the maximum achievable weighted sum-rate. See \cite[Chapter 6]{Tan2014a} for a discussion of these issues for network information theory problems. We have chosen to focus on the weighted sum-rate approach, which has the advantage that we can work with scalar quantities, and we do not need to specify a point on the capacity region boundary. Specifically, for non-negative constants $\alpha_1,\alpha_2$, we define the largest achievable weighted-sum rate as
\be
R^\star_{\alpha_1,\alpha_2}(n,\eps)=\sup\left\{\frac{\alpha_1\log M_1+\alpha_2\log M_2}{n}:\exists \text{ an } (n,M_1,M_2,\eps)\text{ code}\right\}.
\ee
In particular, $R^\star_{1,1}(n,\eps)$ is the largest achievable standard sum rate. Note that for any constant $c$,
\be
R^\star_{c\,\alpha_1,c\,\alpha_2}(n,\eps)=c\,R^\star_{\alpha_1,\alpha_2}(n,\eps).
\ee
Thus, it is enough to consider only pairs $(\alpha_1,\alpha_2)$ where $\max\{\alpha_1,\alpha_2\}=1$.  We also define the weighted-sum capacity as
\be
C_{\alpha_1,\alpha_2}=\sup\{\alpha_1 R_1+\alpha_2 R_2: (R_1,R_2)\in \calC\}.
\ee
Since the capacity region $\calC$ is convex, it is equivalently characterized by $C_{\alpha_1,\alpha_2}$. From the result in \eqref{eq:capacity_region}, it is easy to see that
\begin{align}
C_{\alpha_1,\alpha_2}
&=\sup_{\substack{P_{UXY}:X\perp Y|U,\\ \bbE[ b_1(X)]\le B_1,\\ \bbE[ b_2(Y)]\le B_2}}
\big[\min\{\alpha_1,\alpha_2\}  I(X,Y;Z|U)+|\alpha_1-\alpha_2|^+ I(X;Z|Y,U)+|\alpha_2-\alpha_1|^+ I(Y;Z|X,U)\big].\label{eq:Ca1a2}
\end{align}
Our goal is to prove bounds of the form
\be
R^\star_{\alpha_1,\alpha_2}(n,\eps)\le C_{\alpha_1,\alpha_2}+O\left(\frac{1}{\sqrt{n}}\right).
\ee
Note that if such a bound can be proved in which the implied constant in the $O(1/\sqrt{n})$ term is uniformly bounded over all $\alpha_1,\alpha_2$ where $\max\{\alpha_1,\alpha_2\}=1$, then
\be
\calR(n,\eps)\subseteq \calC+O\left(\frac{1}{\sqrt{n}}\right).
\ee

\section{Wringing Dependence}\label{sec:wringing_dependence}

This section is devoted to defining and characterizing the \emph{wringing dependence}, a new dependence measure that will be critical in our converse proof for the MAC. In Sec.~\ref{sec:motivation}, we first outline Ahlswede's proof of the MAC strong converse from \cite{Ahlswede1982} as motivation for the wringing dependence, and then we define it. The basic properties of wringing dependence are described in Sec.~\ref{sec:props}. The wringing lemma, which is the primary use of wringing dependence in our MAC converse proof, is given in Sec.~\ref{sec:wringing_lemma}. We present some relationships between wringing dependence and other dependence measures---specifically hypercontractivity and maximal correlation---in Sec.~\ref{sec:other_measures}.

\subsection{Motivation and Definition}\label{sec:motivation}

Consider a one-shot MAC given by $W\in\calP(\calX\times\calY\to \calZ)$. Ahlswede's converse proof from \cite{Ahlswede1982}, and ours, involves these basic steps:
\begin{enumerate}
\item given any MAC code, expurgate it by restricting to the subset $\Gamma\subset\calX\times\calY$ of input pairs with limited maximal probability of error,
\item choose sets $\bar\calX\subset\calX,\bar\calY\subset\calY$ so that when the code is restricted to input pairs $(X,Y)\in\Gamma\cap (\bar\calX\times \bar\calY)$, the inputs are close to independent,
\item prove a converse bound on the code restricted to $\Gamma\cap (\bar\calX\times \bar\calY)$,
\item relate this converse bound back to the original code.
\end{enumerate}
Step 2 is called ``wringing,'' as the dependence between $X$ and $Y$ introduced by restricting the code to $\Gamma$ is ``wrung out'' in the choice of $\bar\calX,\bar\calY$. This step is also where our proof deviates most significantly from Ahlswede's. In the wringing step,  choosing the sets $\bar\calX,\bar\calY$ requires trading-off between two objectives: (i) maximizing the probability of the sets $\bar\calX\times\bar\calY$, so that in Step~4, there is limited difference between the subset and the original code; and (ii) minimizing the dependence between the inputs when restricted to $\bar\calX\times\bar\calY$, so that the converse bound proved in Step~3 captures the independence between transmissions that is inherent to the MAC. The key result addressing this trade-off in Ahlswede's proof is \cite[Lemma~4]{Ahlswede1982}; the following is a slight modification of this lemma.\footnote{The main difference is that Ahlswede's lemma has only one sequence $X^n$, even though when the lemma is applied in the converse proof, it is done with two sequences $X^n,Y^n$. Here, we have stated the lemma with two sequences to make the connection to our technique clearer.}

\begin{lemma}\label{lemma:ahlswede}
Let $P_{X^nY^n}\in\calP(\calX^n\times\calY^n)$, $Q_{X^n}\in\calP(\calX^n)$, and $Q_{Y^n}\in\calP(\calY^n)$ be distributions such that
\be\label{eq:ahlswede_renyi}
D_\infty(P_{X^nY^n}\|Q_{X^n}Q_{Y^n})\le \log(1+c).
\ee
For any $0<\gamma<c$, $0<\eps<1$, there exist sets $\bar\calX\subset\calX^n,\bar\calY\subset\calY^n$ such that
\be\label{eq:ahlswede_prob}
P_{X^nY^n}(\bar\calX,\bar\calY)\ge \eps^{c/\gamma}
\ee
and for all $t\in[n]$, $x\in\calX,y\in\calY$
\be\label{eq:ahlswede_indep}
P_{X_tY_t|X^n\in\bar\calX,Y^n\in\bar\calY}(x,y)\le \max\{\eps,(1+\gamma)Q_{X_t|X^n\in\bar\calX}(x)Q_{Y_t|Y^n\in\bar\calY}(y)\}.
\ee
\end{lemma}

In this lemma, one can see the two objectives at play: \eqref{eq:ahlswede_prob} is a bound on the probability of $\bar\calX\times\bar\calY$, and \eqref{eq:ahlswede_indep} is a guarantee on dependence of the channel inputs. The two parameters $\gamma$ and $\eps$ allow one to trade-off between these two objectives; as $\gamma,\eps\to 0$, the guarantee on the probability becomes weaker, while the guarantee on the dependence becomes stronger. In the extreme case that $\gamma=\eps=0$, \eqref{eq:ahlswede_indep} states that $X_t$ and $Y_t$ are independent, whereas \eqref{eq:ahlswede_prob} becomes trivial.

Ahlswede's lemma is proved iteratively. The process is initialized with $\bar\calX=\calX^n,\bar\calY=\calY^n$. At each step, if \eqref{eq:ahlswede_indep} is violated for some $t\in[n]$, $\barx_t\in\calX,\bary_t\in\calY$, then the sets $\bar\calX,\bar\calY$ are revised to
\be\label{eq:bar_sets_reduction}
\bar\calX'=\bar\calX\cap \{x^n:x_t=\barx_t\},
\qquad 
\bar\calY'=\bar\calY\cap \{y^n:y_t=\bary_t\}.
\ee
Because each step involves a violation of \eqref{eq:ahlswede_indep}, at that point
\begin{align}
P_{X_tY_t|X^n\in\bar\calX,Y^n\in\bar\calY}(\barx_t,\bary_t)&>\eps,\label{eq:barxy_prob}\\
\frac{P_{X_tY_t|X^n\in\bar\calX,Y^n\in\bar\calY}(\barx_t,\bary_t)}{Q_{X_t|X^n\in\bar\calX}(\barx_t)Q_{Y_t|Y^n\in\bar\calY}(\bary_t)}
&>1+\gamma.\label{eq:barxy_ratio}
\end{align}
Here, \eqref{eq:barxy_prob} ensures that the probability of the pair $(\barx_t,\bary_t)$ is not too small, while \eqref{eq:barxy_ratio} ensures that each step ``eats into'' the R\'enyi divergence between $P$ and $Q$ from \eqref{eq:ahlswede_renyi} by at least $\log(1+\gamma)$. The latter implies that the number of steps cannot exceed $\frac{\log(1+c)}{\log (1+\gamma)}\le c/\gamma$, which leads to the guarantee on the probability in \eqref{eq:ahlswede_prob}.

To improve on Ahlswede's lemma, we make three principal observations:
\begin{enumerate}
\item Wringing can be done in the one-shot setting.
\item The set reduction steps in \eqref{eq:bar_sets_reduction} need not be limited to individual pairs $(\barx_t,\bary_t)$; we may instead use arbitrary sets $\setA\subset\calX,\setB\subset\calY$, and revise the sets as $\bar\calX'=\bar\calX\cap \setA$, $\bar\calY'=\bar\calY\cap \setB$.
\item The trade-off between the probability as in \eqref{eq:barxy_prob} and the likelihood ratio as in \eqref{eq:barxy_ratio} is most efficient by maximizing
\be\label{eq:ratio_ratio}
\frac{\log \frac{P_{XY}(\setA,\setB)}{Q_X(\setA)Q_Y(\setB)}}{-\log P_{XY}(\setA,\setB)}=\frac{\log Q_X(\setA) Q_Y(\setB)}{\log P_{XY}(\setA,\setB)}-1.
\ee
Note that if the quantity in \eqref{eq:ratio_ratio} is maximized, then neither the likelihood ratio nor the probability of $(\setA,\setB)$ will be too small. Moreover, maximizing this quantity ensures that if a pair $(\setA,\setB)$ has low probability, then the likelihood ratio is larger, ensuring that this step ``eats into'' the R\'enyi divergence by a greater amount.
\end{enumerate}
 We are now ready to give the definition for wringing dependence, in which the quantity in \eqref{eq:ratio_ratio} plays a key role.

\begin{definition}
Given random variables $X,Y$ with joint distribution $P_{XY}$, the wringing dependence between $X$ and $Y$ is given by\footnote{\new{While technically, the wringing dependence is a function of the joint distribution $P_{XY}$ rather than a function of the random variables $X,Y$ themselves, we have chosen to use the notation $\Delta(X;Y)$ wherein the dependence measure is an operator on the random variables. This notational choice is made consistently for all dependence measures in the paper: for example mutual information is $I(X;Y)$, maximal correlation is $\rho_m(X;Y)$, etc. In all cases, the underlying distribution will be clear from context, or specified in a subscript such as $\Delta_P(X;Y)$.}}
\be\label{eq:Delta_def0}
\Delta(X;Y)=\inf_{Q_X,Q_Y}\,\sup_{\setA\subset\calX,\setB\subset\calY}\,\inf\left\{\delta\ge 0:P_{XY}(\setA,\setB)^{1+\delta}\le Q_X(\setA)Q_Y(\setB)\right\}.
\ee
\end{definition}
Note that for any $p,q\in(0,1)$, $\inf\{\delta\ge 0:p^{1+\delta}\le q\}=\left|\frac{\log q}{\log p}-1\right|^+$.
Therefore an alternative definition is
\be\label{eq:Delta_def}
\Delta(X;Y)=\inf_{Q_X,Q_Y}\ \sup_{\substack{\setA\subset\calX,\setB\subset\calY}} 
\left|\frac{\log Q_X(\setA)Q_Y(\setB)}{\log P_{XY}(\setA,\setB)}-1\right|^+
\ee
where $\frac{\log q}{\log p}$ really means $\inf\{\theta:p^\theta\le q\}$, so by convention
\be\label{eq:log_conventions}
\frac{\log q}{\log p}=0\text{ if }p=0\text{ or }q=1,p<1,\quad
\frac{\log q}{\log p}=\infty\text{ if }p=1,q<1,\quad 
\frac{\log 1}{\log 1}=-\infty.
\ee

\new{To compute the wringing dependence given a joint distribution $P_{XY}$ requires optimizing over $Q_X$ and $Q_Y$. In fact, this optimization is convex, as shown as follows. We may write the quantity inside the positive part in \eqref{eq:Delta_def} as
\be
\frac{\log Q_X(\setA)Q_Y(\setB)}{\log P_{XY}(\setA,\setB)}-1
=\frac{\log Q_X(\setA)}{\log P_{XY}(\setA,\setB)}+\frac{\log Q_Y(\setB)}{\log P_{XY}(\setA,\setB)}-1.\label{eq:Delta_convex}
\ee
For fixed sets $\setA,\setB$, $\log P_{XY}(\setA,\setB)\le 0$, which means each of terms in the RHS of \eqref{eq:Delta_convex} is jointly convex in $(Q_X,Q_Y)$. Using the fact that the supremum (or maximum) of convex functions is also convex, this implies that
\be\label{eq:Delta_convex2}
\sup_{\setA\subset\calX,\setB\subset\calY} 
\left|\frac{\log Q_X(\setA)Q_Y(\setB)}{\log P_{XY}(\setA,\setB)}-1\right|^+
\ee
is jointly convex in $(Q_X,Q_Y)$. Thus,}
the wringing dependence can in principle be computed via convex optimization if $\calX$ and $\calY$ are finite sets.
However, this computation quickly becomes impractical as the alphabet sizes grow, since the number of sets $\setA,\setB$ is exponential in the alphabet cardinality.
The following is one example of a simple distribution for which it \emph{can} be computed in closed form.

\begin{figure}
\begin{center}
\includegraphics[width=4in]{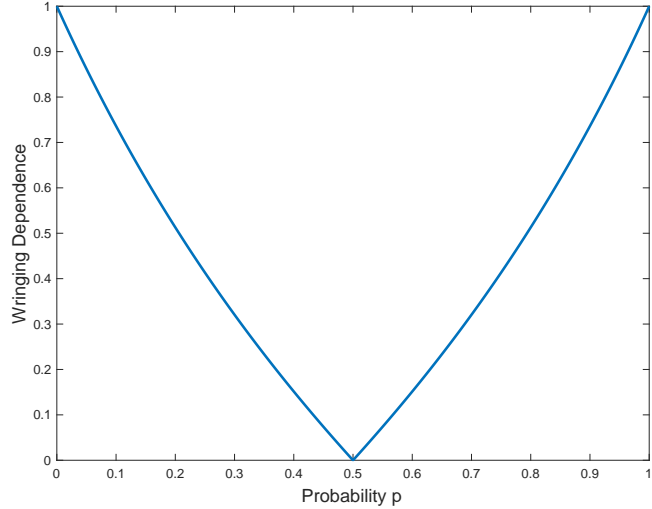}
\end{center}
\caption{The wringing dependence for a doubly symmetric binary source, as a function of the crossover probability $p$.}
\label{fig:DSBS}
\end{figure}

\begin{example}\label{example:DSBS}
Consider a doubly symmetric binary source (DSBS) $(X,Y)$, wherein $X,Y$ are each uniform on $\{0,1\}$, and $P_{XY}(1,1)=P_{XY}(0,0)=\frac{p}{2}$. Since this distribution is symmetric between $X$ and $1-X$, and between $Y$ and $1-Y$, the convexity of \new{\eqref{eq:Delta_convex2}} in $(Q_X,Q_Y)$ means that the optimal $Q_X,Q_Y$ are each uniform on $\{0,1\}$. Thus, if $p\le 1/2$, then $\Delta(X;Y)$ is given by
\begin{align}
\Delta(X;Y)&=\max\left\{0,\frac{\log 1/4}{\log p/2}-1,\frac{\log 1/4}{\log (1-p)/2}-1\right\}
\\&=\frac{\log 4}{\log 2-\log(1-p)}-1
\\&=\frac{1+\log_2(1-p)}{1-\log_2(1-p)}.
\end{align}
Therefore, for any $p$,
\be
\Delta(X;Y)=\frac{1+\log_2\max\{p,1-p\}}{1-\log_2\max\{p,1-p\}}.
\ee
The wringing dependence for a DSBS as a function of $p$ is shown in Fig.~\ref{fig:DSBS}.
\end{example}

\subsection{Properties}\label{sec:props}

The most important property of the wringing dependence is a counterpart of Ahlswede's lemma, which is presented in Sec.~\ref{sec:wringing_lemma}. But before stating this result, we prove some basic properties of the dependence measure. In particular, the following result states that wringing dependence satisfies many properties that one would expect of any dependence measure: it is non-negative, is zero iff $X$ and $Y$ are independent, and satisfies the data processing inequality. Indeed, this result shows that wringing dependence satisfies 6 out of the 7 axioms for dependence measures proposed in \cite{Renyi1959}. (It also satisfies the 7th, which is that for bivariate Gaussians, the wringing dependence equals the correlation coefficient; this fact is established in Sec.~\ref{sec:other_measures}.) The theorem also includes some other properties that will be useful throughout the paper.

\begin{theorem}\label{thm:props}
The wringing dependence $\Delta(X;Y)$ satisfies the following:
\begin{enumerate}
\item $\Delta(X;Y)=\Delta(Y;X)$.
\item $0\le \Delta(X;Y)\le 1$.
\item If $\Delta(X;Y)\le\delta$, then for all $\setA\subset\calX,\setB\subset\calY$,
\begin{align}
P_{XY}(\setA,\setB)&\le (1+2\delta) \left(P_X(\setA)P_Y(\setB)\right)^{1/(1+\delta)},\label{eq:dep_pp_bd}\\
|P_{XY}(\setA,\setB)-P_X(\setA)P_Y(\setB)|&\le 2\delta.\label{eq:gdep_abs_bd}
\end{align}
\item $\Delta(X;Y)=0$ if and only if $X$ and $Y$ are independent.
\item $\Delta(X;Y)=1$ if $X$ and $Y$ are \emph{decomposable}, meaning there exist sets $\setA\subset\calX,\setB\subset\calY$ where $0<P_X(\setA)<1$ and $1(X\in \setA)=1(Y\in \setB)$ almost surely\footnote{Decomposability is equivalent to the G\'acs-K\"orner common information being positive \cite{Gacs1973}.}. Moreover, if $\calX,\calY$ are finite sets and $\Delta(X;Y)=1$, then $X$ and $Y$ are decomposable.
\item For any Markov chain $W-X-Y-Z$, $\Delta(W;Z)\le\Delta(X;Y)$.
\end{enumerate}
\end{theorem}
\begin{IEEEproof}
(1) Symmetry between $X$ and $Y$ follows trivially from the definition.

(2) The fact that $\Delta(X;Y)\ge 0$ follows immediately from the definition. To upper bound $\Delta(X;Y)$, we may take $Q_X=P_X$, $Q_Y=P_Y$, so
\be\label{eq:wringing_simple_upper_bound}
\Delta(X;Y)\le \inf\{\delta\ge 0:P_{XY}(\setA,\setB)^{1+\delta}\le P_X(\setA)P_Y(\setB)\text{ for all }\setA\subset\calX,\setB\subset\calY\}.
\ee
Since $P_{XY}(\setA,\setB)\le P_X(\setA)$ and $P_{XY}(\setA,\setB)\le P_Y(\setB)$, $P_{XY}(\setA,\setB)^2\le P_X(\setA)P_Y(\setB)$ for all $\setA,\setB$. That is, $\delta=1$ is feasible in \eqref{eq:wringing_simple_upper_bound}, so $\Delta(X;Y)\le 1$.

(3) Suppose $\Delta(X;Y)\le \delta$. Thus, for any $\delta'>\delta$, there exist $Q_X,Q_Y$ such that
\be\label{eq:wringing_rearrangement}
P_{XY}(\setA,\setB)^{1+\delta'}\le Q_X(\setA)Q_Y(\setB)\text{ for all }\setA\subset\calX,\setB\subset\calY.
\ee
\new{Consider the function $f(p)=p^{1+\delta'}$ for $p\ge 0$. Since $\delta'>0$, $f$ is convex, so it can be lower bounded by any tangent line. In particular, forming the tangent line around $p=1$ gives
\be\label{eq:tangent_bound}
p^{1+\delta'}=f(p)\ge f(1)+f'(1)(p-1)=1+(1+\delta')(p-1)
=(1+\delta')p-\delta'.
\ee
Using this bound to lower bound the LHS of \eqref{eq:wringing_rearrangement} gives
\be\label{eq:QAB_PAB}
Q_X(\setA)Q_Y(\setB)\ge (1+\delta')P_{XY}(\setA,\setB)-\delta'.
\ee}%
Taking $\setB=\calY$ gives
\be
Q_X(\setA)\ge (1+\delta') P_X(\setA)-\delta'.
\ee
Since this may hold for $\setA^c$ in place of $\setA$, we may write
\begin{align}
Q_X(\setA)&=1-Q_X(\setA^c)
\\&\le 1-(1+\delta')P_X(\setA^c)+\delta'
\\&=(1+\delta')P_X(\setA).\label{eq:gdep_marginalX_upperbd}
\end{align}
By the same argument, for any $\setB\subset\calY$, $Q_Y(\setB)\le (1+\delta')P_Y(\setB)$. Thus
\begin{align}
P_{XY}(\setA,\setB)^{1+\delta'}&\le Q_X(\setA)Q_Y(\setB)
\\&\le (1+\delta')^2 P_X(\setA)P_Y(\setB).
\end{align}
As this holds for all $\delta'>\delta$, we have
\be\label{eq:PXY_delta2}
P_{XY}(\setA,\setB)^{1+\delta}\le (1+\delta)^2 P_X(\setA)P_Y(\setB).
\ee
Thus
\begin{align}
P_{XY}(\setA,\setB)&\le \left[(1+\delta)^2 P_X(\setA)P_Y(\setB)\right]^{1/(1+\delta)}.
\end{align}
Noting that $(1+\delta)^{2/(1+\delta)}\le 1+2\delta$ proves \eqref{eq:dep_pp_bd}. \new{Using again the tangent line bound from \eqref{eq:tangent_bound} to lower bound the LHS of \eqref{eq:PXY_delta2} gives}
\be
(1+\delta)P_{XY}(\setA,\setB)-\delta\le (1+\delta)^2 P_X(\setA)P_Y(\setB).
\ee
Thus
\begin{align}
P_{XY}(\setA,\setB)&\le (1+\delta)P_X(\setA)P_Y(\setB)+\frac{\delta}{1+\delta}\label{eq:gdep_derive2}
\\&\le P_X(\setA)P_Y(\setB)+\delta+\frac{\delta}{1+\delta}
\\&\le P_X(\setA)P_Y(\setB)+2\delta.\label{eq:gdep_derive3}
\end{align}
We prove the corresponding lower bound as follows:
\begin{align}
P_{XY}(\setA,\setB)&=P_X(\setA)-P_{XY}(\setA,\setB^c)
\\&\ge P_X(\setA)-P_X(\setA)P_Y(\setB^c)-2\delta\label{eq:gdep_derive4}
\\&=P_X(\setA)P_Y(\setB)-2\delta\label{eq:gdep_derive5}
\end{align}
where \eqref{eq:gdep_derive4} is simply an application of \eqref{eq:gdep_derive3} with $\setB^c$ swapped with $\setB$.  Combining \eqref{eq:gdep_derive3} and \eqref{eq:gdep_derive5} proves \eqref{eq:gdep_abs_bd}.

(4) If $\Delta(X;Y)=0$, then \eqref{eq:gdep_abs_bd} immediately gives $P_{XY}(\setA,\setB)=P_X(\setA)P_Y(\setB)$  for all $\setA\subset\calX,\setB\subset\calY$; i.e., $X$ and $Y$ are independent. Conversely, suppose $X$ and $Y$ are independent. Thus, if we take $Q_X=P_X,Q_Y=P_Y$, then
\be
P_{XY}(\setA,\setB)\le Q_X(\setA)Q_Y(\setB).
\ee
This proves that $\Delta(X;Y)= 0$ by the definition in \eqref{eq:Delta_def0}.

(5) Assume there exist sets $\setA,\setB$ as stated. Since $1(X\in \setA)=1(Y\in \setB)$ almost surely, $P_{XY}(\setA,\setB)=P_X(\setA)=P_Y(\setB)$, and $P_{XY}(\setA^c,\setB^c)=P_X(\setA^c)=P_Y(\setB^c)$, and also by assumption each of these probabilities is strictly between $0$ and $1$. For convenience let $p=P_{XY}(\setA,\setB)$. Using the definition in \eqref{eq:Delta_def}, we may lower bound the wringing dependence by
\begin{align}
\Delta(X;Y)&\ge \inf_{Q_X,Q_Y}\,\max\left\{\frac{\log Q_X(\setA)Q_Y(\setB)}{\log p},\,\frac{\log Q_X(\setA^c)Q_Y(\setB^c)}{\log (1-p)}\right\}-1\label{eq:full_dependence1}
\\&=\inf_{q\in[0,1]}\, \max\left\{\frac{\log q^2}{\log p},\, \frac{\log (1-q)^2}{\log(1-p)}\right\}-1\label{eq:full_dependence2}
\\&=\max\left\{\frac{\log p^2}{\log p},\, \frac{\log (1-p)^2}{\log(1-p)}\right\}-1\label{eq:full_dependence3}
\\&=1\label{eq:full_dependence4}
\end{align}
where \eqref{eq:full_dependence2} holds since the RHS of \eqref{eq:full_dependence1} is concave in $(Q_X,Q_Y)$ and symmetric between $Q_X(\setA)$ and $Q_Y(\setB)$, so the optimal choice is $Q_X(\setA)=Q_Y(\setB)=q$ for some $q\in[0,1]$; \eqref{eq:full_dependence3} holds since the first term in the max in \eqref{eq:full_dependence2} is decreasing in $q$ while the second term is increasing, so the infimum is achieved when the two terms in the max are equal, which occurs at $q=p$; and \eqref{eq:full_dependence4} holds by the fact that $0<p<1$. Since we know that in general $\Delta(X;Y)\le 1$, this proves $\Delta(X;Y)=1$. For the partial converse, assume $\calX,\calY$ are finite sets, and that $\Delta(X;Y)=1$. This implies that
\be\label{eq:decomposable_bound}
\sup_{\setA\subset\calX,\setB\subset\calY} \frac{\log P_X(\setA)P_Y(\setB)}{\new{\log}\, P_{XY}(\setA,\setB)}=2.
\ee
Since $\calX,\calY$ are finite, the supremum is attained, so there exist sets $\setA,\setB$ where $0<P_{XY}(\setA,\setB)<1$ and
\be
P_X(\setA)P_Y(\setB)=P_{XY}(\setA,\setB)^2.
\ee
This only holds if $P_{XY}(\setA,\setB)=P_X(\setA)=P_Y(\setB)$, which implies that $1(X\in \setA)=1(Y\in \setB)$ almost surely.

(6) The symmetry of the wringing dependence means that it is enough to show $\Delta(X;Z)\le \Delta(X;Y)$. We have
\begin{align}
\Delta(X;Z)&=\inf_{Q_X,Q_Z}\,\sup_{\setA\subset\calX,\setB'\subset\calZ} \left|\frac{\log Q_X(\setA)Q_Z(\setB')}{\log P_{XZ}(\setA,\setB')}-1\right|^+
\\&\le \inf_{Q_X,Q_Y}\,\sup_{\setA\subset\calX,\setB'\subset\calZ} \left|\frac{\log Q_X(\setA)\int dQ_Y(y)P_{Z|Y=y}(\setB')}{\log P_{XZ}(\setA,\setB')}-1\right|^+\label{eq:DPI2}
\\&=\inf_{Q_X,Q_Y}\,\sup_{\substack{\setA\subset\calX,\setB'\subset\calZ}} \left|\frac{\log Q_X(\setA)\int dQ_Y(y)P_{Z|Y=y}(\setB')}{\log \int dP_{XY}(x,y) 1(x\in \setA) P_{Z|Y=y}(\setB')}-1\right|^+\label{eq:DPI3}
\\&\le \inf_{Q_X,Q_Y}\, \sup_{\setA\subset\calX}\, \sup_{g:\calY\to[0,1]} \left| \frac{\log Q_X(\setA)\,\bbE_Q [g(Y)]}{\log \bbE_P [1(X\in \setA) g(Y)]}-1\right|^+\label{eq:DPI4}
\end{align}
where \eqref{eq:DPI2} holds because for any $Q_Y$, $Q_Z=\int dQ_Y(y) P_{Z|Y=y}$ is a valid distribution on $\calZ$, in the denominator of \eqref{eq:DPI3} we have used the fact that $X-Y-Z$ is a Markov chain, and \eqref{eq:DPI4} holds because in \eqref{eq:DPI3} we may take $g(y)=P_{Z|Y=y}(\setB')$ which is feasible for the supremum over $g$ in \eqref{eq:DPI4}. 
For fixed $Q_X$, $Q_Y$, and $\setA$, define
\be
G=\sup_{g:\calY\to[0,1]}\left|\frac{\log Q_X(\setA)\,\bbE_Q [g(Y)]}{\log \bbE_P [1(X\in \setA) g(Y)]}-1\right|^+.
\ee
We may also define
\be\label{eq:Gprime_def}
\new{G'=\sup_{\setB\subset\calY} \left|\frac{\log Q_X(\setA)\,Q_Y(\setB)}{\log P_{XY}(\setA,\setB)}-1\right|^+.}
\ee
\new{To complete the proof, it is enough to show that $G\le G'$.}
Rearranging \eqref{eq:Gprime_def}, for any \new{$\setB\subset\calY$},
\be\label{eq:G_condition}
\new{P_{X,Y}(\setA,\setB)^{1+G'}\le Q_X(\setA)Q_Y(\setB).}
\ee
\new{For any function $g:\calY\to[0,1]$, define the sets $\setB_t=\{y:g(y)<t\}$. Thus
\be
g(y)=\int_{0}^1 1(y\in \setB_t) dt.
\ee
Since $G'\ge 0$, $f(z)=z^{1+G'}$ is a convex function, which allows us to write
\begin{align}
(\bbE_P[1(X\in \setA) g(Y)])^{1+G'}
&=\left(\bbE_P\left[1(X\in \setA) \int_0^1 1(Y\in \setB_t)dt\right]\right)^{1+G'}
\\&\le \int_0^1 dt (\bbE_P[1(X\in \setA)1(Y\in \setB_t)])^{1+G'}\label{eq:DPI_convexity1}
\\&=\int _0^1  P_{XY}(\setA,\setB_t)^{1+G'} dt
\\&\le \int_0^1 Q_X(\setA)Q_Y(\setB_t)dt\label{eq:DPI_convexity}
\\&=Q_X(\setA) \int_0^1 \bbE_Q[Y\in \setB_t]dt
\\&=Q_X(\setA) \bbE_Q [g(Y)]\label{eq:DPI_convexity_end}
\end{align}
where \eqref{eq:DPI_convexity1} follows from Jensen's inequality and the fact that $\int_0^1 dt=1$, and \eqref{eq:DPI_convexity} follows from \eqref{eq:G_condition}. Since \eqref{eq:DPI_convexity_end} holds for all functions $g$, this implies $G\le G'$, which completes the proof.}
\end{IEEEproof}

\subsection{The Wringing Lemma}\label{sec:wringing_lemma}

The following result is our counterpart of Ahlswede's Lemma 4 from \cite{Ahlswede1982}.

\begin{lemma}\label{lemma:wringing}
Let $P_{XY}\in\calP(\calX\times\calY)$, $Q_X\in\calP(\calX)$, and $Q_Y\in\calP(\calY)$ be distributions such that
\be
D_\infty(P_{XY}\|Q_XQ_Y)\le \sigma
\ee
where $\sigma$ is finite. For any $\delta>0$, there exist sets $\bar\calX\subset\calX,\bar\calY\subset\calY$ such that
\be\label{eq:bar_sets_prob_bd}
P_{XY}(\bar\calX,\bar\calY)\ge \exp\left\{-\frac{\sigma}{\delta}\right\}
\ee
and
\be\label{eq:near_independence}
\Delta(\barX;\barY)\le \delta
\ee
where $(\barX,\barY)$ are distributed according to $P_{XY|X\in\bar\calX,Y\in\bar\calY}$.
\end{lemma}

As we outlined in Sec.~\ref{sec:motivation}, Ahlswede's proof of \cite[Lemma~4]{Ahlswede1982} involved iteratively restricting the wringing sets until the desired property is achieved. While a proof of Lemma~\ref{lemma:wringing} along these lines would work for discrete variables, it does not directly generalize to arbitrary variables. Instead, we present a slightly different proof that does work in general.

\begin{IEEEproof}[Proof of Lemma~\ref{lemma:wringing}]
Let $\scA$ be the collection of pairs of sets $(\setA,\setB)$ where $\setA\subset\calX,\setB\subset\calY$ such that $P_{XY}(\setA,\setB)>0$ and
\be\label{eq:scS_def}
P_{XY}(\setA,\setB)^{1+\delta}\ge Q_X(\setA)Q_Y(\setB).
\ee
This set $\scA$ is always non-empty, since it includes $(\setA,\setB)=(\calX,\calY)$. For any $(\setA,\setB)\in\scA$, using the assumption that $P_{XY}(\setA,\setB)>0$, we may rearrange \eqref{eq:scS_def} to write
\begin{align}
P_{XY}(\setA,\setB)&\ge \left(\frac{Q_X(\setA)Q_Y(\setB)}{P_{XY}(\setA,\setB)}\right)^{1/\delta}
\\&\ge\exp\left\{-\frac{\sigma}{\delta}\right\}\label{eq:scS_prob_bd}
\end{align}
where the second inequality follows from the assumption that $D_{\infty}(P_{XY}\|Q_XQ_Y)\le\sigma$.

We proceed to construct a pair of sets $(\bar\calX,\bar\calY)\in\scA$ that satisfy the following property:
\be\label{eq:XbarYbar_prop}
\text{for all }\setA\subset\bar\calX,\setB\subset\bar\calY,\text{ if }P_{XY}(\setA,\setB)<P_{XY}(\bar\calX,\bar\calY)\text{ then }(\setA,\setB)\notin\scA.
\ee
These sets can be easily found if the infimum is attained in
\be\label{eq:inf_attained}
\inf_{(\setA,\setB)\in\scA} P_{XY}(\setA,\setB).
\ee
That is, if there exist $(\bar\calX,\bar\calY)\in\scA$ such that $P_{XY}(\bar\calX,\bar\calY)\le P_{XY}(\setA,\setB)$ for all $(\setA,\setB)\in\scA$, then \eqref{eq:XbarYbar_prop} follows easily. Note that the infimum in \eqref{eq:inf_attained} is always attained if $\calX,\calY$ are finite sets. However, if this infimum is not attained we need a different argument.

We create a sequence of pairs of sets $(\setA_k,\setB_k)\in\scA$ for each non-negative integer $k$, as follows. First let $(\setA_0,\setB_0)=(\calX,\calY)$. For any $k\ge 1$, given $(\setA_{k-1},\setB_{k-1})$, define $(\setA_k,\setB_k)$ as follows. Let
\be\label{eq:pk_def}
p_k=\inf_{\setA\subset \setA_{k-1},\setB\subset \setB_{k-1}:(\setA,\setB)\in\scA} P_{XY}(\setA,\setB).
\ee
Let $\setA_k\subset \setA_{k-1},\setB_k\subset \setB_{k-1}$ be such that $(\setA_k,\setB_k)\in\scA$ and
\be\label{eq:AkBk_def}
P_{XY}(\setA_k,\setB_k)\le p_k+\frac{1}{k}.
\ee
This iteratively defines the sets $\setA_k,\setB_k$ for all $k$. We now define
\be
\bar\calX=\bigcap_{k\ge 0} \setA_k,\quad \bar\calY=\bigcap_{k\ge 0} \setB_k.
\ee
We need to prove that $(\bar\calX,\bar\calY)\in\scA$ and that \eqref{eq:XbarYbar_prop} is satisfied. By the dominated convergence theorem,
\be
P_{XY}(\bar\calX,\bar\calY)=\lim_{k\to\infty} P_{XY}(\setA_k,\setB_k),
\quad
Q_X(\bar\calX)=\lim_{k\to\infty} Q_X(\setA_k),
\quad
Q_Y(\bar\calY)=\lim_{k\to\infty} Q_Y(\setB_k).
\ee
These limits imply that $\bar\calX,\bar\calY$ satisfy \eqref{eq:scS_def}. Moreover, since $(\setA_k,\setB_k)\in\scA$ for each $k$, the lower bound in \eqref{eq:scS_prob_bd} implies that $P_{XY}(\setA_k,\setB_k)\ge \exp\{-\frac{\sigma}{\delta}\}$, so $P_{XY}(\bar\calX,\bar\calY)$ is bounded away from $0$. Thus $(\bar\calX,\bar\calY)\in\scA$. To prove \eqref{eq:XbarYbar_prop}, consider any $\setA\subset\bar\calX,\setB\subset\bar\calY$ where $P_{XY}(\setA,\setB)<P_{XY}(\bar\calX,\bar\calY)$. Note that 
\be\label{eq:k_limit}
\lim_{k\to\infty} \left[P_{XY}(\setA_k,\setB_k)-\frac{1}{k}\right]=P_{XY}(\bar\calX,\bar\calY).
\ee
Thus, there exists a finite $k$ such that $P_{XY}(\setA,\setB)<P_{XY}(\setA_k,\setB_k)-\frac{1}{k}$. By \eqref{eq:AkBk_def}, this implies that $P_{XY}(\setA,\setB)<p_k$, which means $\setA,\setB$ cannot be feasible for the infimum defining $p_k$ in \eqref{eq:pk_def}. In particular, since $\setA\subset\bar\calX\subset \setA_{k-1}$ and $\setB\subset\bar\calY\subset \setB_{k-1}$, it must be that $(\setA,\setB)\notin\scA$. This proves the desired property of $(\bar\calX,\bar\calY)$ in \eqref{eq:XbarYbar_prop}.

Given \eqref{eq:XbarYbar_prop}, we now complete the proof. Since $(\bar\calX,\bar\calY)\in\scA$, we immediately have the probability bound in \eqref{eq:bar_sets_prob_bd}. We now need to prove the bound on the wringing dependence in \eqref{eq:near_independence}. To show that $\Delta(\barX;\barY)\le\delta$, it is enough to show that for all $\setA\subset\calX,\setB\subset\calY$,
\be\label{eq:PQ_XbarYbar}
P_{XY|X\in\bar\calX,Y\in\bar\calY}(\setA,\setB)^{1+\delta}\le Q_{X|X\in\bar\calX}(\setA)\,Q_{Y|Y\in\bar\calY}(\setB).
\ee
Letting $\setA'=\setA\cap\bar\calX,\setB'=\setB\cap\bar\calY$, we have
\be
P_{XY|X\in\bar\calX,Y\in\bar\calY}(\setA,\setB)=\frac{P_{XY}(\setA',\setB')}{P_{XY}(\bar\calX,\bar\calY)},
\qquad Q_{X|X\in\bar\calX}(\setA)=\frac{Q_X(\setA')}{Q_X(\bar\calX)},
\qquad Q_{Y|Y\in\bar\calY}(\setB)=\frac{Q_Y(\setB')}{Q_Y(\bar\calX)}.
\ee
Consider the case that $P_{XY}(\setA',\setB')=P_{XY}(\bar\calX,\bar\calY)$. Since $\setA'\subset\bar\calX,\setB'\subset\bar\calY$, we must have $P_{XY}((\bar\calX\times\bar\calY)\setminus(\setA'\times \setB'))=0$. By the assumption that $\sigma$ is finite, $P_{XY}\ll Q_XQ_Y$, so in particular $Q_XQ_Y((\bar\calX\times\bar\calY)\setminus(\setA'\times \setB'))=0$, and thus $Q_X(\setA')Q_Y(\setB')=Q_X(\bar\calX)Q_Y(\bar\calY)$. Thus, each side of \eqref{eq:PQ_XbarYbar} equals $1$, so the inequality holds. Now consider the case that $P_{XY}(\setA',\setB')=0$. This implies that the LHS of \eqref{eq:PQ_XbarYbar} is $0$, so it holds trivially. 

The remaining case is when $0<P_{XY}(\setA',\setB')<P_{XY}(\bar\calX,\bar\calY)$. By the key property of $(\bar\calX,\bar\calY)$ in \eqref{eq:XbarYbar_prop}, we must have $(\setA',\setB')\notin\scA$. Thus
\begin{align}
P_{XY|X\in\bar\calX,Y\in\bar\calY}(\setA,\setB)^{1+\delta}
&=\frac{P_{XY}(\setA',\setB')^{1+\delta}}{P_{XY}(\bar\calX,\bar\calY)^{1+\delta}}
\\&<\frac{Q_X(\setA')Q_Y(\setB')}{P_{XY}(\bar\calX,\bar\calY)^{1+\delta}}\label{eq:AB_derive1}
\\&\le \frac{Q_X(\setA')Q_Y(\setB')}{Q_X(\bar\calX)Q_Y(\bar\calY)}\label{eq:AB_derive2}
\\&=Q_{X|X\in\bar\calX}(\setA)\,Q_{Y|Y\in\bar\calY}(\setB)
\end{align}
where \eqref{eq:AB_derive1} follows because $(\setA',\setB')\notin\scA$ and $P_{XY}(\setA',\setB')>0$, which imply that \eqref{eq:scS_def} must be violated; and \eqref{eq:AB_derive2} follows because $(\bar\calX,\bar\calY)\in\scA$. This proves \eqref{eq:PQ_XbarYbar} for all $\setA\subset\calX,\setB\subset\calY$.
\end{IEEEproof}

\subsection{Relationship to Other Dependence Measures}\label{sec:other_measures}

\subsubsection{Hypercontractivity}\label{sec:hypercontractivity}

One of the first uses of hypercontractivity in information theory was \cite{Ahlswede1976a}, wherein Ahlswede and G\'acs were interested in establishing conditions under which random variables $X,Y$ satisfy
\be\label{eq:sigma_tau}
P_{XY}(\setA,\setB)\le P_X(\setA)^{\sigma} P_Y(\setB)^{\tau}\text{ for all }\setA\subset\calX,\setB\subset\calY.
\ee
To establish this inequality, they actually proved something stronger, namely
\be\label{eq:hyper0}
\bbE [f(X)g(Y)]\le \|f(X)\|_{1/\sigma} \|g(Y)\|_{1/\tau}\text{ for all }f:\calX\to\bbR,g:\calY\to\bbR
\ee
where for a real-valued \new{random} variable $Z$, $\|Z\|_r=(\bbE [|Z|^r])^{1/r}$. By optimizing over $f$, one finds that \eqref{eq:hyper0} is equivalent to
\be\label{eq:hyper1}
\|\bbE[g(Y)|X]\|_{1/(1-\sigma)}\le \|g(Y)\|_{1/\tau}\text{ for all }g:\calY\to\bbR.
\ee
Such an inequality is known as \emph{hypercontractivity}. If the inequality is reversed, it is known \emph{reverse hypercontractivity} \cite{Mossel2013}. The advantage of working with hypercontractivity rather than the more operationally meaningful inequality \eqref{eq:sigma_tau} is that hypercontractivity tensorizes: that is, if \eqref{eq:hyper1} holds for $X,Y$, then it also holds for $X^n,Y^n$ where $(X_t,Y_t)$ are i.i.d. with the same distribution as $X,Y$.

The relationship between hypercontractivity and wringing dependence is apparent from \eqref{eq:sigma_tau}; namely this inequality is identical to the inequality defining the wringing dependence in \eqref{eq:Delta_def0} but with $Q_X=P_X,Q_Y=P_Y$, and $\sigma=\tau=1/(1+\delta)$. We make this relationship precise as follows.

For a pair of random variables $X,Y$, \cite{Kamath2012} defined the \emph{hypercontractivity ribbon} $\calR_{X;Y}$ as the set of pairs $(r,s)$ where one of the following hold:
\begin{itemize}
\item $1\le s\le r$, and for all $g:\calY\to\bbR$,
\be\label{eq:hyper}
\|\bbE[g(Y)|X]\|_r\le \|g(Y)\|_s,
\ee
\item $1\ge s\ge r$, and for all $g:\calY\to\bbR_+$,
\be\label{eq:reverse_hyper}
\|\bbE[g(Y)|X]\|_r\ge \|g(Y)\|_s.
\ee
\end{itemize}
The second condition concerns reverse hypercontractivity, which does not appear to be related to the wringing dependence, but we have included it for completeness. The following proposition, which is proved in Appendix~\ref{appendix:hyper} connects the wringing dependence to the hypercontractivity ribbon.

\begin{proposition}\label{prop:hypercontractivity}
Given random variables $X,Y$, let
\be\label{eq:Delta_hyp_def}
\Delta_{\text{hyp}}(X;Y)=\inf\{\delta\in[0,1]: (1+1/\delta,1+\delta)\in\calR_{X;Y}\}.
\ee
Then
\be\label{eq:hyp_upper_bd}
\Delta(X;Y)\le\Delta_{\text{hyp}}(X;Y).
\ee
Moreover, if we let $X^n,Y^n$ be jointly i.i.d.\ where $P_{X_tY_t}=P_{XY}$ for each $t\in[n]$, then $\Delta(X^n;Y^n)$ is a non-decreasing sequence such that
\be\label{eq:hyp_limit}
\lim_{n\to\infty} \Delta(X^n;Y^n)=\Delta_{\text{hyp}}(X;Y).
\ee
\end{proposition}

Note that the quantity $\Delta_{\text{hyp}}(X;Y)$ defined in \eqref{eq:Delta_hyp_def} involves checking whether $(r,s)\in\calR_{X;Y}$ where $r=1+1/\delta$ and $s=1+\delta$ for some $\delta\in[0,1]$; this is the regime where $1\le s\le r$, which corresponds to hypercontractivity rather than reverse hypercontractivity. The proof of the upper bound on wringing dependence in \eqref{eq:hyp_upper_bd} follows from essentially the same argument as the one \cite{Ahlswede1976a} used to establish inequalities of the form \eqref{eq:sigma_tau} via hypercontractivity. The limiting behavior of the wringing dependence in \eqref{eq:hyp_limit} is proved by an argument very similar to that of \cite{Nair2014}, which gives several equivalent characterizations of the hypercontractivity ribbon.

We illustrate Prop.~\ref{prop:hypercontractivity} with two examples: the doubly-symmetric binary source, and bivariate Gaussians. For the DSBS, $\Delta_{\text{hyp}}(X;Y)$ is shown to be strictly larger than the wringing dependence, and so \eqref{eq:Delta_hyp_def} is a loose bound. For bivariate Gaussians, \eqref{eq:Delta_hyp_def} gives a tight bound. In fact, the wringing dependence for bivariate Gaussians is quite difficult to compute directly from the definition, but Prop.~\ref{prop:hypercontractivity} allows us to find it exactly: for bivariate Gaussians with correlation coefficient $\rho$, $\Delta(X;Y)=|\rho|$. This establishes that the last of R\'enyi's axioms from \cite{Renyi1959} holds for wringing dependence.

\begin{example}[DSBS]
Let $(X,Y)$ be a DSBS with parameter $p$ as in Example~\ref{example:DSBS}. In \cite{Kamath2012}, it was established that the hypercontractivity ribbon consists of the pairs $(r,s)$ where either $(1-2p)^2 (r-1)+1\le s\le r$ or $r\le s\le (1-2p)^2(r-1)+1$. In particular, $(1+1/\delta,1+\delta)\in\calR_{X;Y}$ iff
\be
(1-2p)^2 \frac{1}{\delta}+1\le 1+\delta
\ee
which holds if $\delta\ge |1-2p|$. Therefore, $\Delta_{\text{hyp}}(X;Y)=|1-2p|$. Note that this quantity is strictly smaller than the the wringing dependence as calculated in Example~\ref{example:DSBS}, except for the trivial cases where $p\in\{0,1/2,1\}$.
\end{example}

\begin{example}[Bivariate Gaussians] Let $(X,Y)$ have a bivariate Gaussian distribution with correlation coefficient $\rho$. We claim that $\Delta(X;Y)=|\rho|$. Without loss of generality, we may assume that $X,Y$ each have zero mean, and covariance matrix
\be
\left[\begin{array}{cc} 1 & \rho \\ \rho & 1\end{array}\right].
\ee
We may assume that $\rho \ge 0$, since if not we may simply replace $Y$ with $-Y$. We upper bound $\Delta(X;Y)$ via Prop.~\ref{prop:hypercontractivity}. A result originally by Nelson \cite{Nelson1973}, which is also a consequence of the Gaussian log-Sobolev inequality \cite{Gross1975}, is that for any function $g:\bbR\to\bbR$, \eqref{eq:hyper} holds for $r\ge s\ge 1$ if $\rho\le \sqrt{(s-1)/(r-1)}$. (See \cite[Sec.~3.2]{Raginsky2013} for an information-theoretic treatment of this inequality.) Thus, with $r=1+1/\delta$ and $s=1+\delta$, $(r,s)\in\calR_{X;Y}$ if $\rho\le \delta$. Therefore $\Delta_{\text{hyp}}(X;Y)\le\rho$, and so $\Delta(X;Y)\le \rho$ by Prop.~\ref{prop:hypercontractivity}.

We now show that $\Delta(X;Y)\ge\rho$. If $\rho=1$, then $X=Y$, so $\Delta(X;Y)=1$. Now suppose that $\rho<1$. Let $\delta=\Delta(X;Y)$. Applying \eqref{eq:dep_pp_bd} from Thm.~\ref{thm:props}, for any $\setA,\setB\subset\bbR$
\be\label{eq:gaussian_pp_bd}
P_{XY}(\setA,\setB)\le (1+2\delta)(P_X(\setA)P_Y(\setB))^{1/(1+\delta)}.
\ee
In particular, for a parameter $a\ge 0$ (we will eventually take the limit $a\to\infty$), we may choose $\setA=\setB=[a,a+1]$. Let $\phi(x)$ be the standard Gaussian PDF. Since $\phi(x)$ is decreasing for $x\in[a,a+1]$, we have
\be
P_X(\setA)=P_Y(\setB)=\int_{a}^{a+1} \phi(x)dx\le \phi(a).
\ee
The joint PDF of $(X,Y)$ is
\be
f_{XY}(x,y)=\frac{1}{2\pi\sqrt{1-\rho^2}} \exp\left\{-\frac{x^2+y^2-2\rho xy}{2(1-\rho^2)}\right\}.
\ee
In particular, $f_{XY}(x,y)$ is decreasing in $x$ and $y$ if $x\ge \rho y$ and $y\ge \rho x$. From the assumption that $\rho<1$, these conditions hold for all $x,y\in[a,a+1]$ for sufficiently large $a$. Thus
\be
P_{XY}(\setA,\setB)=\int_{a}^{a+1}dx\int_{a}^{a+1}dy\, f_{XY}(x,y)\ge f_{XY}(a+1,a+1).
\ee
Plugging into \eqref{eq:gaussian_pp_bd} gives
\be
\frac{1}{2\pi\sqrt{1-\rho^2}} \exp\left\{-\frac{(a+1)^2(1-\rho)}{1-\rho^2}\right\}
\le (1+2\delta)\exp\left\{-\frac{a^2}{1+\delta}\right\}.
\ee
Thus
\be
-\frac{(a+1)^2}{1+\rho}-\log(2\pi\sqrt{1-\rho^2})\le -\frac{a^2}{1+\delta}+\log(1+2\delta).
\ee
Dividing by $a^2$ and taking a limit as $a\to\infty$ gives $\rho\le\delta$. That is, $\Delta(X;Y)\ge\rho$.
\end{example}

\subsubsection{Maximal Correlation}

The maximal correlation, which was introduced in \cite{Hirschfeld1935,Gebelein1941} and further studied in \cite{Renyi1959}, is given by
\be\label{eq:max_correlation_def}
\rho_m(X;Y)=\sup_{f,g} \rho(f(X);g(Y))
\ee
where the supremum is over all real-valued functions $f:\calX\to\bbR$ and $g:\calY\to\bbR$ such that $f(X)$ and $g(Y)$ have finite, non-zero variances, and $\rho(\cdot;\cdot)$ is the correlation coefficient. The maximal correlation shares much in common with the wringing dependence: in particular, both satisfy all 7 axioms from \cite{Renyi1959}. Moreover, the maximal correlation provides a simple bound on the hypercontractivity ribbon (see \cite{Kamath2012}); this implies that $\Delta_\text{hyp}(X;Y)\ge \rho_m(X;Y)$, where $\Delta_{\text{hyp}}$ is defined in \eqref{eq:Delta_hyp_def}. The following result, proved in Appendix~\ref{appendix:max_corr}, shows that if the wringing dependence is small, then the maximal correlation is also small. 

\begin{lemma}\label{lemma:max_corr}
If $\Delta(X;Y)\le\delta$, then the maximal correlation is bounded by
\be
\rho_m(X;Y)\le O(\delta\log \delta^{-1}).
\ee
\end{lemma}

This result will be particularly useful when addressing the Gaussian MAC; see Sec.~\ref{sec:gaussian}. Unfortunately, the bound in Lemma~\ref{lemma:max_corr} is not linear; in fact, no universal bound of the form $\rho_m(X;Y)\le K\,\Delta(X;Y)$ is possible.\footnote{If there were such a bound, analyzing the Gaussian MAC would dramatically simplify.} 
This is illustrated in the following example. This example also shows that Lemma~\ref{lemma:max_corr} is order-optimal; in fact, for any $0<c<1$ and any $\delta>0$, there exists a distribution $P_{XY}$ where $\Delta(X;Y)\le\delta$ and
\be\label{eq:binary_example_bd}
\rho_m(X;Y)\ge c\, \delta\log \delta^{-1}.
\ee

\begin{figure}
\begin{center}
\includegraphics[width=4in]{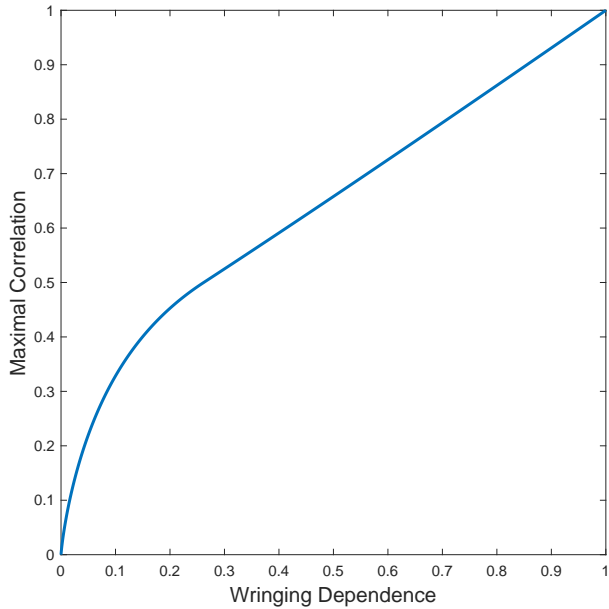}
\caption{The relationship between wringing dependence and maximal correlation for Example~\ref{example:max_corr}, plotted across the range of $a\in[0,1/2]$. Of particular note about this example is that, in the vicinity of the point $(0,0)$, the slope of the curve is infinite.}
\label{fig:max_corr_example}
\end{center}
\end{figure}

\begin{example}\label{example:max_corr}
For any $a\in[0,1/2]$, let $X,Y$ be binary variables with joint PMF given by
\vspace{1ex}
\begin{center}
\begin{tabular}{c|c c}
\diagbox{$Y$}{$X$} & $0$ & $1$\\ \hline
$0$ & $1-2a$ & $a$\\ 
$1$ & $a$ & $0$
\end{tabular}
\end{center}
\vspace{1ex}
Note that $P_X=P_Y=\text{Ber}(a)$. We first calculate the maximal correlation. Since $X,Y$ are both binary, the only nontrivial functions of them are the identity function and its complement, so
\be
\rho_m(X;Y)=|\rho(X;Y)|=\frac{|\bbE[ XY]-\bbE [X]\,\bbE [Y]|}{\sqrt{\var(X)\var(Y)}}=\frac{a^2}{a(1-a)}=\frac{a}{1-a}.
\ee
To compute the wringing dependence, recall that the function of $(Q_X,Q_Y)$ in the definition in \eqref{eq:Delta_def} is concave. Since $X$ and $Y$ have the same distribution, the optimal choice has $Q_X=Q_Y$. If we let $Q_X=Q_Y=\text{Ber}(q)$, then we see that wringing dependence between $X$ and $Y$ is 
\be
\Delta(X;Y)=\inf_{q\in[0,1]} \max\left\{\frac{\log q(1-q)}{\log a},\,\frac{\log(1-q)^2}{\log(1-2a)}\right\}-1.
\ee
While there is no simpler closed-form expression, this quantity can be easily computed. Fig.~\ref{fig:max_corr_example} shows the relationship between maximal correlation and wringing dependence across the range of $a$. To analytically establish that this example satisfies the claim \eqref{eq:binary_example_bd}, we may upper bound the wringing dependence by plugging in $q=a$, to find
\begin{align}
\Delta(X;Y)&\le \max\left\{\frac{\log a(1-a)}{\log a},\frac{\log(1-a)^2}{\log(1-2a)}\right\}-1
\\&=\frac{\log a(1-a)}{\log a}-1
\\&=\frac{\log (1-a)}{\log a}.
\end{align}
Thus
\begin{align}
\lim_{a\to 0} \frac{\Delta(X;Y)\log \Delta(X;Y)^{-1}}{\rho_m(X;Y)}
&\le \lim_{a\to 0} \frac{1-a}{a}\, \frac{\log(1-a)}{\log a} \log\left( \frac{\log a}{\log(1-a)}\right) 
\\&=\lim_{a\to 0} (1-a)\cdot \frac{-\log(1-a)}{a}\cdot \frac{\log(-\log a)-\log(-\log(1-a))}{-\log a}.\label{eq:binary_example_limit}
\end{align}
\new{We proceed to show that the limit as $a\to 0$ of each of the three multiplied terms in \eqref{eq:binary_example_limit} is $1$. The limit of the first term is certainly $1$; the limit of the second term can be seen to be $1$ by an application of L'Hopital's rule. For the third term, we have
\begin{align}
\lim_{a\to 0}\frac{\log(-\log a)-\log(-\log(1-a))}{-\log a}
&=\lim_{a\to 0} \frac{\frac{1}{a\log a}+\frac{1}{(1-a)\log(1-a)}}{-1/a}\label{eq:lhopital1}
\\&=\lim_{a\to 0} \left[\frac{-1}{\log a}-\frac{a}{(1-a)\log(1-a)}\right]\label{eq:lhopital2}
\\&=\lim_{a\to 0}\frac{-a}{(1-a)\log(1-a)}\label{eq:lhopital3}
\\&=\lim_{a\to 0}\frac{-1}{-\log(1-a)-1}\label{eq:lhopital4}
\\&=1\label{eq:lhopital5}
\end{align}
where \eqref{eq:lhopital1} and \eqref{eq:lhopital4} follow from L'Hopital's rule, and \eqref{eq:lhopital3} holds since $\log a\to -\infty$.}
Therefore, for any $0<c<1$, there exists a sufficiently small $a$ such that \eqref{eq:binary_example_bd} holds.
\end{example}

Another interesting fact is that while Lemma~\ref{lemma:max_corr} upper bounds the maximal correlation by a function of the wringing dependence, no lower bound is possible. The follow example illustrates that the maximal correlation can be arbitrarily close to $0$ while the wringing dependence is arbitrarily close to $1$.

\begin{example}
Given parameter $a$, let $X,Y$ be binary variables with joint PMF given by
\vspace{1ex}
\begin{center}
\begin{tabular}{c|c c}
\diagbox{$Y$}{$X$} & $0$ & $1$\\ \hline
$0$ & $a$ & $a\log a^{-1}$\\ 
$1$ & $a\log a^{-1}$ & $1-a-2a\log a^{-1}$
\end{tabular}
\end{center}
\vspace{1ex}
We claim that as $a\to 0$, $\rho_m(X;Y)\to 0$ while $\Delta(X;Y)\to 1$. The maximal correlation can be computed as
\be
\rho_m(X;Y)
=\frac{a-(a+a\log a^{-1})^2}{(a+a\log a^{-1})(1-a-a\log a^{-1})}
=\frac{a-o(a)}{a\log a^{-1}+o(a\log a^{-1}))}=\frac{1-o(1)}{\log a^{-1}}
\ee
which vanishes as $a\to 0$. We may lower bound the wringing dependence by
\begin{align}
\Delta(X;Y)&\ge \inf_q \max\left\{\frac{\log q^2}{\log a},\, \frac{\log (1-q)^2}{\log (1-a-2a\log a^{-1})}\right\}-1\label{eq:ex_min_max0}
\\&=\sup_q \min\left\{\frac{\log q^2}{\log a},\, \frac{\log (1-q)^2}{\log (1-a-2a\log a^{-1})}\right\}-1\label{eq:ex_min_max}
\end{align}
where \eqref{eq:ex_min_max} holds since the first function inside the maximum in \eqref{eq:ex_min_max0} is decreasing in $q$ while the second function is increasing. We may now lower bound \eqref{eq:ex_min_max} by choosing $q=2a\log a^{-1}$, which gives
\be
\frac{\log q^2}{\log a}
=\frac{2\log (2a\log a^{-1})}{\log a}
=\frac{2\log a+2\log (2\log a^{-1})}{\log a}
=2-O\left(\frac{\log \log a^{-1}}{\log a^{-1}}\right)
\ee
and
\be
\frac{\log (1-q)^2}{\log (1-a-2a\log a^{-1})}
=\frac{2\log (1-2a\log a^{-1})}{\log (1-a-2a\log a^{-1})}
=\frac{4a\log a^{-1}+O(a^2\log^2 a^{-1})}{2a\log a^{-1}+O(a)}
= 2-O\left(\frac{1}{\log a^{-1}}\right).
\ee
Therefore, in the limit as $a\to 0$, \eqref{eq:ex_min_max} approaches $1$.
\end{example}

\section{Finite Blocklength Converse Bound}\label{sec:fbl}

Before stating our main finite blocklength bound, we need the following definition. Given distributions $P,Q_1,\ldots,Q_k$ on alphabet $\calX$, we define the achievable region for a hypothesis test between a simple hypothesis $P$ and the composite hypothesis $\{Q_1,\ldots,Q_k\}$ by the set
\be\label{eq:composite_hypothesis_testing}
\bbeta_\alpha(P,Q_1,\ldots,Q_k)=\bigcup_{\substack{T:\calX\to[0,1],\\ \bbE_P [T(X)]\ge\alpha}}\{(\beta_1,\ldots,\beta_k)\in[0,1]^k:\bbE_{Q_i}[T(X)]\le\beta_i\text{ for }i=1,\ldots,k\}.
\ee
The following is our finite blocklength converse bound for the MAC. It follows the same core steps as Ahlswede's proof from \cite{Ahlswede1982}, while using wringing dependence in the wringing step, and is also written in a one-shot manner.

\begin{theorem}\label{thm:FBL}
Suppose there exists an $(M_1,M_2,\eps)$ code for the one-shot MAC $W\in\calP(\calX\times\calY\to\calZ)$. For any $\lambda>\eps$, $\delta>0$, there exists a distribution $P_{XY}\in\calP(\calX\times\calY)$ where $\Delta(X;Y)\le\delta$, and for any $Q_Z\in\calP(\calZ),Q_{Z|Y}\in\calP(\calZ|\calY),Q_{Z|X}\in\calP(\calZ|\calX)$,
\begin{align}
\frac{1}{M_1M_2}&\ge \left(1-\frac{\eps}{\lambda}\right)^{1+1/\delta}
\bbE [\beta_{12}(X,Y)],\label{eq:FBL_MN}\\
\frac{1}{M_1}&\ge \left(1-\frac{\eps}{\lambda}\right)^{1+1/\delta}
\bbE[ \beta_{1}(X,Y)],\label{eq:FBL_M}\\
\frac{1}{M_2}&\ge \left(1-\frac{\eps}{\lambda}\right)^{1+1/\delta}
\bbE[ \beta_{2}(X,Y)]\label{eq:FBL_N}
\end{align}
where the expectations are with respect to $P_{XY}$, and for each $x,y$,
\be\label{eq:beta_condition}
(\beta_{12}(x,y),\beta_1(x,y),\beta_2(x,y))\in\bbeta_{1-\lambda}(W_{xy},Q_Z,Q_{Z|Y=y},Q_{Z|X=x}).
\ee
\end{theorem}

\begin{IEEEproof}
Consider a (stochastic) code given by encoders $P_{X|I_1}\in\calP([M_1]\to\calX)$ and $P_{Y|I_2}\in\calP([M_2]\to\calY)$, and decoder $P_{\hatI_1,\hatI_2|Z}\in\calP(\calZ\to [M_1]\times[M_2])$ with average probability of error at most $\eps$. Let $Q_X$ be the distribution induced on $X$ assuming $I_1$ is uniform on $[M_1]$; \new{i.e.,
\be\label{eq:QX_written_out}
Q_X(\setA)=\frac{1}{M_1}\sum_{i_1=1}^{M_1} P_{X|I_1=i_1}(\setA).
\ee
Let} $Q_Y$ be the \new{corresponding} distribution induced on $Y$ assuming $I_2$ is uniform on $[M_2]$. Also let $Q_{XY}=Q_XQ_Y$ be the product distribution. Let $\calE$ be the error event, that is
\be
\calE=\{(\hatI_1,\hatI_2)\ne (I_1,I_2)\}.
\ee
Given any $\lambda>\eps$, we may define the expurgation set by
\be
\Gamma=\{(x,y)\in\calX\times\calY:\bbP(\calE|X=x,Y=y)\le\lambda\}.
\ee
That is, $\Gamma$ is the set of transmitted pairs $(x,y)$ that give probability of error at most $\lambda$. From the assumption that the probability of error is at most $\eps$,
\begin{align}
\eps&\ge \bbP(\calE)
\\&\ge \bbP(\calE,(X,Y)\notin\Gamma)
\\&\ge(1-Q_{XY}(\Gamma))\lambda
\end{align}
so
\be\label{eq:Q_Gamma_bd}
Q_{XY}(\Gamma)\ge1-\frac{\eps}{\lambda}.
\ee
Let $P_{X'Y'}=Q_{XY|(X,Y)\in\Gamma}.$ We may bound the R\'enyi divergence between these two distributions by
\begin{align}
D_\infty(P_{X'Y'}\|Q_{XY})&=\sup_{F\subset\calX\times\calY} \log \frac{P_{X'Y'}(F)}{Q_{XY}(F)}
\\&=\sup_{F\subset\calX\times\calY} \log \frac{Q_{XY}(F\cap\Gamma)}{Q_{XY}(\Gamma)Q_{XY}(F)}
\\&\le -\log Q_{XY}(\Gamma)
\\&\le -\log\left(1-\frac{\eps}{\lambda}\right).
\end{align}
We may now apply Lemma~\ref{lemma:wringing} with $\sigma=-\log(1-\eps/\lambda)$ and any fixed $\delta>0$, to find sets $\bar\calX\subset\calX,\bar\calY\subset\calY$. Let $P_{XY}=P_{X'Y'|X'\in\bar\calX,Y'\in\bar\calY}$. From the lemma,
\begin{align}
\Delta(X;Y)&\le\delta,\\
P_{X'Y'}(\bar\calX,\bar\calY)&\ge \exp\{-\sigma/\delta\}.
\end{align}
Using an identical calculation to the earlier bound on R\'enyi divergence,
\begin{align}
D_\infty(P_{XY}\|Q_{XY})&\le -\log Q_{XY}(\Gamma\cap \bar\calX\times\bar\calY)
\\&=-\log Q_{XY}(\Gamma) P_{X'Y'}(\bar\calX,\bar\calY)
\\&\le \sigma+ \frac{\sigma}{\delta}
\\&=-\left(1+\frac{1}{\delta}\right)\log \left(1-\frac{\eps}{\lambda}\right).
\end{align}
Thus 
\be\label{eq:radon_nikodym_bd}
\frac{dP_{XY}}{dQ_{XY}}(x,y)\le \exp\{D_\infty(P_{XY}\|Q_{XY})\}\le \left(1-\frac{\eps}{\lambda}\right)^{-1-1/\delta}.
\ee
We now define a hypothesis testing function $T:\calX\times\calY\times\calZ\to[0,1]$ given by
\be
T(x,y,z)=\bbP(\calE^c|(X,Y,Z)=(x,y,z)).
\ee
From the definition of $\Gamma$, for any $(x,y)\in\Gamma$,
\be
\int dW_{xy}(z) T(x,y,z)=\bbP(\calE^c|(X,Y)=(x,y))\ge 1-\lambda.
\ee
Thus, by the definition of the hypothesis testing quantity in \eqref{eq:composite_hypothesis_testing}, for any $Q_Z,Q_{Z|Y},Q_{Z|X}$, \eqref{eq:beta_condition} holds with
\begin{align}
\beta_{12}(x,y)&=\int dQ_Z(z)T(x,y,z),\\
\beta_1(x,y)&=\int dQ_{Z|Y=y}(z)T(x,y,z),\\
\beta_2(x,y)&=\int dQ_{Z|X=x}(z)T(x,y,z).
\end{align}
Thus
\begin{align}
\bbE[ \beta_{12}(X,Y)]
&=\int dP_{XY}(x,y) dQ_Z(z) T(x,y,z)
\\&\le \int dP_{XY}(x,y) dQ_Z(z)  \bbP(\calE^c|(X,Y,Z)=(x,y,z))
\\&\le \left(1-\frac{\eps}{\lambda}\right)^{-1-1/\delta} \int dQ_X(x)dQ_Y(y) dQ_Z(z) \bbP(\calE^c|(X,Y,Z)=(x,y,z))\label{eq:MN_indep0}
\\&\le  \left(1-\frac{\eps}{\lambda}\right)^{-1-1/\delta} \frac{1}{M_1M_2}\label{eq:MN_indep}
\end{align}
where \eqref{eq:MN_indep0} holds by the bound on the R\'enyi divergence from \eqref{eq:radon_nikodym_bd}, and \eqref{eq:MN_indep} holds because if $(X,Y,Z)\sim Q_XQ_YQ_Z$, then $(I_1,I_2)$ are uniformly random on $[M_1]\times [M_2]$ and $(\hatI_1,\hatI_2)$ are independent from them, so the probability of correct decoding is at most $\frac{1}{M_1M_2}$. Rearranging \eqref{eq:MN_indep} yields \eqref{eq:FBL_MN}.
By a nearly identical argument,
\begin{align}
\bbE[\beta_1(X,Y)]
&=\int dP_{XY}(x,y)dQ_{Z|Y=y}(z) T(x,y,z)
\\&\le\left(1-\frac{\eps}{\lambda}\right)^{-1-1/\delta} \int dQ_X(x) dQ_{Y}(y) dQ_{Z|Y=y}(z) \bbP(\calE^c|(X,Y,Z)=(x,y,z))
\\&\le \left(1-\frac{\eps}{\lambda}\right)^{-1-1/\delta} \frac{1}{M_1}\label{eq:M_indep}
\end{align}
where \eqref{eq:M_indep} holds because if $(X,Y,Z)\sim Q_X Q_Y Q_{Z|Y}$, then $I_1$ and $\hatI_1$ are independent. Rearranging yields \eqref{eq:FBL_M}. The same calculation for $\bbE[\beta_2(X,Y)]$ yields \eqref{eq:FBL_N}.
\end{IEEEproof}

\section{Asymptotic Results}\label{sec:asymptotics}

We present two asymptotic results, each characterizing the second-order rate as $O(1/\sqrt{n})$ under certain assumptions on the channel. The first result \new{(Thm.~\ref{thm:most_general}) aims to bound the second-order rate with minimal assumptions on the channel, while giving the simplest possible proof of the result. In particular, Thm.~\ref{thm:most_general} avoids an assumption on the third-moment of the information density. The second result (Thm.~\ref{thm:DMC_asymptotic}) applies only to MACs with finite alphabets, but it gives a substantially tighter bound on the second-order rate for these channels. Thm.~\ref{thm:DMC_asymptotic} is intended to give the tightest possible bound on the second-order rate, at the cost of a more complicated proof.}
We state both results first, and then prove them in Secs.~\ref{sec:general_proof} and~\ref{sec:dmc_proof}. \new{Sec.~\ref{sec:maximal} provides some discussion of the maximal probability of error case.}

For $\alpha_1\ge \alpha_2\ge 0$, and any $\delta\ge 0$, define
\begin{align}
C_{\alpha_1,\alpha_2}(\delta)
&=\sup_{\substack{P_{UXY}:\Delta(X;Y|U=u)\le\delta\text{ for all }u,\\ \bbE[ b_1(X)]\le B_1,\\ \bbE[ b_2(Y)]\le B_2}}
\big[\alpha_2 I(X,Y;Z|U)
+(\alpha_1-\alpha_2) I(X;Z|Y,U)\big].\label{eq:Cdelta_def}
\end{align}
For $\alpha_2\ge \alpha_1\ge 0$, we define $C_{\alpha_1,\alpha_2}(\delta)$ similarly, except there is a term with $I(Y;Z|X,U)$ in place of the $I(X;Z|Y,U)$ term. Note that $C_{\alpha_1,\alpha_2}(0)=C_{\alpha_1,\alpha_2}$. Also let $C'_{\alpha_1,\alpha_2}(\delta)$ be the derivative of $C_{\alpha_1,\alpha_2}(\delta)$ with respect to $\delta$. Since $C_{\alpha_1,\alpha_2}(\delta)$ is non-decreasing in $\delta$, $C'_{\alpha_1,\alpha_2}(\delta)$ is well-defined, although it may be infinite. Let
\be\label{eq:Vmax_def}
V_{\max}=\sup_{\substack{P_{UXY}:\\ \bbE[ b_1(X)]\le B_1, \\ \bbE[ b_2(Y)]\le B_2}}
\max\big\{V(W\|P_{Z|U}|P_{UXY}),\,V(W\|P_{Z|YU}|P_{UXY}),\,V(W\|P_{Z|XU}|P_{UXY})\big\}
\ee
where $P_{Z|U},P_{Z|YU},P_{Z|XU}$ are the induced distributions from $P_{UXY}$.
Note that in this definition, there is no independence constraint on $P_{UXY}$.

\begin{theorem}\label{thm:most_general}
For any $\alpha_1,\alpha_2$ where $\max\{\alpha_1,\alpha_2\}=1$, and any $\eps\in(0,1)$,
\be\label{eq:thm:most_general}
R_{\alpha_1,\alpha_2}^\star(n,\eps)\le C_{\alpha_1,\alpha_2}+\min_{\lambda\in(\eps,1)}\left[ 2\sqrt{C_{\alpha_1,\alpha_2}'(0)\log\frac{\lambda}{\lambda-\eps}}+\sqrt{\frac{V_{\max}}{1-\lambda}}\right]\frac{1}{\sqrt{n}}+o\left(\frac{1}{\sqrt{n}}\right).
\ee
\end{theorem}

The proof of this result, found in Sec.~\ref{sec:general_proof}, applies an Augustin-type argument (cf. \cite{Augustin1966}), wherein Chebyshev's inequality is used to bound the hypothesis testing fundamental limit. Thus, the bound is only meaningful if the second moment statistic $V_{\max}$ is finite, but there is no requirement on the third moment, which allows Thm.~\ref{thm:most_general} to hold in a great deal of generality, although it can typically be improved with more careful analysis. The following corollary comes by plugging in, for example, $\lambda=\frac{\eps+1}{2}$ into \eqref{eq:thm:most_general}.
\begin{corollary}\label{cor:regularity}
If (i) $V_{\max}<\infty$, and (ii) $C_{\alpha_1,\alpha_2}'(0)$ is uniformly bounded for all $\alpha_1,\alpha_2$ where $\max\{\alpha_1,\alpha_2\}=1$, then for any $\eps\in(0,1)$,
\be
\calR(n,\eps)\subseteq \calC+O\left(\frac{1}{\sqrt{n}}\right).
\ee
\end{corollary}

As seen from Corollary~\ref{cor:regularity}, the second-order coding rate is $O(1/\sqrt{n})$ as long as two regularity conditions hold. The condition on $V_{\max}$ is not surprising, as any result of this form requires that the information density has a finite second moment. One slight complication arises from the fact that, in the definition of $V_{\max}$ in \eqref{eq:Vmax_def}, one cannot choose the output distribution $P_{Z|U}$ separately from the input distribution. That is, even though in Thm.~\ref{thm:FBL} the distribution $Q_Z$ (and $Q_{Z|Y},Q_{Z|X}$) is a free choice, we select only the induced output distribution. This complicates the analysis for some channels; for example, for the Gaussian point-to-point channel, in the second-order converse bound one typically chooses an i.i.d. Gaussian for the output distribution, as in \cite[Sec.~III-J]{Polyanskiy2010a}. By contrast, here that choice is not available. This difficulty is addressed for the Gaussian MAC in Appendix~\ref{appendix:gaussian}.

The second regularity condition, on the boundedness of $C_{\alpha_1,\alpha_2}'(0)$, wherein the wringing dependence appears, is more particular to our method. Verifying this condition requires analyzing the effect of the wringing dependence between the two inputs on the maximum achievable weighted-sum-rate. In the sequel, we establish that this condition holds in two cases: for any discrete-memoryless channel, as shown in Thm.~\ref{thm:DMC_asymptotic}, and for the Gaussian MAC, as discussed in Sec.~\ref{sec:gaussian} with the proof in Appendix~\ref{appendix:gaussian}.

We now state a more precise result for discrete-memoryless channels, which will require a few new definitions. Let $\calP_{\alpha_1,\alpha_2}^{\text{in}}$ be the set of distributions $P_{UXY}$ satisfying the supremum in the characterization of $C_{\alpha_1,\alpha_2}$ in \eqref{eq:Ca1a2}. For any $\alpha\in[0,1]$, let
\be\label{eq:Vplus_def}
V_{1,\alpha}^+=\sup_{P_{UXY\in\calP_{1,\alpha}^{\text{in}}}}
\left[\alpha \sqrt{V(W\|P_{Z|U}|P_{UXY})}+(1-\alpha)\sqrt{V(W\|P_{Z|YU}|P_{UXY})}\right]^2
\ee
where $P_{Z|U}$ and $P_{Z|YU}$ are the induced distributions from $P_{UXY}$. Also let
\be\label{eq:Vminus_def}
V_{1,\alpha}^{-}=\inf_{P_{UXY},P_{X'Y'|U}}
\left[\alpha \sqrt{V(W\|P_{Z|U}|P_{UX'Y'})}+(1-\alpha)\sqrt{V(W\|P_{Z|YU}|P_{UX'Y'})}\right]^2
\ee
where the infimum is over all $P_{UXY}\in\calP_{1,\alpha}^{\text{in}}$ and $P_{X'Y'|U}$ satisfying
\be\label{eq:Vminus_feasibility}
\alpha D(W\|P_{Z|U}|P_{UX'Y'})+(1-\alpha) D(W\|P_{Z|YU}|P_{UX'Y'})=C_{1,\alpha}.
\ee
Define $V_{\alpha,1}^-$ and $V_{\alpha,1}^+$ analogously. For any $\alpha_1,\alpha_2$ where $\max\{\alpha_1,\alpha_2\}=1$ and any $\lambda\in(0,1)$, let
\be
V_{\alpha_1,\alpha_2}^{\lambda}=\begin{cases} V_{\alpha_1,\alpha_2}^{-}, & \lambda<1/2 \\ V_{\alpha_1,\alpha_2}^{+}, & \lambda\ge 1/2.\end{cases}
\ee
\begin{theorem}\label{thm:DMC_asymptotic}
If $\calX,\calY,\calZ$ are finite sets, then both regularity conditions in Corollary~\ref{cor:regularity} are satisfied. In addition, for any $\alpha_1,\alpha_2$ where $\max\{\alpha_1,\alpha_2\}=1$, and any $\eps\in(0,1)$,
\be\label{eq:DMC_asymptotic}
R_{\alpha_1,\alpha_2}^\star(n,\eps)\le \left(C_{\alpha_1,\alpha_2}+\min_{\lambda\in(\eps,1)}\left[ 2\sqrt{C'_{\alpha_1,\alpha_2}(0)\log\frac{\lambda}{\lambda-\eps}} -\sqrt{V_{\alpha_1,\alpha_2}^{\lambda}}\,\mathsf{Q}^{-1}(\lambda)\right]\frac{1}{\sqrt{n}}\right)^{**}+o\left(\frac{1}{\sqrt{n}}\right)
\ee
where $\mathsf{Q}$ is the Gaussian complementary CDF and $\mathsf{Q}^{-1}$ is its inverse function, and $(\cdot)^{**}$ represents the lower convex envelope as a function of $(\alpha_1,\alpha_2)$.
\end{theorem}

Note that $V_{\alpha_1,\alpha_2}^+$ and $V_{\alpha_1,\alpha_2}^-$ are not quite complementary. In particular, $V_{\alpha_1,\alpha_2}^-$ is in general smaller than the quantity obtained by simply replacing the supremum with an infimum in \eqref{eq:Vplus_def}. However, for at least some channels of interest, such as the binary additive erasure channel (see Sec.~\ref{sec:bamac}), all of these divergence variance quantities are equal.

Thm.~\ref{thm:DMC_asymptotic} settles the question, at least for some discrete channels, of whether the maximum achievable rates approach the capacity region from below or above for sufficiently small probability of error. We state this precisely in the following corollary.
\begin{corollary}
Let $\calX,\calY,\calZ$ be finite sets. If $V_{\alpha_1,\alpha_2}^{-}>0$, then for sufficiently small $\eps$ and sufficiently large $n$,
\be
R^\star_{\alpha_1,\alpha_2}(n,\eps)<C_{\alpha_1,\alpha_2}.
\ee
\end{corollary}
This corollary is proved by choosing, for example, $\lambda=2\eps$ in \eqref{eq:DMC_asymptotic} and taking $\eps$ to be sufficiently small.

\subsection{Proof of Thm.~\ref{thm:most_general}}\label{sec:general_proof}

Consider any $(n,M_1,M_2,\eps)$ code for the $n$-length product channel. We consider $(\alpha_1,\alpha_2)=(1,\alpha)$ where $\alpha\in[0,1]$. The alternative case is proved identically. We apply Thm.~\ref{thm:FBL} wherein the one-shot input alphabets $\calX,\calY$ are replaced by the cost-constrained input sets
\be\label{eq:cost_constrained_inputs}
\left\{x^n\in\calX^n:\sum_{t=1}^n b_1(x_t)\le nB_1\right\},\quad
\left\{y^n\in\calY^n:\sum_{t=1}^n b_2(y_t)\le nB_2\right\}.
\ee
Thus, for any $\lambda>\eps,\delta>0$, there exists a distribution $P_{X^nY^n}$ such that $X^n$ and $Y^n$ fall into the sets in \eqref{eq:cost_constrained_inputs} almost surely, $\Delta(X^n;Y^n)\le\delta$, and
\begin{align}
\log (M_1M_2)&\le  -\log \bbE\big[\beta_{1-\lambda}(W_{X^nY^n},
{\textstyle\prod_{t=1}^n P_{Z_t}})\big]+\left(\frac{1}{\delta}+1\right)\log \frac{\lambda}{\lambda-\eps},\label{eq:nlength_FBL_bd}\\
\log M_1&\le  -\log \bbE\big[\beta_{1-\lambda}(W_{X^nY^n},
{\textstyle\prod_{t=1}^n P_{Z_t|Y_t=Y_t}})\big]+\left(\frac{1}{\delta}+1\right)\log \frac{\lambda}{\lambda-\eps},\label{eq:nlength_FBL_bd1}\\
\log M_2&\le  -\log \bbE\big[\beta_{1-\lambda}(W_{X^nY^n},
{\textstyle\prod_{t=1}^n P_{Z_t|X_t=X_t}})\big]+\left(\frac{1}{\delta}+1\right)\log \frac{\lambda}{\lambda-\eps}.\label{eq:nlength_FBL_bd2}
\end{align}
Here, we have relaxed Thm.~\ref{thm:FBL} by noting that if $(\beta_{1},\ldots,\beta_k)\in\bbeta_{1-\lambda}(P,Q_1,\ldots,Q_k)$, then $\beta_{i}\ge \beta_{1-\lambda}(P,Q_{i})$ for each $i\in[k]$. We have also chosen the induced product distributions for $Q_Z,Q_{Z|Y},Q_{Z|X}$. Since by Thm.~\ref{thm:props}, wringing dependence satisfies the data processing inequality, $\Delta(X_t;Y_t)\le\delta$ for any $t\in[n]$. We will make use of the $\eps$-information spectrum divergence (cf. \cite{Han2003,Tan2014a}), which is given by
\be
D_s^{\eps}(P\|Q)=\sup\left\{R\in\bbR: P\left(\log \frac{dP}{dQ}(Z)\le R\right)\le\eps\right\}.
\ee
The hypothesis testing quantity can be related to the information spectrum divergence as
\be
-\log \beta_{1-\lambda}(P,Q)\le \inf_{0<\eta<1-\lambda} \left[D_s^{\lambda+\eta}(P\|Q)-\log\eta\right].
\ee
Using Chebyshev's inequality, the information spectrum divergence may in turn be bounded by (see e.g., \cite[Prop.~2.2]{Tan2014a})
\be
D_s^{\eps}(P\|Q)\le D(P\|Q)+\sqrt{\frac{V(P\|Q)}{1-\eps}}
\ee
and so
\be\label{eq:beta_chebyshev}
-\log \beta_{1-\lambda}(P,Q)\le D(P\|Q)+\inf_{0<\eta<1-\lambda} \left(\sqrt{\frac{V(P\|Q)}{1-\lambda-\eta}}-\log\eta\right).
\ee
Applying \eqref{eq:beta_chebyshev} to the bound in \eqref{eq:nlength_FBL_bd} gives, for any $0<\eta<1-\lambda$,
\begin{align}
&\log(M_1M_2)-\left(\frac{1}{\delta}+1\right)\log\frac{\lambda}{\lambda-\eps}
\\&\le -\log \int dP_{X^nY^n}(x^n,y^n) \exp\Bigg\{-\sum_{t=1}^n D(W_{x_ty_t}\|P_{Z_t})
-\sqrt{\frac{1}{1-\lambda-\eta} \sum_{t=1}^n V(W_{x_ty_t}\|P_{Z_t})}+\log \eta\Bigg\}
\\&\le \sum_{t=1}^n D(W\|P_{Z_t}|P_{X_tY_t})+\sqrt{\frac{1}{1-\lambda-\eta}\sum_{t=1}^n V(W\|P_{Z_t}|P_{X_tY_t})}-\log\eta\label{eq:beta_bound2}
\\&= n D(W\|P_{Z|U}|P_{XYU})+\sqrt{\frac{n}{1-\lambda-\eta}V(W\|P_{Z|U}|P_{XYU})}-\log\eta\label{eq:beta_bound3}
\\&\le nI(XY;Z|U)+\sqrt{\frac{nV_{\max}}{1-\lambda-\eta}}-\log\eta\label{eq:beta_bound4}
\end{align}
where \eqref{eq:beta_bound2} holds by convexity of the exponential and concavity of the square root; in \eqref{eq:beta_bound3} we have let $U\sim\text{Unif}[n]$, $X=X_U,Y=Y_U,Z=Z_U$; and \eqref{eq:beta_bound4} follows from the definition of $V_{\max}$ in \eqref{eq:Vmax_def}. Applying the same derivation to \eqref{eq:nlength_FBL_bd1} gives
\be\label{eq:beta_bound5}
\log M_1-\left(\frac{1}{\delta}+1\right)\log\frac{\lambda}{\lambda-\eps}
\le nI(X;Z|YU)+\sqrt{\frac{nV_{\max}}{1-\lambda-\eta}}-\log\eta.
\ee
Recall that for each $t\in[n]$, $\Delta(X_t;Y_t)\le\delta$, which means that for each $u$, $\Delta(X;Y|U=u)\le\delta$. Moreover, by the fact that $X^n,Y^n$ fall into the cost-constrained sets in \eqref{eq:cost_constrained_inputs},
\begin{align}
\bbE[ b_1(X)]&=\frac{1}{n}\sum_{t=1}^n \bbE [b_1(X_t)]\le B_1,\\
\bbE[ b_2(Y)]&=\frac{1}{n}\sum_{t=1}^n \bbE [b_2(Y_t)]\le B_2.
\end{align}
Thus, from the definition of $C_{1,\alpha}(\delta)$ in \eqref{eq:Cdelta_def},
\be\label{eq:beta_bound6}
\alpha I(XY;Z|U)+(1-\alpha)I(X;Z|Y,U)\le C_{1,\alpha}(\delta)= C_{1,\alpha}+C_{1,\alpha}'(0)\,\delta+o(\delta)
\ee
where the equality follows from the definition of the derivative. We may combine \eqref{eq:beta_bound4} and \eqref{eq:beta_bound5}, then plug in \eqref{eq:beta_bound6} to find
\be
\log M_1+\alpha \log M_2\le nC_{1,\alpha}+nC_{1,\alpha}'(0)\delta+o(n\delta)+\sqrt{\frac{nV_{\max}}{1-\lambda-\eta}}-\log\eta+\left(\frac{1}{\delta}+1\right)\log \frac{\lambda}{\lambda-\eps}.\label{eq:pre_lambda_choice}
\ee
Recall that $\delta$ is a free parameter. The optimal choice (ignoring the $o(n\delta)$ term) is $\delta=\sqrt{\frac{\log\frac{\lambda}{\lambda-\eps}}{nC'_{1,\alpha}(0)}}$ which gives
\be\label{eq:post_delta_choice}
\log M_1+\alpha \log M_2
\le nC_{1,\alpha}+2\sqrt{nC'_{1,\alpha}(0)\log \frac{\lambda}{\lambda-\eps}}+\sqrt{\frac{nV_{\max}}{1-\lambda-\eta}}-\log\eta+\log \frac{\lambda}{\lambda-\eps}+o(\sqrt{n})
\ee

We now distinguish two cases. If $V_{\max}>0$, then the optimal value of $\lambda$ in the minimization in \eqref{eq:thm:most_general} is bounded away from $1$. Let $\lambda$ take on this optimal value, and we choose  $\eta=1/\sqrt{n}$ to give
\begin{align}
\log M_1+\alpha \log M_2
&\le 
nC_{1,\alpha}+2\sqrt{nC'_{1,\alpha}(0)\log\frac{\lambda}{\lambda-\eps}}+\sqrt{\frac{nV_{\max}}{1-\lambda}}+o(\sqrt{n}).
\end{align}
If alternatively $V_{\max}=0$, then the optimal value of $\lambda$ in the minimization in \eqref{eq:thm:most_general} is $\lambda=1$, but plugging $\lambda=1$ into \eqref{eq:post_delta_choice} does not quite work, because of the requirement that $\eta<1-\lambda$. Instead we may choose $\lambda=1-2/n$ and $\eta = 1/n$ to give
\begin{align}
\log M_1+\alpha \log M_2
&\le nC_{1,\alpha}+2\sqrt{nC'_{1,\alpha}(0)\log(1-\eps)^{-1}}+o(\sqrt{n}).
\end{align}

\subsection{Proof of Thm.~\ref{thm:DMC_asymptotic}} \label{sec:dmc_proof}

We will need the following lemma, which is proved in Appendix~\ref{appendix:dmc}.

\begin{lemma}\label{lemma:DMCs}
Consider a MAC where $\calX,\calY,\calZ$ are finite sets. Let $W_{\min}$ be the smallest non-zero value of $W_{xy}(z)$. Consider any random variables $X,Y$ with distribution $P_{XY}$ where $\Delta(X;Y)\le\delta$. Let $(\tilX,\tilY,\tilZ)\sim P_X P_Y W$. Then
\begin{align}
I(X,Y;Z)&\le I(\tilX,\tilY;\tilZ)
+\bigg[8\min\{|\calX|,|\calY|\}
+|\calZ|\left((1-\log W_{\min})e^{-1}+4e^{-2}\right)
+2\min\{|\calX|,|\calY|\}\log|\calZ|\bigg]\delta+O(\delta^2),\label{eq:MI_DMC_bound}
\\I(X;Z|Y)&\le I(\tilX;\tilZ|\tilY)
+\bigg[8\min\{|\calX|,|\calY|\}
+|\calY|\cdot|\calZ|\left((1-\log W_{\min})e^{-1}+4e^{-2}\right)
+2\min\{|\calX|,|\calY|\}\log|\calZ|\bigg]\delta+O(\delta^2),\label{eq:MIX_DMC_bound}
\\I(Y;Z|X)&\le I(\tilX;\tilZ|\tilY)
+\bigg[8\min\{|\calX|,|\calY|\}
+|\calX|\cdot|\calZ|\left((1-\log W_{\min})e^{-1}+4e^{-2}\right)
+2\min\{|\calX|,|\calY|\}\log|\calZ|\bigg]\delta+O(\delta^2).\label{eq:MIY_DMC_bound}
\end{align}
\end{lemma}

Lemma~\ref{lemma:DMCs} immediately gives that $C'_{\alpha_1,\alpha_2}(0)$ is uniformly bounded for any $\alpha_1,\alpha_2$ with $\max\{\alpha_1,\alpha_2\}=1$. To prove that $V_{\max}<\infty$, we 
note that for any distribution $P_{XY}$ and its induced distribution $P_Z$
\begin{align}
V(W\|P_Z|P_{XY})&\le \bbE \left[\log^2 \frac{W_{XY}(Z)}{P_Z(Z)}\right]
\\&\le \left(\sqrt{\bbE \left[\log^2 W_{XY}(Z)\right]}+\sqrt{\bbE [\log^2 P_Z(Z)]}\right)^2
\\&\le \left(2\sqrt{4e^{-2} |\calZ|}\right)^2
\\&=16 e^{-2}|\calZ|
\end{align}
where we have used the fact that $p\log^2 p\le 4e^{-2}$. By the same argument, $V(W\|P_{Z|Y}\|P_{XY}),V(W\|P_{Z|X}\|P_{XY})$ are also bounded by $16 e^{-2}|\calZ|$.

Recall that $R^\star_{\alpha_1,\alpha_2}(n,\eps)$, as defined in \eqref{eq:Rstar_def}, is the supremum of linear functions in $(\alpha_1,\alpha_2)$, so it is convex in $(\alpha_1,\alpha_2)$. Thus, to prove the theorem it is enough to show \eqref{eq:DMC_asymptotic} but without the lower convex envelope. We assume that $(\alpha_1,\alpha_2)=(1,\alpha)$ for $\alpha\in[0,1]$. We proceed with with the first step as in the proof of Thm.~\ref{thm:most_general}; namely from Thm.~\ref{thm:FBL} we derive \eqref{eq:nlength_FBL_bd}--\eqref{eq:nlength_FBL_bd2}. Combining \eqref{eq:nlength_FBL_bd} and \eqref{eq:nlength_FBL_bd1}, and using the fact that $p^\alpha q^{1-\alpha}$ is concave in $(p,q)$, gives
\be\label{eq:M1M2_combined_beta}
\log M_1+\alpha \log M_2\le 
 -\log \bbE \left[\left(\beta_{1-\lambda}(W_{X^nY^n},
{\textstyle\prod_{t=1}^n P_{Z_t}})\right)^\alpha \left(\beta_{1-\lambda}(W_{X^nY^n},
{\textstyle\prod_{t=1}^n P_{Z_t|Y_t=Y_t}})\right)^{1-\alpha}\right]+\left(\frac{1}{\delta}+1\right)\log \frac{\lambda}{\lambda-\eps}
\ee
\new{Since we will apply a Berry-Esseen bound to the hypothesis testing quantities, rather than a Chebyshev bound as in Thm.~\ref{thm:most_general}, we need to avoid some potentially badly-behaving $(x^n,y^n)$ sequences. In particular, define the set 
\be\label{eq:Omega_0_def}
\Omega_0=\left\{(x^n,y^n):P_{X_tY_t}(x_t,y_t)\le \frac{1}{n^2}\text{ for some }t\in[n]\right\}.
\ee
Let $p_0=P_{X^nY^n}(\Omega_0)$. By the union bound,
\begin{align}
p_0&\le \sum_{t=1}^n \bbP\left(P_{X_tY_t}(X_t,Y_t)\le \frac{1}{n^2}\right)
\\&= \sum_{t=1}^n \sum_{x,y} P_{X_tY_t}(x,y)\,1\left(P_{X_tY_t}(x,y)\le \frac{1}{n^2}\right)
\\&\le \frac{|\calX|\,|\calY|}{n}.\label{eq:p0_bound}
\end{align}
From the fact that the $\beta$ quantities are non-negative, we may further bound \eqref{eq:M1M2_combined_beta} by
\begin{align}
\log M_1+\alpha\log M_2
&\le  -\log \bbE \left[1((X^n,Y^n)\in\Omega_0^c)\big(\beta_{1-\lambda}(W_{X^nY^n},
{\textstyle\prod_{t=1}^n P_{Z_t}})\big)^\alpha \big(\beta_{1-\lambda}(W_{X^nY^n},
{\textstyle\prod_{t=1}^n P_{Z_t|Y_t=Y_t}})\big)^{1-\alpha}\right]\nonumber
\\&\qquad+\left(\frac{1}{\delta}+1\right)\log \frac{\lambda}{\lambda-\eps}.\label{eq:M1M2_combined_beta2}
\end{align}}
We now use the Berry-Esseen theorem via \cite[Prop.~2.1]{Tan2014a} to bound each of the hypothesis testing quantities in \eqref{eq:M1M2_combined_beta2}. Specifically, for any $x^n,y^n$
\new{\be\label{eq:beta_berry_esseen}
-\log \beta_{1-\lambda}(W_{x^ny^n},{\textstyle\prod_{t=1}^n P_{Z_t}})
\le \inf_{0<\eta\le 1-\lambda} nD_n-\sqrt{nV_n}\,\rvQ^{-1}\left(\lambda+\eta+\frac{6T_n}{\sqrt{nV_n^3}}\right)-\log\eta
\ee
where
\begin{align}
D_n&=\frac{1}{n}\sum_{t=1}^n D(W_{x_ty_t}\|P_{Z_t}),\\
V_n&=\frac{1}{n}\sum_{t=1}^n V(W_{x_ty_t}\|P_{Z_t}),\\
T_n&=\frac{1}{n}\sum_{t=1}^n T(W_{x_ty_t}\|P_{Z_t}).
\end{align}
For any $(x^n,y^n)\in\Omega_0^c$, any $t\in[n]$, and any $z\in\calZ$,
\begin{align}
\log \frac{W_{x_ty_t}(z)}{P_{Z_t}(z)}
&=\log \frac{W_{x_ty_t}(z)}{\sum_{x,y} P_{X_tY_t}(x,y) W_{xy}(z)}
\\&\le \log \frac{1}{P_{X_tY_t}(x_t,y_t)}
\\&\le 2\log n
\end{align}
where the last inequality follows from the definition of $\Omega_0$ in \eqref{eq:Omega_0_def}. (In fact, this is the purpose of the set the set $\Omega_0$ in the first place.)
We may prove a simple lower bound by, for any $z$ where $W_{x_ty_t}(z)>0$,
\be
\log \frac{W_{x_ty_t}(z)}{P_{Z_t}(z)}\ge \log W_{x_ty_t}(z)
\ge  \log W_{\min}.
\ee
where $W_{\min}=\min_{x,y,z:W_{xy}(z)>0} W_{xy}(z)$. For any fixed channel with finite alphabets, $W_{\min}>0$. Thus, for sufficiently large $n$,
\be
\left|\log \frac{W_{x_ty_t}(z)}{P_{Z_t}(z)}\right|\le 2\log n.
\ee
This implies that $0\le D(W_{x_ty_t}\|P_{Z_t})\le 2\log n$, so we have
\be
\left|\log \frac{W_{x_ty_t}(z)}{P_{Z_t}(z)}-D(W_{x_ty_t}\|P_{Z_t})\right|
\le 2\log n-\log W_{\min}\le 3\log n
\ee
where the last inequality holds for sufficiently large $n$. Thus, for any $(x^n,y^n)\in\Omega_0^c$,
\be
T_n\le \max_{t\in [n]} T(W_{x_ty_t}\|P_{Z_t})
\le (3\log n)^3.\label{eq:Tn_bound}
\ee
By the same argument, $V_n\le (3\log n)^3$.
Applying the upper bound on $T_n$ in \eqref{eq:Tn_bound} to the bound on the hypothesis testing quantity from \eqref{eq:beta_berry_esseen} and selecting $\eta=\min\{1/\sqrt{n},1-\lambda\}$, for any $(x^n,y^n)\in\Omega_0^c$ we have
\be
-\log \beta_{1-\lambda}(W_{x^ny^n},{\textstyle\prod_{t=1}^n P_{Z_t}})
\le  nD_n-\sqrt{nV_n}\,\rvQ^{-1}\left(\lambda+\frac{1}{\sqrt{n}}+\frac{6(3\log n)^3}{\sqrt{nV_n^3}}\right)+\frac{1}{2}\log n
\ee
where we adopt the convention that $\rvQ^{-1}(p)=-\infty$ if $p\ge 1$. We now consider two cases. Consider first the case that $V_n\ge n^{-1/4}$. This implies $\sqrt{nV_n^3}\ge n^{1/8}$, so in particular $\sqrt{nV_n^3}\to\infty$. Thus, applying a Taylor expansion to the $\rvQ^{-1}$ function, there exists a constant $c_0$ depending only on $\lambda$ such that, for sufficiently large $n$,
\begin{align}
\sqrt{nV_n}\,\rvQ^{-1}\left(\lambda+\frac{1}{\sqrt{n}}+\frac{6(3\log n)^3}{\sqrt{nV_n^3}}\right)
&\ge\sqrt{nV_n}\left[\rvQ^{-1}(\lambda)-c_0\left(\frac{1}{\sqrt{n}}+\frac{6(3\log n)^3}{\sqrt{nV_n^3}}\right)\right]
\\&=\sqrt{nV_n}\,Q^{-1}(\lambda)-c_0\left(\sqrt{V_n}+\frac{6(3\log n)^3}{V_n}\right)
\\&\ge \sqrt{nV_n}\,Q^{-1}(\lambda)-c_0\left((3\log n)^{3/2}+162\,n^{1/4}\log^3 n\right).
\end{align}
Now consider the case that $V_n\le n^{-1/4}$. Then we apply the simpler Chebyshev bound of \cite[Prop.~2.2]{Tan2014a} on the hypothesis testing quantity to write
\begin{align}
-\log \beta_{1-\lambda}(W_{x^ny^n},{\textstyle\prod_{t=1}^n P_{Z_t}})
&\le \inf_{0<\eta\le 1-\lambda} nD_n+\frac{\sqrt{nV_n}}{1-\lambda-\eta}-\log\eta
\\&\le nD_n+\frac{2\sqrt{nV_n}}{1-\lambda}-\log \frac{1-\lambda}{2}\label{eq:beta_cheb2}
\\&= nD_n-\sqrt{nV_n}\,\rvQ^{-1}(\lambda)+\sqrt{nV_n}\left(\rvQ^{-1}(\lambda)+\frac{2}{1-\lambda}\right)-\log \frac{1-\lambda}{2}\label{eq:beta_cheb3}
\\&\le nD_n-\sqrt{nV_n}\,\rvQ^{-1}(\lambda)+n^{3/8}\left(\left|\rvQ^{-1}(\lambda)\right|^++\frac{2}{1-\lambda}\right)-\log \frac{1-\lambda}{2}\label{eq:beta_cheb4}
\end{align}
where in \eqref{eq:beta_cheb2} we have selected $\eta=\frac{1-\lambda}{2}$.
Thus, in all cases, if $(x^n,y^n)\in\Omega_0^c$, then for sufficiently large $n$,
\be
-\log \beta_{1-\lambda}(W_{x^ny^n},{\textstyle\prod_{t=1}^n P_{Z_t}})
\le nD_n-\sqrt{nV_n}\,\rvQ^{-1}(\lambda)+a_n
\ee
where
\be
a_n=\max\left\{c_0\left((3\log n)^{3/2}+162\,n^{1/4}\log^3 n\right),\,n^{3/8}\left(\left|\rvQ^{-1}(\lambda)\right|^++\frac{2}{1-\lambda}\right)-\log \frac{1-\lambda}{2}\right\}.
\ee
Note that the constants in the definition of $a_n$ depend only on $\lambda$, and that for any $\lambda>0$, $a_n=o(\sqrt{n})$.
By a similar argument, if $(x^n,y^n)\in\Omega_0^c$, then for sufficiently large $n$
\be
-\log \beta_{1-\lambda}(W_{x^ny^n},
{\textstyle\prod_{t=1}^n P_{Z_t|Y_t=y_t}})
\le \sum_{t=1}^n D(W_{x_ty_t}\|P_{Z_t|Y_t=y_t})
-\sqrt{\sum_{t=1}^n V(W_{x_ty_t}\|P_{Z_t|Y_t=y_t})}\,\rvQ^{-1}(\lambda)+a_n
\ee
Applying both bounds to \eqref{eq:M1M2_combined_beta2} gives}
\begin{multline}\label{eq:dmc_stat_bound}
\log M_1+\alpha \log M_2\le 
-\log \bbE\left[\new{1((X^n,Y^n)\in\Omega_0^c)}\exp\left\{-nD(X^n,Y^n)+\sqrt{nV(X^n,Y^n)}\,\rvQ^{-1}(\lambda)-\new{a_n}\right\}\right]
\\+\left(\frac{1}{\delta}+1\right)\log \frac{\lambda}{\lambda-\eps}
\end{multline}
where we have defined the statistics
\begin{align}
D(x^n,y^n)&=\frac{1}{n}\sum_{t=1}^n \big[\alpha D(W_{x_ty_t}\|P_{Z_t})+(1-\alpha) D(W_{x_ty_t}\|P_{Z_t|Y_t=y_t})\big],\\
V(x^n,y^n)&=\left(\alpha \sqrt{\frac{1}{n}\sum_{t=1}^n V(W_{x_ty_t}\|P_{Z_t})}+(1-\alpha)\sqrt{\frac{1}{n}\sum_{t=1}^n V(W_{x_ty_t}\|P_{Z_t|Y_t=y_t})}\right)^2.
\end{align}

Consider any $\lambda\ge 1/2$. From \eqref{eq:dmc_stat_bound}, by the convexity of the exponential, we have
\new{
\begin{align}
\log M_1+\alpha \log M_2
&\le \frac{1}{1-p_0}\bbE\left[1((X^n,Y^n)\in\Omega_0^c)\left(nD(X^n,Y^n)-\sqrt{nV(X^n,Y^n)}\,\rvQ^{-1}(\lambda)\right)\right]+a_n-\log(1-p_0)\nonumber
\\&\qquad+\left(\frac{1}{\delta}+1\right)\log \frac{\lambda}{\lambda-\eps}
\\&\le \frac{1}{1-p_0} \bbE\left[nD(X^n,Y^n)-\sqrt{nV(X^n,Y^n)}\,\rvQ^{-1}(\lambda)\right]+a_n-\log(1-p_0)+\left(\frac{1}{\delta}+1\right)\log \frac{\lambda}{\lambda-\eps}
\end{align}
where we have used the facts that $D(x^n,y^n)$ and $V(x^n,y^n)$ are non-negative, and since $\lambda\ge 0$, $\rvQ^{-1}(\lambda)\le 0$.}
Note that
\begin{align}
\bbE[ D(X^n,Y^n)]&=\frac{1}{n}\sum_{t=1}^n \left[\alpha D(W\|P_{Z_t}|P_{X_tY_t})+(1-\alpha) D(W\|P_{Z_t|Y_t}|P_{X_tY_t})\right]
\\&=\frac{1}{n}\sum_{t=1}^n \left[\alpha I(X_t,Y_t;Z_t)+(1-\alpha)I(X_t;Z_t|Y_t)\right]
\\&=\alpha I(X,Y;Z|U)+(1-\alpha) I(X;Z|Y,U)
\end{align}
where in the last equality we have defined $U\sim\text{Unif}[n]$ and $X=X_U,Y=Y_U,Z=Z_U$. 
Moreover, by concavity of the square root,
\begin{align}
\bbE\left[ \sqrt{V(X^n,Y^n)}\right]
&\le \alpha \sqrt{\frac{1}{n}\sum_{t=1}^n V(W\|P_{Z_t}|P_{X_tY_t})}+(1-\alpha)\sqrt{\frac{1}{n}\sum_{t=1}^n V(W\|Y_{Z_t|Y_t}|P_{X_tY_t})}
\\&=\alpha \sqrt{V(W\|P_{Z|U}|P_{UXY})}+(1-\alpha)\sqrt{V(W\|P_{Z|YU}|P_{UXY})}.
\end{align}
Thus, since $\rvQ^{-1}(\lambda)\le 0$,
\new{
\begin{align}
&\log M_1+\alpha\log M_2\nonumber
\\&\le \frac{1}{1-p_0}\bigg[n(\alpha I(X,Y;Z|U)+(1-\alpha) I(X;Z|Y,U))\nonumber
\\&\qquad-\sqrt{n}\left(\alpha \sqrt{V(W\|P_{Z|U}|P_{UXY})}+(1-\alpha)\sqrt{V(W\|P_{Z|YU}|P_{UXY})}\right)\rvQ^{-1}(\lambda)\bigg]\nonumber
\\&\qquad+a_n-\log(1-p_0)+\left(\frac{1}{\delta}+1\right)\log \frac{\lambda}{\lambda-\eps}\label{eq:multi_term_bound}
\\&\le n(\alpha I(X,Y;Z|U)+(1-\alpha) I(X;Z|Y,U))\nonumber
\\&\qquad-\sqrt{n}\left(\alpha \sqrt{V(W\|P_{Z|U}|P_{UXY})}+(1-\alpha)\sqrt{V(W\|P_{Z|YU}|P_{UXY})}\right)\rvQ^{-1}(\lambda)
+\left(\frac{1}{\delta}+1\right)\log \frac{\lambda}{\lambda-\eps}+o(\sqrt{n})\label{eq:dmc_lambda_big}
\end{align}
where we have used the facts that $a_n=o(\sqrt{n})$, $p_n= O(1/n)$, and that the quantity inside the square brackets in \eqref{eq:multi_term_bound} is at most $n\log|\calZ|-\sqrt{nV_{\max}}\,\rvQ^{-1}(\lambda)$.}
From the cost-constraint assumptions on the code, we also have $\bbE[b_1(X)]\le B_1$ and $\bbE[b_2(Y)]\le B_2$. By Carath\'edory's theorem, we may reduce the cardinality of $\calU$ to $|\calU|\le 6$ while preserving the following values:
\be
\alpha I(X,Y;Z|U)+(1-\alpha) I(X;Z|Y,U),\,
V(W\|P_{Z|U}|P_{UXY}),\,
V(W\|P_{Z|YU}|P_{UXY}),\,
\bbE[b_1(X)],\, \bbE[ b_2(Y)].
\ee
Choosing $\delta=O(n^{-1/2})$ allows us to derive the crude bound
\be
\log M_1+\alpha \log M_2\le n(\alpha I(X,Y;Z|U)+(1-\alpha) I(X;Z|Y,U))+O(\sqrt{n}).
\ee
Define $\tilX,\tilY,\tilZ$ where
\be
P_{\tilX\tilY\tilZ|U=u}(x,y,z)=P_{X|U=u}(x)P_{Y|U=u}(y)W_{xy}(z).
\ee
By Lemma~\ref{lemma:DMCs},
\be
\log M_1+\alpha \log M_2
\le n(\alpha I(\tilX,\tilY;\tilZ,U)+(1-\alpha)I(\tilX;\tilZ|\tilY,U))+O(\sqrt{n}).
\ee
Our goal is to prove that
\be\label{eq:Vplus_goal}
\log M_1+\alpha \log M_2\le nC_{1,\alpha}
+2\sqrt{n C'_{1,\alpha}(0)\log\frac{\lambda}{\lambda-\eps}}-\sqrt{nV_{1,\alpha}^+}\rvQ^{-1}(\lambda)+o(\sqrt{n}).
\ee
Since $Q^{-1}(\lambda)\le 0$, we may assume that
\be
\log M_1+\alpha \log M_2\ge nC_{1,\alpha}
\ee
or else there is nothing to prove. Thus
\be\label{eq:tilde_constraint1}
\alpha I(\tilX,\tilY;\tilZ,U)+(1-\alpha)I(\tilX;\tilZ|\tilY,U)\ge C_{1,\alpha}-O\left(\frac{1}{\sqrt{n}}\right).
\ee
Noting that the mutual information is continuous over distributions with finite alphabets, by the definition of $C_{1,\alpha}$, \eqref{eq:tilde_constraint1} implies that there exists a distribution $P^\star_{UXY}\in\calP_{1,\alpha}^{\text{in}}$ where $d_{TV}(P_{U\tilX\tilY},P^\star_{UXY})\le o(1)$. Since $\Delta(X;Y|U=u)\le\delta$, from Thm.~\ref{thm:props} we have
\be
|P_{XY|U=u}(x,y)-P_{\tilX\tilY|U=u}(x,y)|\le 2\delta.
\ee
As we have taken $\delta=O(1/\sqrt{n})$, then $d_{TV}(P_{UXY},P_{U\tilX\tilY})\le o(1)$. Thus by the triangle inequality, $d_{TV}(P_{UXY},P^\star_{UXY})\le o(1)$. Since the dispersion variance is also is a continuous function of $P_{UXY}$ (again for finite alphabets), we must have 
\begin{align}
&\alpha \sqrt{V(W\|P_{Z|U}|P_{UXY})}+(1-\alpha)\sqrt{V(W\|P_{Z|YU}|P_{UXY})}
\\&\le \alpha \sqrt{V(W\|P^\star_{Z|U}|P^\star_{UXY})}+(1-\alpha)\sqrt{V(W\|P^\star_{Z|YU}|P^\star_{UXY})}+o(1)
\\&\le V_{1,\alpha}^++o(1)
\end{align}
where the second inequality holds since $P^\star_{UXY}\in\calP_{1,\alpha}^{\text{in}}$ and by the definition of $V_{1,\alpha}^+$ in \eqref{eq:Vplus_def}. Now returning to the bound in \eqref{eq:dmc_lambda_big},
\begin{align}
\log M_1+\alpha \log M_2
&\le 
n(\alpha I(X,Y;Z|U)+(1-\alpha) I(X;Z|Y,U))-\sqrt{nV_{1,\alpha}^+}\rvQ^{-1}(\lambda)+\left(\frac{1}{\delta}+1\right)\log\frac{\lambda}{\lambda-\eps}+o(\sqrt{n})
\\&\le n C_{1,\alpha}(\delta)-\sqrt{n V_{1,\alpha}^{+}}\, \mathsf{Q}^{-1}(\lambda)+\left(\frac{1}{\delta}+1\right)\log\frac{\lambda}{\lambda-\eps}+o(\sqrt{n})\label{eq:Vplus3}
\\&=nC_{1,\alpha}+C'_{1,\alpha}(0)\delta+o(n\delta)-\sqrt{n V_{1,\alpha}^{+}}\, \mathsf{Q}^{-1}(\lambda)+\left(\frac{1}{\delta}+1\right)\log\frac{\lambda}{\lambda-\eps}+o(\sqrt{n})\label{eq:Vplus4}
\end{align}
\eqref{eq:Vplus3} holds by the definition of $C_{1,\alpha}(\delta)$; and \eqref{eq:Vplus4} follows by the definition of the derivative. Selecting $\delta=\sqrt{\frac{\log\frac{\lambda}{\lambda-\eps}}{C'_{1,\alpha}(0)}}$, we derive the desired bound in \eqref{eq:Vplus_goal}.

Now consider any $\lambda<1/2$. Our goal is to show that
\be\label{eq:V_minus_goal}
\log M_1+\alpha \log M_2\le nC_{1,\alpha}(\delta)-\sqrt{nV_{1,\alpha}^-}\,\mathsf{Q}^{-1}(\lambda)+\left(\frac{1}{\delta}+1\right)\log\frac{\lambda}{\lambda-\eps}+o(\sqrt{n})
\ee
where eventually we will choose $\delta=O(n^{-1/2})$. Thus, we may assume
\be\label{eq:Omega_assumption}
\log M_1+\alpha \log M_2\ge nC_{1,\alpha}-\sqrt{nV_{1,\alpha}^-}\,\mathsf{Q}^{-1}(\lambda) +\left(\frac{1}{\delta}+1\right)\log\frac{\lambda}{\lambda-\eps}
\ee
or else we are done. Now let
\begin{align}
\Omega_1&=\left\{(x^n,y^n):nD(x^n,y^n)\le nC_{1,\alpha}-\sqrt{nV_{1,\alpha}^{-}}\,\mathsf{Q}^{-1}(\lambda)-\new{a_n-\log n}\right\},\\
\Omega_2&=\left\{(x^n,y^n):nD(x^n,y^n)\ge nC_{1,\alpha}(\delta)+\log n\right\}
\end{align}
and let \new{$p_i=P_{X^nY^n}(\Omega_i\cap\Omega_0^c)$ } for $i=1,2$. To upper bound $p_1$, beginning from the bound in \eqref{eq:dmc_stat_bound} we may write
\begin{align}
&\log M_1+\alpha \log M_2\new{+\log(1-p_0)}-\left(\frac{1}{\delta}+1\right)\log\frac{\lambda}{\lambda-\eps}
\\&\le -\log \sum_{(x^n,y^n)\in\Omega_1\new{\cap\Omega_0^c}} P_{X^nY^n}(x^n,y^n) 
\exp\left\{-nD(x^n,y^n)+\sqrt{nV(x^n,y^n)}\,\mathsf{Q}^{-1}(\lambda)-\new{a_n}\right\}
\\&\le -\log \sum_{(x^n,y^n)\in\Omega_1\new{\cap\Omega_0^c}} P_{X^nY^n}(x^n,y^n) 
\exp\left\{-nC_{1,\alpha}+\sqrt{nV_{1,\alpha}^{-}}\,\mathsf{Q}^{-1}(\lambda)+\log n\right\}\label{eq:Omega1_bd1}
\\&=-\log p_1 +nC_{1,\alpha}-\sqrt{nV_{1,\alpha}^{-}}\,\mathsf{Q}^{-1}(\lambda)-\log n\label{eq:Omega1_bd2}
\end{align}
where in \eqref{eq:Omega1_bd1} we have used the definition of $\Omega_1$, and the fact that $\mathsf{Q}^{-1}(\lambda)\ge 0$ since $\lambda<1/2$; and \eqref{eq:Omega1_bd2} holds by the definition of $p_1$. Thus by the assumption in \eqref{eq:Omega_assumption}
\new{\be
p_1\le \frac{1}{(1-p_0)n}=O\left(\frac{1}{n}\right)
\ee
since $p_0=O(1/n)$.
}

Let
\be\label{eq:O1O2}
V'=\min\{V(x^n,y^n):(x^n,y^n)\in(\Omega_1\cup\Omega_2)^c\}.
\ee
We will prove that $V'\ge V_{1,\alpha}^--o(1)$. Fix $(x^n,y^n)\in (\Omega_1\cup\Omega_2)^c$. By the definitions of $\Omega_1,\Omega_2$, \new{since $a_n=o(\sqrt{n})$ we have
\be
C_{1,\alpha}-O(n^{-1/2})\le D(x^n,y^n)\le C_{1,\alpha}(\delta)+\log n.
\ee
Since $\delta=O(n^{-1/2})$, by Taylor's theorem and the fact from Lemma~\ref{lemma:DMCs} that $C'_{1,\alpha}(0)$ is bounded, $C_{1,\alpha}(\delta)=C_{1,\alpha}+O(n^{-1/2})$. Thus }
$|D(x^n,y^n)-C_{1,\alpha}|\le O(n^{-1/2})$. 
If we again let $U\sim\text{Unif}[n]$, and
\be\label{eq:XprimeYprime_def}
P_{X'Y'|U=t}(x,y)=1(x=x_t,y=y_t)
\ee
then we may write
\begin{align}
D(x^n,y^n)&=\alpha D(W\|P_{Z|U}|P_{U\barX\barY})+(1-\alpha) D(W\|P_{Z|YU}|P_{UX'Y'}),\\
\sqrt{V(x^n,y^n)}&=\alpha \sqrt{V(W\|P_{Z|U}|P_{UX'Y'})}+(1-\alpha)\sqrt{V(W\|P_{Z|YU}|P_{UX'Y'})}.
\end{align}
Also note that $\bbE [b_1(X')]=\frac{1}{n}\sum_{t=1}^n b_1(x_t)\le B_1$, and similarly $\bbE [b_2(Y')]\le B_2$. We 
may perform a dimensionality reduction on $\calU$ where $|\calU|\le 9$ to preserve the following values:
\begin{gather}
\alpha I(X,Y;Z|U)+(1-\alpha) I(X;Z|Y,U),\\
\alpha D(W\|P_{Z|U}|P_{UX'Y'})+(1-\alpha) D(W\|P_{Z|YU}|P_{UX'Y'}),\\
V(W\|P_{Z|U}|P_{UX'Y'}),\,
V(W\|P_{Z|YU}|P_{UX'Y'}),\\
\bbE [b_1(X)],\,\bbE[ b_2(Y)],\,\bbE[ b_1(X')],\,\bbE[ b_2(Y')].
\end{gather}
Note that this is not the same dimensionality reduction as above; in particular, this one depends on $x^n,y^n$. Since $\delta=O(n^{-1/2})$, by the same argument as above, there exists $P^\star_{UXY}\in\calP_{1,\alpha}^{\text{in}}$ where $d_{TV}(P_{UXY},P^\star_{UXY})\le o(1)$. Since $|D(x^n,y^n)-C_{1,\alpha}|\le o(1)$, by continuity of the relative entropy (for finite alphabets) there exists a distribution $P^\star_{X'Y'|U}$ such that $d_{TV}(P_{UX'Y'},P^\star_{UX'Y'})\le o(1)$ and
\be
\alpha D(W\|P^\star_{Z|U}|P^\star_{UX'Y'})+(1-\alpha) D(W\|P^\star_{Z|YU}|P^\star_{UX'Y'})=C_{1,\alpha}.
\ee
That is, $(P^\star_{UXY},P^\star_{X'Y'|U})$ satisfy the feasibility condition for the definition of $V_{1,\alpha}^{-}$ in \eqref{eq:Vminus_feasibility}. By continuity of the divergence variance, this implies that $V(x^n,y^n)\ge V_{1,\alpha}^{-}-o(1)$. This proves that $V'\ge V_{1,\alpha}^{-}-o(1)$. Now we \new{may lower bound the expectation in \eqref{eq:dmc_stat_bound} by}
\begin{align}
&\bbE\left[\new{1((X^n,Y^n)\in\Omega_0^c)}
\exp\left\{-n D(X^n,Y^n)+\sqrt{n V(X^n,Y^n)}\,\rvQ^{-1}(\lambda)-\new{a_n}\right\}\right]
\\&\ge \sum_{(x^n,y^n)\in (\new{\Omega_0\cup}\Omega_1\cup\Omega_2)^c} P_{X^nY^n}(x^n,y^n) \exp\left\{-n D(x^n,y^n)+\sqrt{nV'}\,\mathsf{Q}^{-1}(\lambda)-\new{a_n}\right\} \label{eq:dmc_final1}
\\&\ge\sum_{(x^n,y^n)\in \new{(\Omega_0\cup\Omega_1)}^c} P_{X^nY^n}(x^n,y^n) \exp\left\{-n D(x^n,y^n)+\sqrt{nV'}\,\mathsf{Q}^{-1}(\lambda)-\new{a_n}\right\} \nonumber
\\&\qquad-\sum_{(x^n,y^n)\in \new{\Omega_0^c\cap}\Omega_2} P_{X^nY^n}(x^n,y^n) \exp\left\{-n D(x^n,y^n)+\sqrt{nV'}\,\mathsf{Q}^{-1}(\lambda)-\new{a_n}\right\} \label{eq:dmc_final2}
\\&\ge (1\new{-p_0}-p_1) \exp\left\{-\frac{1}{1\new{-p_0}-p_1} \sum_{(x^n,y^n)\in\new{(\Omega_0\cup\Omega_1)}^c}P_{X^nY^n}(x^n,y^n) nD(x^n,y^n)+\sqrt{nV'}\,\mathsf{Q}^{-1}(\lambda)-\new{a_n}\right\} \nonumber
\\&\qquad-p_2 \exp\left\{-nC_{1,\alpha}(\delta)+\sqrt{nV'}\,\mathsf{Q}^{-1}(\lambda)-\new{\log n-a_n}\right\}\label{eq:dmc_final3}
\\&\ge (1\new{-p_0}-p_1) \exp\left\{-\frac{1}{1\new{-p_0}-p_1} n\bbE [D(X^n,Y^n)]+\sqrt{nV'}\,\mathsf{Q}^{-1}(\lambda)-\new{a_n}\right\} \nonumber
\\&\qquad- \exp\left\{-nC_{1,\alpha}(\delta)+\sqrt{nV'}\,\mathsf{Q}^{-1}(\lambda)-\new{\log n-a_n})\right\}\label{eq:dmc_final4}
\\&\ge (1\new{-p_0}-p_1) \exp\left\{-\frac{1}{1\new{-p_0}-p_1} nC_{1,\alpha}(\delta)+\sqrt{nV'}\,\mathsf{Q}^{-1}(\lambda)-\new{a_n}\right\} \nonumber
\\&\qquad- \exp\left\{-nC_{1,\alpha}(\delta)+\sqrt{nV'}\,\mathsf{Q}^{-1}(\lambda)-\new{\log n-a_n}\right\}\label{eq:dmc_final5}
\\&=\exp\left\{-nC_{1,\alpha}(\delta)+\sqrt{nV'}\,\mathsf{Q}^{-1}(\lambda)-\new{a_n}\right\}
\left(\exp\left\{\log(1\new{-p_0}-p_1)-\frac{\new{p_0+}p_1}{1\new{-p_0}-p_1}nC_{1,\alpha}(\delta)-O(1)\right\}-\new{\frac{1}{n}}\right)\label{eq:dmc_final6}
\\&= \exp\left\{-nC_{1,\alpha}(\delta)+\sqrt{nV'}\,\mathsf{Q}^{-1}(\lambda)-\new{a_n}\right\}O(1)\label{eq:dmc_final7}
\\&\ge\exp\left\{-nC_{1,\alpha}(\delta)+\sqrt{nV_{1,\alpha}^{-}}\,\mathsf{Q}^{-1}(\lambda)-o(\sqrt{n})\right\}\label{eq:dmc_final8}
\end{align}
where \eqref{eq:dmc_final1} holds by the definition of $V'$, \eqref{eq:dmc_final3} holds by the definition of $\Omega_2$ and by convexity of the exponential, \eqref{eq:dmc_final4} holds by extending the sum over all $(x^n,y^n)$, \eqref{eq:dmc_final5} holds since $\bbE [D(X^n,Y^n)]=\alpha I(X,Y;Z|U)+(1-\alpha) I(X;Z|Y,U)\le C_{1,\alpha}(\delta)$; \eqref{eq:dmc_final7} holds \new{since $p_0+p_1=O(1/n)$, which implies that $\log(1-p_0-p_1)=-O(1/n)$ and $\frac{(p_0+p_1)n}{1-p_0-p_1}=O(1)$, and we also use the fact that $C_{1,\alpha}(\delta)\le\log|\calZ|$;}%
 and \eqref{eq:dmc_final8} holds since $V'\ge V_{1,\alpha}^{-}-o(1)$ \new{and $a_n=o(\sqrt{n})$}. This proves \eqref{eq:V_minus_goal}. Again using the definition of the derivative, and choosing $\delta$ optimally (this involves $\delta=O(n^{-1/2})$ as promised) completes the proof.

\new{
\subsection{Discussion of the Maximal Error Case}
\label{sec:maximal}

While the results in this paper focus on the average error probability criterion, an important variant of the problem is the one using maximal error probability. In a sense, the maximal error variant is an easier problem, because it imposes a stronger condition on each message pair. Unfortunately, as originally shown in \cite{Dueck1978}, the capacity regions for the two problem variants can differ, and in general the capacity region of the maximal error case (with deterministic encoders) is not even known. 

A second-order converse bound for the maximal-error case was presented in \cite{Moulin2013}; however, the proof of the main result of \cite{Moulin2013} appears to have a gap (namely, the derivation of equation (28)). The recent work \cite{Wei2021} used a wringing-based proof (following a similar approach as this paper) to derive a similar bound to that claimed in \cite{Moulin2013}. The result derived in \cite{Wei2021} is as follows. Let $R^{\star,\max}_{\alpha_1,\alpha_2}(n,\eps)$ be the largest achievable weighted-sum rate for a length-$n$ code with maximal probability of error $\eps$. Consider a discrete-memoryless MAC such that there is a unique optimal input distribution for the standard sum-rate; i.e. $\calP_{1,1}^{\text{in}}$ contains a single distribution $P_{X}^\star P_Y^\star$. Then \cite{Wei2021} shows that
\be\label{eq:max_err_bound}
R^{\star,\max}_{1,1}(n,\eps)\le C_{1,1}-\sqrt{\frac{V^\star}{n}}\,\rvQ^{-1}(\eps)+o\left(\frac{1}{\sqrt{n}}\right)
\ee
where $V^\star=V(W\|P_Z^\star|P_X^\star P_Y^\star)$ where $P_Z^\star$ is the induced output distribution from $P_X^\star P_Y^\star$. This constitutes a tighter bound on the sum-rate than Thm.~\ref{thm:DMC_asymptotic}. However, note that in \eqref{eq:max_err_bound}, $C_{1,1}$ is the average-case sum-capacity, which may not be the same as the maximal-error sum-capacity, and indeed the maximal-error sum-capacity may not even be known. Thus, for many channels the gap between the best-known achievability and converse bounds for the maximal-error case is $O(1)$, as opposed to $O(1/\sqrt{n})$ for the average-error case.
}

\section{Example Multiple-Access Channels}\label{sec:examples}

\subsection{Binary Additive Erasure Channel}\label{sec:bamac}

Let $\calX\in\{0,1\}$, $\calY\in\{0,1\}$, $\calZ=\{0,1,2,\rve\}$. Given $(X,Y)=(x,y)$, $Z=\rve$ with probability $\gamma$, and $Z=x+y$ with probability $\bar\gamma=1-\gamma$. The capacity region for this channel is the pentagonal region
\be
\calC=\left\{(R_1,R_2):R_1+R_2\le \frac{3}{2}\bar\gamma\log 2,\, R_1\le \bar\gamma\log 2,\, R_2\le \bar\gamma\log 2.\right\}
\ee
Thus the weighted-sum-capacity is
\be
C_{\alpha_1,\alpha_2}=\left(\max\{\alpha_1,\alpha_2\}+\frac{1}{2}\min\{\alpha_1,\alpha_2\}\right)\bar\gamma\log 2.
\ee

In order to apply Thm.~\ref{thm:DMC_asymptotic}, we need to find $C'_{\alpha_1,\alpha_2}(0)$, $V_{\alpha_1,\alpha_2}^+$, and $V_{\alpha_1,\alpha_2}^-$. First we compute $C_{\alpha_1,\alpha_2}(\delta)$. Since the channel is symmetric between the two inputs, $C_{\alpha_1,\alpha_2}(\delta)=C_{\alpha_2,\alpha_1}(\delta)$.
Let $(\alpha_1,\alpha_2)=(1,\alpha)$ for $\alpha\in[0,1]$. Since this channel has no cost constraints, the time sharing variable $U$ can be eliminated in the definition of $C_{\alpha_1,\alpha_2}(\delta)$ in \eqref{eq:Cdelta_def}. Thus
\begin{align}
C_{1,\alpha}(\delta)&=\max_{P_{XY}:\Delta(X;Y)\le\delta} \big[\alpha I(X,Y;Z)+(1-\alpha) I(X;Z|Y)\big]
\\&=\max_{P_{XY}:\Delta(X;Y)\le\delta} \bar\gamma\left[\alpha H(X+Y)+(1-\alpha) H(X|Y)\right].
\end{align}
To lower bound $C_{1,\alpha}(\delta)$, we may take $P_{XY}$ to be a DSBS with parameter $p\le 1/2$. Recalling  the calculation from Example~\ref{example:DSBS}, $\Delta(X;Y)=\frac{1+\log_2(1-p)}{1-\log_2(1-p)}$, so
\begin{align}
C_{1,\alpha}(\delta)&\ge \max_{p\le 1/2: \frac{1+\log_2(1-p)}{1-\log_2(1-p)}\le\delta} \bar\gamma\left[\alpha(H_b(p)+(1-p)\log 2)+(1-\alpha) H_b(p)\right]
\\&=\begin{cases} \bar\gamma \left[ H_b(2^{1-2/(1+\delta)})+\alpha 2^{1-2/(1+\delta)}\log 2\right], 
& \delta< \frac{1-\log_2(1+2^{-\alpha})}{1+\log_2(1+2^{-\alpha})},
\\ \bar\gamma [\log (1+2^{-\alpha})+\alpha\log 2], 
& \delta\ge \frac{1-\log_2(1+2^{-\alpha})}{1+\log_2(1+2^{-\alpha})}\end{cases}\label{eq:BAMAC_lb}
\end{align}
where \eqref{eq:BAMAC_lb} follows from a straightforward entropy calculation. In fact, this lower bound is tight, although the proof is a little more difficult. The following proposition is proved in Appendix~\ref{appendix:bamac}.

\begin{proposition}\label{prop:bamac}
For any $\alpha\in[0,1]$ and $\delta\in[0,1]$, $C_{1,\alpha}(\delta)$ is equal to the expression in \eqref{eq:BAMAC_lb}.
\end{proposition}

Given the expression for $C_{1,\alpha}(\delta)$ in \eqref{eq:BAMAC_lb}, the first-order Taylor expansion is given by
\be
C_{1,\alpha}(\delta)=\bar\gamma \left(1+\frac{\alpha}{2}\right)\log 2+\bar\gamma \alpha (\log^2 2)\delta+O(\delta^2).
\ee
In particular, $C'_{1,\alpha}(0)=\bar\gamma\alpha \log^2 2$.

\begin{figure}
\begin{minipage}{.49\textwidth}
\centering
\includegraphics[width=3.5in]{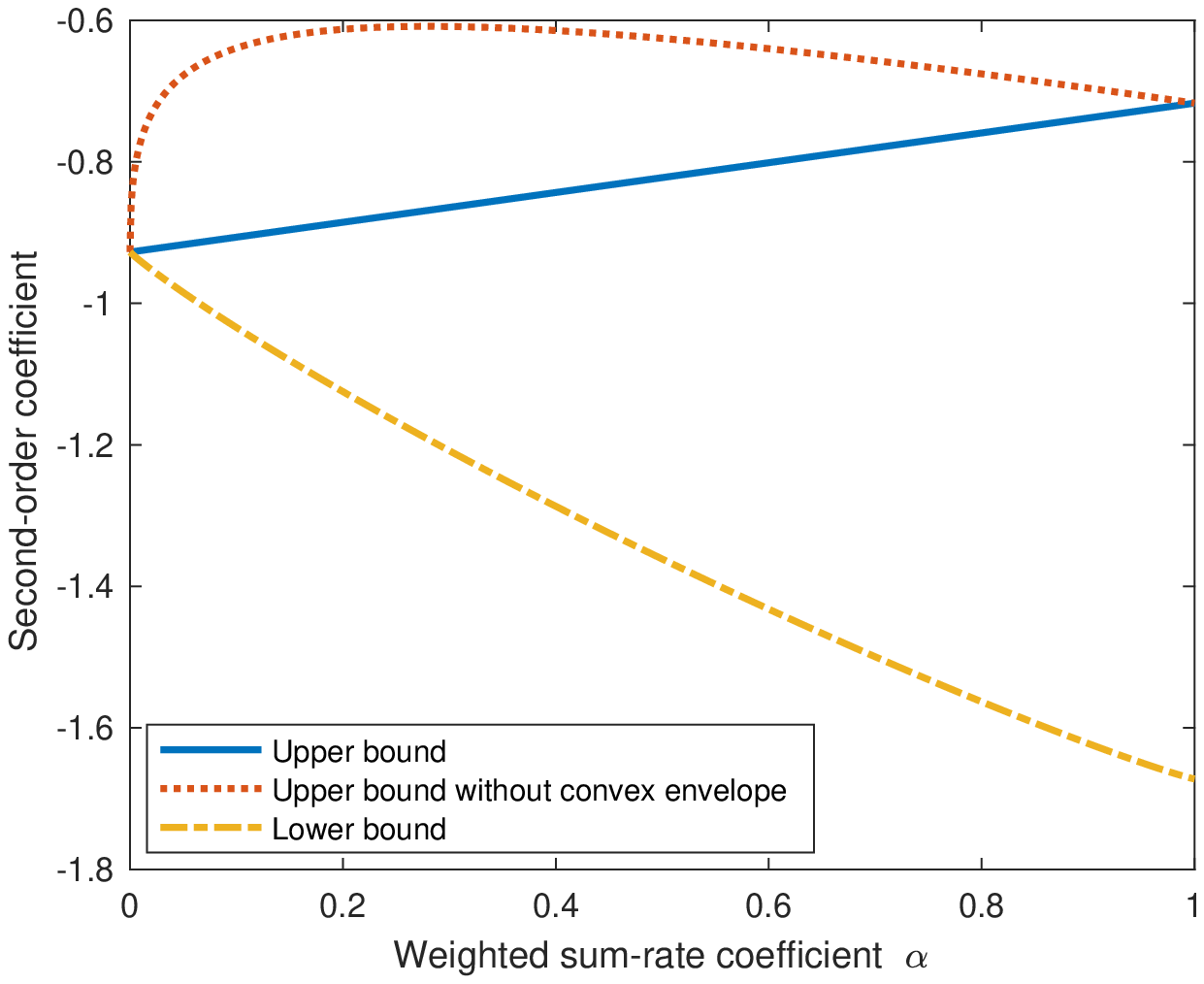}\\
{\small (a)}
\end{minipage}
\hfill
\begin{minipage}{.49\textwidth}
\centering
\includegraphics[width=3.5in]{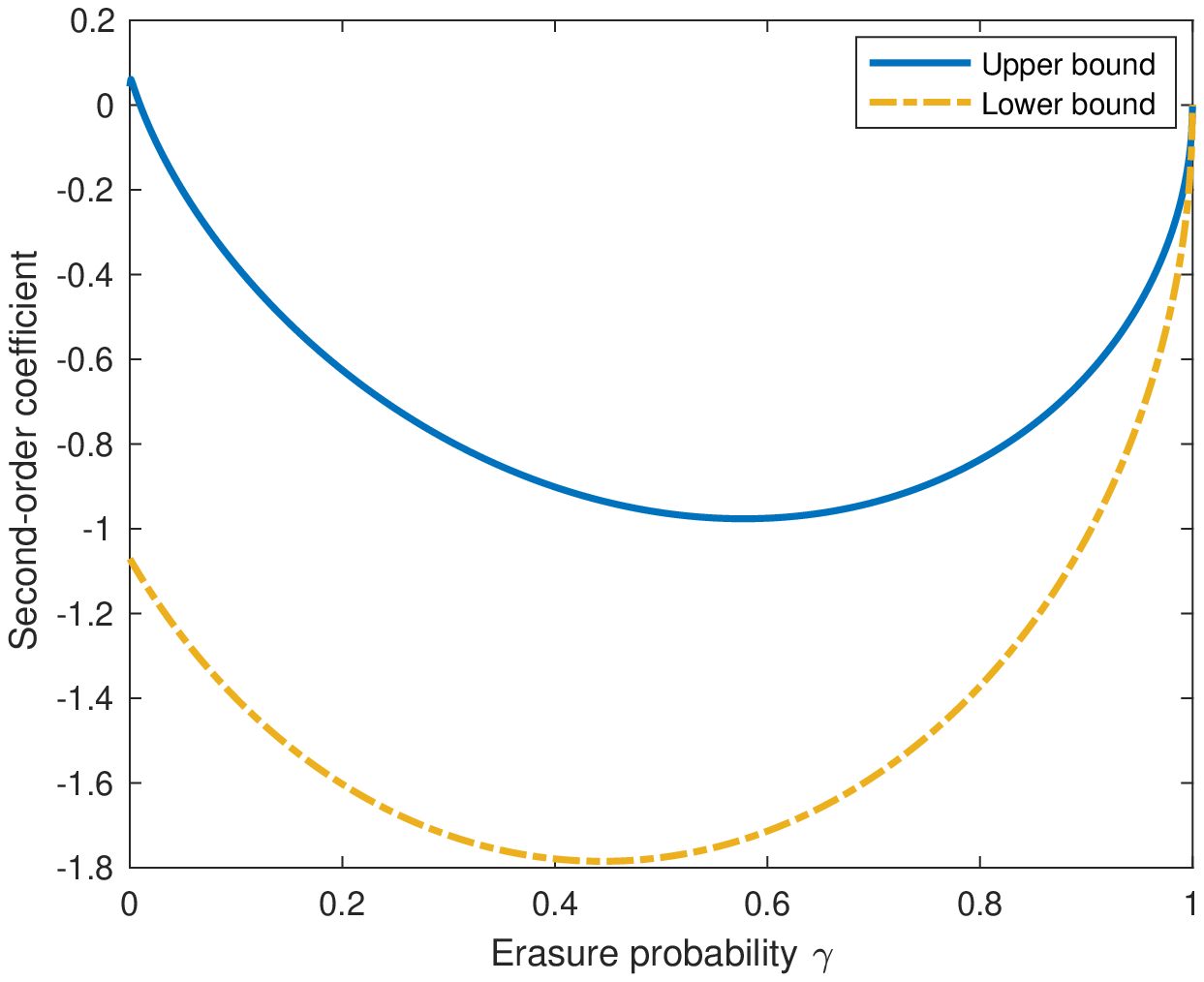}\\
{\small (b)}
\end{minipage}
\caption{Upper and lower bounds on the second-order coefficient for the binary additive erasure channel. Subfigure (a) shows the second-order bounds for the maximum achievable weighted-sum-rate $R^\star_{1,\alpha}(n,\eps)$ as a function of $\alpha\in[0,1]$ for erasure probability $\gamma=0.25$ and probability of error $\eps=10^{-3}$. Subfigure (b) shows second-order bounds for the standard sum-rate $R^\star_{1,1}(n,\eps)$ as a function of $\gamma\in[0,1]$ for $\eps=10^{-3}$. The lower bound is from prior work \cite{Tan2014,Haim2012,Huang2012,MolavianJazi2012,Scarlett2015a}, while the upper bound is our contribution. In subfigure (a), along with the upper bound from \eqref{eq:convexified_baec}, we also show the weaker upper bound found by not taking the lower convex envelope in \eqref{eq:unconvex_baec}. Note that the stronger bound is simply the lower convex envelope of the weaker bound.}
\label{fig:baec_bounds}
\end{figure}

We now calculate the dispersion variance quantities $V_{\alpha_1,\alpha_2}^+,V_{\alpha_1,\alpha_2}^-$. For any\footnote{The $\alpha=0$ case allows other optimal input distributions, although this case is somewhat trivial, as is reduces to a point-to-point binary erasure channel.} $\alpha\in(0,1]$, $\calP_{1,\alpha}^{\text{in}}$ is the set of distributions $P_{UXY}$ where $P_{XY|U=u}$ is uniform on $\{0,1\}^2$.  That is, $(X,Y)$ are independent of $U$, so we may ignore $U$. Taking $P_Z,P_{Z|Y}$ to be the induced distributions from the unique optimal input distribution, we may calculate
\begin{align}
D(W_{xy}\|P_Z)&=(1+1(x=y))\bar\gamma\log 2,\\
D(W_{xy}\|P_{Z|Y=y})&=\bar\gamma\log 2.
\end{align}
Note that $\alpha D(W\|P_Z|P_{X'Y'})+(1-\alpha) D(W\|P_{Z|Y}|P_{X'Y'})=C_{1,\alpha}$ iff $P_{X'Y'}(0,0)+P_{X'Y'}(1,1)=1/2$. Moreover,
\begin{align}
V(W_{xy}\|P_Z)&=\gamma\bar\gamma (1+4\cdot 1(x=y))\log^2 2,\\
V(W_{xy}\|P_{Z|Y=y})&=\gamma\bar\gamma \log^2 2.
\end{align}
Thus
\be
V_{1,\alpha}^-=\gamma\bar\gamma\left(\alpha\sqrt{\frac{5}{2}}+1-\alpha\right)^2.
\ee
Moreover, $V_{1,\alpha}^+$ is the same quantity. Thm.~\ref{thm:DMC_asymptotic} now gives
\be\label{eq:unconvex_baec}
R^\star_{1,\alpha}(n,\eps)
\le \bar\gamma\left(1+\frac{\alpha}{2}\right)\log 2
+\left(\min_{\lambda\in(\eps,1)} 2\sqrt{\bar\gamma\alpha \log\frac{\lambda}{\lambda-\eps}}-\sqrt{\gamma\bar\gamma}\left(\alpha\sqrt{\frac{5}{2}}+1-\alpha\right) \mathsf{Q}^{-1}(\lambda)\right)^{**}\frac{\log 2}{\sqrt{n}}+o\left(\frac{1}{\sqrt{n}}\right).
\ee
In fact, the quantity inside the $(\cdot)^{**}$ is concave (see Fig.~\ref{fig:baec_bounds}), so it is equivalent to simply take the convex combination of the points at $\alpha=0$ and $\alpha=1$. At $\alpha=0$ one can see that it is optimal to choose $\lambda=\eps$. Thus
\be\label{eq:convexified_baec}
R^\star_{1,\alpha}(n,\eps)
\le \bar\gamma\left(1+\frac{\alpha}{2}\right)\log 2
+\left[(1-\alpha)\sqrt{\gamma\bar\gamma}\,\rvQ^{-1}(\eps)
+\min_{\lambda\in(\eps,1)} \alpha \left( 2\sqrt{\bar\gamma\log \frac{\lambda}{\lambda-\eps}}-\sqrt{\gamma\bar\gamma\frac{5}{2}}\,\rvQ^{-1}(\lambda)\right)\right]\frac{\log 2}{\sqrt{n}}
+o\left(\frac{1}{\sqrt{n}}\right).
\ee

The corresponding achievability bound from any of \cite{Tan2014,Haim2012,Huang2012,MolavianJazi2012,Scarlett2015a}\footnote{The achievable bound from \cite{Scarlett2015a} is in general the strongest, but for this channel these all produce the same bound.} is
\be
R_{1,\alpha}^{\star}(n,\eps)
\ge \bar\gamma\left(1+\frac{\alpha}{2}\right)\log 2
+L(\alpha,\eps)\log 2
-o\left(\frac{1}{\sqrt{n}}\right)
\ee
where
\begin{align}
L(\alpha,\eps)&=\sup\{\alpha s_1+(1-\alpha) s_2:\bbP(S_1\ge s_1,S_2\ge s_2)\ge 1-\eps\}
\end{align}
and $(S_1,S_2)$ are jointly Gaussian with zero mean and covariance matrix
\be
\gamma\bar\gamma\left[\begin{array}{cc} 5/2 & 3/2 \\ 3/2 & 1\end{array}\right].
\ee
Fig.~\ref{fig:baec_bounds} illustrates the upper and lower bounds on the coefficient in the $O(1/\sqrt{n})$ term. The figure shows bounds on the second-order coefficient for $R_{1,\alpha}(n,\eps)$ for $\gamma=0.25,\eps=10^{-3}$, and also bounds on $R_{1,1}(n,\eps)$---i.e., the standard sum-rate---for all $\gamma\in[0,1]$ and $\eps=10^{-3}$. Unfortunately, the upper and lower bounds only match for essentially trivial cases: when $\alpha=0$, wherein the problem reduces to the point-to-point binary erasure channel, and when $\gamma=1$, wherein the output is independent from the inputs so no communication is possible.

\subsection{Gaussian MAC}\label{sec:gaussian}

In the Gaussian MAC, $X,Y,Z$ are all real-valued, the output is $Z=X+Y+N$, where $N\sim\calN(0,1)$, and the input sequences $X^n,Y^n$ are subject to power constraints $\sum_{t=1}^n X_t^2\le nS_1$ and $\sum_{t=1}^n Y_t^2\le nS_2$.  The following result, proved in Appendix~\ref{appendix:gaussian}, states that the Gaussian MAC satisfies the conditions of Corollary~\ref{cor:regularity}, and so its second-order rate is $O(1/\sqrt{n})$. 

\begin{theorem}\label{thm:gaussian}
For the Gaussian MAC, $C_{\alpha_1,\alpha_2}'(0)$ is uniformly bounded for all $\alpha_1,\alpha_2$ where $\max\{\alpha_1,\alpha_2\}=1$, and $V_{\max}<\infty$.
\end{theorem}

In the statement of this theorem, we have omitted any specific bound on $C'_{\alpha_1,\alpha_2}(0)$ or $V_{\max}$. While such bounds can be extracted from the proof, we have sought clarity of the proof over optimality of the bounds\footnote{The length and complexity of the proof in Appendix~\ref{appendix:gaussian} may make you skeptical of this claim, but it's true!}, and so we have elected to highlight the order of the bound on the second-order rate, rather than the coefficient.

\section{Conclusion}\label{sec:conclusion}

The main result of this paper is that, for most multiple-access channels of interest, under the average probability of error constraint the second-order coding rate is $O(1/\sqrt{n})$ bits per channel use. Along the way, we introduced and characterized the wringing dependence, which was a critical element in the proof of the main results.

Possible future work includes extensions to more than two transmitters, or applying similar techniques to other network information theory problems (the interference channel with strong interference should be a straightforward extension). Moreover,  there are a number of ways that our results could potentially be improved even for the two-user MAC. First, the regularity conditions given in Corollary~\ref{cor:regularity}, under which we are able to prove the second-order bound of $O(1/\sqrt{n})$, are quite difficult to verify for non-discrete channels. The only continuous channel for which we have successfully verified the conditions is the Gaussian MAC; the proof of this in Appendix~\ref{appendix:gaussian} is quite technical, as well as being very specific to the Gaussian channel. It would be advantageous to find conditions that are easier to verify under which the second-order bound holds.

A second potential improvement has to do with the quantity $V^{\lambda}_{\alpha_1,\alpha_2}$ in Thm.~\ref{thm:DMC_asymptotic}. Specifically, the form of $V_{\alpha_1,\alpha_2}^{-}$ in \eqref{eq:Vminus_def} is not especially natural; it may be possible to improve the result so that this quantity is complementary to $V_{\alpha_1,\alpha_2}^{+}$; that is, \eqref{eq:Vplus_def} with an infimum instead of a supremum. In addition,  Thm.~\ref{thm:DMC_asymptotic} could be strengthened using dispersion quantities extracted from multi-dimensional Gaussian CDFs, along the lines of the achievable bounds in \cite{Tan2014,Haim2012,Huang2012,MolavianJazi2012,Scarlett2015a,MolavianJazi2015}. One may also wish to prove something similar to Thm.~\ref{thm:DMC_asymptotic} for non-discrete channels.

Of course, the ultimate goal would be to determine the second-order coefficient exactly. Even if the above improvements could be made, there would remain a gap between achievability and converse bounds for almost all channels, \new{including} such simple examples as the deterministic binary additive channel. It appears that new ideas are required in order to close the gap completely. One possible direction of improvement, which the method used here fails to address, is the following. Consider the distribution of the error probability conditioned on the message pair. That is, let $\eps(i_1,i_2)$ be the error probability given message pair $(i_1,i_2)$. Taking $(I_1,I_2)$ to be uniformly random over the message sets, it is critical to characterize the distribution of the random variable $\eps(I_1,I_2)$ in any MAC converse proof. In our proof, we do not use anything about the distribution of $\eps(I_1,I_2)$ beyond that its expected value is the overall error probability. In particular, the proof would allow $\eps(I_1,I_2)$ to take values only $\{0,\lambda\}$ for some $\lambda$. Intuitively, no good code could give rise to such a distribution on $\eps(I_1,I_2)$. Indeed, existing achievable bounds produce distributions on $\eps(I_1,I_2)$ that are close to Gaussian---very different from a distribution taking only two values. The independence of the messages would seem to impose certain restrictions on the distribution of this variable, but the precise nature of these restrictions remains elusive.

Another intriguing area of inquiry relates to hypercontractivity. As discussed in Sec.~\ref{sec:other_measures}, the wringing dependence can be upper bounded by a quantity related to hypercontractivity. However, this upper bound did not actually help in the converse proof. A \emph{lower} bound on wringing dependence could help establish that the regularity conditions of Corollary~\ref{cor:regularity} are satisfied, as one must show that the information capacity region does not grow too much by allowing a small wringing dependence between the channel inputs. It is unclear whether there is some alternative method of wringing that uses hypercontractivity more directly. Another question along these lines is whether there is any connection between the technique used here and that of \cite{Liu2019}, which proves second-order converses for a variety of problems via \emph{reverse} hypercontractivity.

\appendices

\section{Proof of Proposition~\ref{prop:hypercontractivity}}\label{appendix:hyper}

To prove \eqref{eq:hyp_upper_bd}, we take $\delta\in[0,1]$ to be such that $(1+1/\delta,1+\delta)\in\calR_{X;Y}$, and we will show $\Delta(X;Y)\le\delta$. Let $r=1+1/\delta$ and $s=1+\delta$. It was found in \cite{Kamath2012} that an equivalent condition for $(r,s)\in\calR_{X;Y}$ is that, for all $f:\calX\to\bbR$, $g:\calY\to\bbR$,
\be
\bbE [f(X)g(Y)]\le \|f(X)\|_{r'}\|g(Y)\|_s,
\ee
where $r'$ is the H\"older conjugate of $r$, defined by $\frac{1}{r}+\frac{1}{r'}=1$. In this case, since $r=1+1/\delta$, $r'=1+\delta$. Thus, for all real-valued functions $f$ and $g$,
\be\label{eq:ribbon_condition}
\bbE [f(X)g(Y)]\le \|f(X)\|_{1+\delta}\|g(Y)\|_{1+\delta}.
\ee
Given any $\setA\subset\calX,\setB\subset\calY$, let $f(x)=1(x\in \setA)$ and $g(y)=1(y\in \setB)$. Thus
\begin{align}
P_{XY}(\setA,\setB)&=\bbE [f(X)g(Y)]
\\&\le \|f(X)\|_{1+\delta}\|g(Y)\|_{1+\delta}\label{eq:ribbon2}
\\&= \left(\bbE \left[f(X)^{1+\delta}\right]\,\bbE \left[g(Y)^{1+\delta}\right]\right)^{1/(1+\delta)}
\\&= \left(P_X(\setA)P_Y(\setB)\right)^{1/(1+\delta)}.\label{eq:ribbon4}
\end{align}
Therefore, $\delta$ satisfies the feasibility condition in \eqref{eq:Delta_def0} with $Q_X=P_X,Q_Y=P_Y$, so $\Delta(X;Y)\le\delta$.

It follows from the data processing inequality for wringing dependence that $\Delta(X^n;Y^n)$ is non-decreasing in $n$. We now prove the limiting behavior in \eqref{eq:hyp_limit}. Due to the tensorization property of hypercontractivity (cf.~\cite{Mossel2013}), $\calR_{X^n;Y^n}=\calR_{X;Y}$, and so $\Delta_{\text{hyp}}(X^n;Y^n)=\Delta_{\text{hyp}}(X;Y)$. From the upper bound we have already proved, $\Delta(X^n;Y^n)\le \Delta_{\text{hyp}}(X;Y)$ for any $n$.
Now it is enough to show
\be\label{eq:n_hyp_lower_bd}
\lim_{n\to\infty} \Delta(X^n;Y^n)\ge \Delta_{\text{hyp}}(X;Y).
\ee
To prove this lower bound, suppose first that $\calX,\calY$ are finite sets; we will later relax this assumption. We will need some results from the method of types. In particular, let $\calP_n(\calX)$ be the set of $n$-length types on alphabet $\calX$; that is, distributions $P\in\calP(\calX)$ where $P(x)$ is a multiple of $1/n$ for each $x\in\calX$. For a sequence $x^n$, let $P_{x^n}\in\calP_n(\calX)$ be its type:
\be
P_{x^n}(x)=\frac{|\{t:x_t=x\}|}{n}.
\ee
Fix a finite alphabet $\calU$, and a conditional distribution $P_{U|XY}$. Let $P_{UXY}=P_{XY}P_{U|XY}$. For each integer $n$, let $P_{UXY}^{(n)}$ be the element of $\calP_n(\calU\times\calX\times\calY)$ closest in total variational distance to $P_{UXY}$. Note that $d_{TV}(P^{(n)}_{UXY},P_{UXY})\to 0$ as $n\to\infty$. Define the type class
\be
T(X)=\{x^n:P_{x^n}=P_X^{(n)}\};
\ee
$T(U),T(XY)$, etc. are defined similarly. Given a sequence $u^n\in T(U)$, define the conditional type class
\begin{align}
T(X|u^n)&=\{x^n:P_{u^nx^n}=P_{UX}^{(n)}\};
\end{align}
again $T(Y|u^n),T(XY|u^n)$ are defined similarly. A basic result from the method of types (see e.g. \cite[Chap.~11]{Cover1991}) is that
\be
\frac{1}{(n+1)^{|\calX|\cdot|\calU|}} \exp\{nH(X|U)\}\le|T(X|u^n)|\le \exp\{nH(X|U)\}
\ee
where the conditional entropy is with respect to $P^{(n)}_{UXY}$. Moreover, for any $x^n\in T(X|u^n)$,
\be
P_{X^n}(x^n)=\exp\{-n(H(X)+D(P_X^{(n)}\|P_X)\}.
\ee
Similar facts hold for $T(Y|u^n),T(XY|u^n)$. We may now lower bound $\Delta(X^n;Y^n)$ by restricting $\setA$ and $\setB$ to the sets $T(X|u^n)$ and $T(Y|u^n)$ respectively, for some $u^n\in T(U)$. Thus
\begin{align}\label{eq:XnYn_lower_bound}
\Delta(X^n;Y^n)
&\ge \inf_{Q_{X^n},Q_{Y^n}}\ \max_{u^n\in T(U)}
\frac{\log Q_{X^n}(T(X|u^n))Q_{Y^n}(T(Y|u^n))}{\log P_{X^nY^n}(T(X|u^n),T(Y|u^n))}-1.
\end{align}
In this expression, $Q_{X^n}$ is only evaluated on sequences $x^n\in T(X)$. Moreover, the objective function is symmetric among the sequences $x^n$ in this type class. Similar facts hold for $Q_{Y^n}$. Thus, by the convexity of the expression in \eqref{eq:XnYn_lower_bound} in $(Q_{X^n},Q_{Y^n})$, the optimal choices of $Q_{X^n}$ and $Q_{Y^n}$ are uniform over $T(X)$ and $T(Y)$ respectively. Thus, for any $u^n\in T(U)$,
\be
Q_{X^n}(T(X|u^n))=\frac{|T(X|u^n)|}{|T(X)|}
\le (n+1)^{|\calX|} \exp\{-n I(U;X)\}.
\ee
Similarly
\be
Q_{Y^n}(T(Y|u^n))\le (n+1)^{|\calY|} \exp\{-n I(U;Y)\}.
\ee
We may also write
\begin{align}
P_{X^nY^n}(T(X|u^n),T(Y|u^n))
&\ge P_{X^nY^n}(T(XY|u^n))
\\&= |T(XY|u^n)| \exp\{-n(H(XY)+D(P_{XY}^{(n)}\|P_{XY})\}
\\&\ge \frac{1}{(n+1)^{|\calX|\cdot|\calY|\cdot|\calU|}} \exp\{-n (I(U;XY)+D(P_{XY}^{(n)}\|P_{XY}))\}.
\end{align}
Thus
\begin{align}
\Delta(X^n;Y^n)&\ge 
\frac{-n(I(U;X)+I(U;Y))+(|\calX|+|\calY|)\log (n+1)}{-n(I(U;XY)+D(P_{XY}^{(n)}\|P_{XY}))-(|\calX|\cdot|\calY|\cdot|\calU|)\log(n+1)}-1
\end{align}
By the continuity of Kullback-Leibler divergence for finite alphabets, $D(P_{XY}^{(n)}\|P_{XY})\to 0$ as $n\to\infty$. Thus, if we take a limit as $n\to\infty$, we find
\be\label{eq:hyp_lower_bound_MI_form}
\lim_{n\to\infty}\Delta(X^n;Y^n)
\ge \sup_{U} \frac{I(U;X)+I(U;Y)}{I(U;XY)}-1
\ee
where we have taken a supremum over all finite alphabets $\calU$ and all conditional distributions $P_{U|XY}$, and now the mutual informations are with respect to $P_{UXY}$. 

We now show that the RHS of \eqref{eq:hyp_lower_bound_MI_form} is lower bounded by $\Delta_{\text{hyp}}(X;Y)$. As shown in \cite{Nair2014}, for any $r\ge s\ge 1$, $(r,s)\in\calR_{X;Y}$ if and only if
\be\label{eq:ribbon_MI_form}
s\ge \sup_{U} \frac{r I(U;Y)}{rI(U;XY)-(r-1)I(U;X)}
\ee
where the supremum is over variables $U$ with finite alphabets. (In fact, an alphabet of size $2$ is enough.) Consider any $\delta<\Delta_{\text{hyp}}(X;Y)$. By the definition of $\Delta_{\text{hyp}}$ in \eqref{eq:Delta_hyp_def}, it must be that $(1+1/\delta,1+\delta)\notin\calR_{X;Y}$. By the equivalent characterization of $\calR_{X;Y}$ in \eqref{eq:ribbon_MI_form}, this implies there exists a variable $U$ such that
\be
1+\delta<\frac{(1+\frac{1}{\delta})I(U;Y)}{(1+\frac{1}{\delta})I(U;XY)-\frac{1}{\delta}I(U;X)}.
\ee
Rearranging gives
\be
\delta < \frac{I(U;Y)+I(U;X)}{I(U;XY)}-1.
\ee
As this holds for any $\delta<\Delta_{\text{hyp}}(X;Y)$, the RHS of \eqref{eq:hyp_lower_bound_MI_form} is indeed lower bounded by $\Delta_{\text{hyp}}(X;Y)$.

While the above argument only applies for finite alphabets, for infinite alphabets we may apply a quantization argument as follows. Let $[X],[Y]$ be finite quantizations of $X,Y$. We write $[X]^n=([X_1],\ldots,[X_n])$ where each $[X_t]$ is the quantization of $X_t$ using the same quantization. By the data processing inequality and the  fact that we have already proved the lower bound in \eqref{eq:n_hyp_lower_bd} for finite alphabets,
\be
\lim_{n\to\infty}\Delta(X^n;Y^n)
\ge \lim_{n\to\infty}  \Delta([X]^n;[Y]^n)
\ge \Delta_{\text{hyp}}([X];[Y]).
\ee
We may take a supremum on the RHS over all finite quantizations, so it is enough to show that this supremum equals $\Delta_{\text{hyp}}(X;Y)$. Some equivalent forms for $\Delta_{\text{hyp}}$ are as follows:
\begin{align}
\Delta_{\text{hyp}}(X;Y)&=\inf\{\delta\ge 0: \bbE [f(X)g(Y)]\le \|f(X)\|_{1+\delta}\|g(Y)\|_{1+\delta}\text{ for all }f,g\}\\
&=\sup\{\delta\ge 0: \bbE [f(X)g(Y)]> \|f(X)\|_{1+\delta}\|g(Y)\|_{1+\delta}\text{ for some }f,g\}.
\end{align}
Recalling the definition of a \emph{simple} function as one that takes on only finitely many values, we may write
\begin{align}\label{eq:finite_quantizations}
\sup_{\text{finite quantizations }[X],[Y]}
\Delta_{\text{hyp}}([X];[Y])
=\sup\{\delta\ge 0:\bbE [f(X)g(Y)]> \|f(X)\|_{1+\delta}\|g(Y)\|_{1+\delta}\text{ for some simple }f,g\}.
\end{align}
By the usual definition of the Lebesgue integral, if there exist functions $f,g$ such that $\bbE[ f(X)g(Y)]> \|f(X)\|_{1+\delta}\|g(Y)\|_{1+\delta}$, then there also exist simple functions satisfying the same inequality. This proves that the quantity in \eqref{eq:finite_quantizations} equals $\Delta_{\text{hyp}}(X;Y)$.

\section{Proof of Lemma~\ref{lemma:max_corr}}\label{appendix:max_corr}

Assume $\Delta(X;Y)\le\delta$. One way to express the maximal correlation is
\be
\rho_m(X;Y)=\sup_{\substack{f,g:\\ \bbE[ f(X)]=\bbE [g(Y)]=0,\\ \var (f(X))=\var (g(Y))=1}} \bbE [f(X)g(Y)].
\ee
Take any $f,g$ such that $f(X),g(Y)$ have zero mean and unit variance. We wish to show that $\bbE [f(X)g(Y)]\le O(\delta\log\delta^{-1})$. We may define $X'=f(X)$ and $Y'=g(Y)$. By the fact that $\Delta$ satisfies the data processing inequality, $\Delta(X';Y')\le\delta$. To simplify notation, we drop the primes, and assume that $X$ and $Y$ are themselves real-valued random variables with zero mean and unit variance. Now it is enough to show that $\bbE [XY]\le O(\delta\log\delta^{-1})$.

We upper bound $\bbE [XY]$ by breaking into pieces as follows:
\begin{multline}\label{eq:XY_four_parts}
\bbE [XY]=\bbE [XY1(X>0,Y>0)]+\bbE [XY1(X>0,Y<0)]
 +\bbE [XY1(X<0,Y>0)]+\bbE [XY1(X<0,Y<0)].
\end{multline}
We will proceed to show that
\be\label{eq:XY_bound_goal}
\left|\bbE [XY1(X>0,Y>0)]-\bbE [X1(X>0)]\,\bbE [Y1(Y>0)]\right|\le O(\delta\log \delta^{-1}).
\ee
This is enough to prove the lemma, since each term in \eqref{eq:XY_four_parts} can be bounded using \eqref{eq:XY_bound_goal} by swapping $X$ with $-X$ and/or $Y$ with $-Y$. The primary tool we use to prove \eqref{eq:XY_bound_goal} is the consequence of $\Delta(X;Y)\le\delta$ in \eqref{eq:dep_pp_bd}, which upper bounds a joint probability over $P_{XY}$ in terms of the marginal probabilities raised to the power $1/(1+\delta)$. To apply this fact to bound the expectation requires writing the expectation in terms of probabilities, which can be done as follows:
\be
\bbE [XY1(X>0,Y>0)]=\int_0^\infty dx \int_0^\infty dy \bbP(X>x,Y>y)\label{eq:integral_expansion}.
\ee
We may now apply \eqref{eq:dep_pp_bd} to the probability $\bbP(X>x,Y>y)$ to derive the upper bound
\begin{align}
\bbE [XY1(X>0,Y>0)]
&\le (1+2\delta) \int_0^\infty \bbP(X>x)^{1/(1+\delta)} dx \int_0^\infty \bbP(Y>y)^{1/(1+\delta)} dy.\label{eq:two_integrals}
\end{align}
We may now bound one of the integrals in \eqref{eq:two_integrals} by writing
\begin{align}
\int_0^\infty \bbP(X>x)^{1/(1+\delta)}dx-\bbE [X1(X>0)]
&=\int_0^\infty \left[\bbP(X>x)^{1/(1+\delta)}-\bbP(X>x)\right]dx
\\&\le \int_0^\infty \left[\bbP(X>x)^{1/(1+\delta)}-\frac{1}{1+\delta}\bbP(X>x)\right]dx\label{eq:chebyshev0}
\\&\le \int_0^1 \frac{\delta}{1+\delta}dx+\int_1^\infty \left[\left(\frac{1}{x^2}\right)^{1/(1+\delta)}-\frac{1}{(1+\delta)x^2}\right]dx\label{eq:chebyshev1}
\\&=\frac{4\delta}{1-\delta^2}\label{eq:chebyshev2}
\\&=O(\delta)\label{eq:chebyshev4}
\end{align}
where \eqref{eq:chebyshev1} holds because the function $p\mapsto p^{1/(1+\delta)}-\frac{p}{1+\delta}$ is an increasing function for any $\delta$ with a maximum value of $\frac{\delta}{1+\delta}$, and since $\bbP(X>x)\le 1/x^2$ from the assumption that $\bbE [X^2]=1$ and Chebyshev's inequality.
 Since the same argument holds for the integral over $y$ in \eqref{eq:two_integrals}, we have
\begin{align}
\bbE [XY1(X>0,Y>0)]&\le (1+2\delta)\left(\bbE [X1(X>0)]+O(\delta)\right)\left(\bbE [Y1(Y>0)]+O(\delta)\right)
\\&\le \bbE [X1(X>0)]\,\bbE [Y1(Y>0)]+O(\delta)\label{eq:EXY_upper_bd}
\end{align}
where we have used the fact that
\be
\bbE [X1(X>0)]\le \sqrt{\bbE [X^2 1(X>0)]}\le \sqrt{\bbE [X^2]}\le 1
\ee
and the same holds for $Y$.

We now lower bound $\bbE [XY1(X>0,Y>0)]$.  Again using the integral expansion in \eqref{eq:integral_expansion}, we may do so by lower bounding $\bbP(X>x,Y>y)$.  It will be convenient to define the function
\be\label{eq:k_delta_def}
k_\delta(p)=\begin{cases}(1+2\delta) p^{1/(1+\delta)}-p, & p\le 1\\ 2\delta, & p>1\end{cases}.
\ee
For $p\ge 0$, $k_\delta(p)$ is non-decreasing, concave, and $0\le k_\delta(p)\le 2\delta$. For any $x\ge 0,y\ge 0$,
\begin{align}
\bbP(X>x,Y>y)&=\bbP(X>x)-\bbP(X>x,Y\le y)\label{eq:PXY_lower_bd0}
\\&\ge \bbP(X>x)-(1+2\delta)\left[\bbP(X>x)\bbP(Y\le y)\right]^{1/(1+\delta)}\label{eq:PXY_lower_bd1}
\\&= \bbP(X>x)\bbP(Y>y)+\bbP(X>x)\bbP(Y\le y)-(1+2\delta)\left[\bbP(X>x)\bbP(Y\le y)\right]^{1/(1+\delta)}
\\&=\bbP(X>x)\bbP(Y>y)-k_\delta(\bbP(X>x,Y\le y))\label{eq:PXY_lower_bd2}
\\&\ge \bbP(X>x)\bbP(Y>y)-k_\delta(\bbP(X>x))\label{eq:PXY_lower_bd}
\end{align}
where in \eqref{eq:PXY_lower_bd1} we have again applied \eqref{eq:dep_pp_bd}, in \eqref{eq:PXY_lower_bd2} we have used the definition of $k_\delta$, and in \eqref{eq:PXY_lower_bd} we have used the fact that $k_\delta$ is non-decreasing. We may now bound
\begin{align}
&\bbE [X1(X>0)]\,\bbE [Y1(Y>0)]-\bbE [XY1(X>0,Y>0)]
\\&=\int_0^\infty dx \int_0^\infty dy \left[ \bbP(X>x)\bbP(Y>y)-\bbP(X>x,Y>y)\right]
\\&\le \int_0^\infty dx \int_0^\infty dy \min\{\bbP(X>x)\bbP(Y>y),\, k_\delta(\bbP(X>x)),\, k_\delta(\bbP(Y>y))\label{eq:3min}
\end{align}
where \eqref{eq:3min} holds by three upper bounds on $\bbP(X>x)\bbP(Y>y)-\bbP(X>x,Y>y)$: the fact that $\bbP(X>x,Y>y)\ge 0$, the bound in \eqref{eq:PXY_lower_bd}, and the bound in \eqref{eq:PXY_lower_bd} with $X$ and $Y$ swapped. To further upper bound \eqref{eq:3min}, we separate the integral over $x$ and $y$ into three regions: when $x,y\ge \delta^{-1/2}$, we upper bound the integrand by $\bbP(X>x)\bbP(Y>y)$; when $y\le x$ and $y \le \delta^{-1/2}$, we upper bound the integrand by $k_\delta(\bbP(X>x))$; when $x\le y$ and $x\le\delta^{-1/2}$, we upper bound the integrand by $k_\delta(\bbP(Y>y))$. Thus \eqref{eq:3min} is at most
\begin{multline}\label{eq:three_terms}
\int_{\delta^{-1/2}}^\infty \bbP(X>x) dx \int_{\delta^{-1/2}}^\infty \bbP(Y>y) dy
 +\int_0^\infty dx \int_0^{\min\{x,\delta^{-1/2}\}}dy k_\delta(\bbP(X>x))
\\ +\int_0^\infty dy \int_0^{\min\{y,\delta^{-1/2}\}}dx k_\delta(\bbP(Y>y)).
\end{multline}
We now bound each term in \eqref{eq:three_terms} in turn. In the first term in \eqref{eq:three_terms}, Chebyshev's inequality gives
\begin{align}
\int_{\delta^{-1/2}}^\infty \bbP(X>x)dx
\le \int_{\delta^{-1/2}}^\infty \frac{1}{x^2}dx
=\sqrt{\delta}.
\end{align}
The same calculation holds for $Y$, so the first term in \eqref{eq:three_terms} is at most $\delta$. The second term in \eqref{eq:three_terms} may be bounded by
\begin{align}
&\int_0^\infty \min\{x,\delta^{-1/2}\} k_\delta(\bbP(X>x))dx
\\&=\int_0^{\delta^{-1/2}} x\, k_\delta(\bbP(X>x))dx+\delta^{-1/2}\int_{\delta^{-1/2}}^\infty k_\delta(\bbP(X>x))dx
\\&\le \frac{1}{2\delta} \int_0^{\delta^{-1/2}} 2\delta x\, k_\delta(\bbP(X>x))dx+\delta^{-1/2}\int_{\delta^{-1/2}}^\infty k_\delta(1/x^2)dx
\label{eq:integral1z}
\\&\le \frac{1}{2\delta} k_\delta\left(\int_0^{\delta^{-1/2}} 2\delta x \bbP(X>x)\right)+\delta^{-1/2}\left(\frac{(1+2\delta)(1+\delta)}{1-\delta} \delta^{\frac{1-\delta}{2(1+\delta)}}-\delta^{1/2}\right)\label{eq:integral1a}
\\&\le\frac{1}{2\delta}k_\delta(\delta)+\frac{(1+2\delta)(1+\delta)}{1-\delta} \delta^{-\delta/(1+\delta)}-1\label{eq:integral1b}
\\&=\frac{1}{2}\left((1+2\delta)\delta^{-\delta/(1+\delta)}-1\right)+\frac{(1+2\delta)(1+\delta)}{1-\delta} \delta^{-\delta/(1+\delta)}-1\
\label{eq:integral1c}
\\&=O(-\delta\log\delta)\label{eq:integral1d}
\end{align}
where \eqref{eq:integral1z} holds by Chebyshev's inequality and the fact that $k_\delta$ is increasing; \eqref{eq:integral1a} holds since $k_\delta$ is concave and $\int_0^{\delta^{-1/2}} 2\delta x=1$; \eqref{eq:integral1b} holds since
\be
\int_0^{\delta^{-1/2}} 2 x \bbP(X>x)\le \int_0^\infty 2x \bbP(X>x)= \bbE [X^2]=1
\ee
and \eqref{eq:integral1d} holds since $\delta^{-\delta/(1+\delta)}=1-\delta\log\delta+O(\delta^2\log^2\delta)$. The third term in \eqref{eq:three_terms} may be bounded by an identical calculation. This completes the proof of \eqref{eq:XY_bound_goal}, which therefore proves the lemma.

\section{Proof of Lemma~\ref{lemma:DMCs}}\label{appendix:dmc}

Given that $\Delta(X;Y)\le\delta$, 
\begin{align}
d_{TV}(P_{XY},P_XP_Y)
&=\sum_{x,y} |P_{XY}(x,y)-P_X(x)P_Y(y)|^+
\\&=\sum_x \sum_{y:P_{XY}(x,y)>P_X(x)P_Y(y)} (P_{XY}(x,y)-P_X(x)P_Y(y))
\\&\le \sum_x 2\delta\label{eq:dtv_independence_bd}
\\&=2\delta|\calX|
\end{align}
where in \eqref{eq:dtv_independence_bd} we have applied \eqref{eq:gdep_abs_bd} from Thm.~\ref{thm:props} with the particularizations $\setA=\{x\}$ and $\setB=\{y:P_{XY}(x,y)>P_X(x)P_Y(y)\}$. Applying the same argument swapping $X$ and $Y$ gives
\be
d_{TV}(P_{XY},P_XP_Y)\le 2\delta\min\{|\calX|,|\calY|\}.
\ee
Since $Z$ is the output of the channel with $X,Y$ as the inputs, while $\tilZ$ is the output of the channel with $\tilX,\tilY$ as the inputs, this also means that $d_{TV}(P_{XYZ},P_{\tilX\tilY\tilZ})\le 2\delta\min\{|\calX|,|\calY|\}$.

We may relate the conditional entropies as follows:
\begin{align}
H(Z|X,Y)
&=\sum_{x,y} P_{XY}(x,y) H(Z|X=x,Y=y)
\\&\ge \sum_{x,y} P_X(x)P_Y(y)H(Z|X=x,Y=y)-\sum_{x,y} |P_{XY}(x,y)-P_X(x)P_Y(y)|^+ H(Z|X=x,Y=y)
\\&\ge H(\tilZ|\tilX,\tilY)-2\delta\min\{|\calX|,|\calY|\}\log|\calZ|.\label{eq:conditional_bound}
\end{align}

To complete the proof of the lemma, we must bound $H(Z)$, $H(Z|X)$, and $H(Z|Y)$. The main difficulty is that the entropy is not Lipschitz continuous, so the fact that the total variational distance is $O(\delta)$ does not immediately imply that the entropies differ by $O(\delta)$. We circumvent this problem using the stronger consequence of $\Delta(X;Y)\le\delta$ in \eqref{eq:dep_pp_bd} from Thm.~\ref{thm:props}. We first bound $H(Z)$. Let $z\in\calZ$ be such that $P_{\tilZ}(z)\ge 1/4$. Then by the total variational bound,
\be
P_Z(z)\ge P_{\tilZ}(z)-2\delta \min\{|\calX|,|\calY|\} \ge e^{-2}
\ee
where the second inequality holds for sufficiently small $\delta$, and since $e^{-2}<1/4$. Consider the function $f(p)=-p\log p$. Since $f'(p)=-\log p-1$, if $p\ge e^{-2}$ then
\be
|f'(p)|\le 1.
\ee
Since we have established that $P_Z(z),P_{\tilZ}(z)\ge  e^{-2}$, and $|P_Z(z)-P_{\tilZ}(z)|\le 2 \delta\min\{|\calX|,|\calY|\}$, we have
\be
-P_Z(z)\log P_Z(z)\le -P_{\tilZ}(z)\log P_{\tilZ}(z)+2\min\{|\calX|,|\calY|\}\delta.
\ee
Note there are at most $4$ values of $z$ where $P_{\tilZ}(z)\ge 1/4$, so
\be
\sum_{z:P_{\tilZ}(z)\ge 1/4}\left[-P_Z(z)\log P_Z(z)+P_Z(z)\log P_{\tilZ}(z)\right]\le 8\min\{|\calX|,|\calY|\}\delta.
\ee

Now suppose $z\in\calZ$ is such that $P_{\tilZ}(z)<1/4$. Let $r_z=\sum_{x,y} W(z|x,y)$. Assume without loss of generality that all letters in $\calZ$ are reachable (i.e. $W(z|x,y)>0$ for some $x,y$). Thus $r_z\ge W_{\min}$. We may now bound
\begin{align}
P_Z(z)&=\sum_{x,y} P_{XY}(x,y) W(z|x,y)
\\&\le \sum_{x,y} (1+2\delta)(P_X(x)P_Y(y))^{1/(1+\delta)} W(z|x,y)\label{eq:PtilZ0b}
\\&=(1+2\delta)r_z \sum_{x,y} \frac{W(z|x,y)}{r_z} (P_X(x)P_Y(y))^{1/(1+\delta)}
\\&\le (1+2\delta) r_z \left(\sum_{x,y} \frac{W(z|x,y)}{r_z} P_X(x)P_Y(y)\right)^{1/(1+\delta)}\label{eq:PtilZ0}
\\&=(1+2\delta) r_z^{-\delta/(1+\delta)} P_{\tilZ}(z)^{1/(1+\delta)}
\\&\le (1+2\delta) W_{\min}^{-\delta/(1+\delta)} P_{\tilZ}(z)^{1/(1+\delta)}
\\&\le (1+2\delta)(1-\delta\log W_{\min}+O(\delta^2)) P_{\tilZ}(z)^{1/(1+\delta)}\label{eq:PtilZ0a}
\end{align}
where \eqref{eq:PtilZ0b} follows from \eqref{eq:dep_pp_bd}, and \eqref{eq:PtilZ0} holds by the definition of $r_z$ and by the concavity of the function $p^{1/(1+\delta)}$.
By the assumption that $P_{\tilZ}(z)<1/4$, for sufficiently small $\delta$,  \eqref{eq:PtilZ0a} is less than $e^{-1}$. Thus, we are in the increasing regime of the function $-p\log p$. In particular
\begin{align}
-P_Z(z)\log P_Z(z)&\le- \left[(1+2\delta) (1-\delta\log W_{\min}+O(\delta^2)) P_{\tilZ}(z)^{1/(1+\delta)}\right]\nonumber
\\&\qquad\cdot \log\left[(1+2\delta) (1-\delta\log W_{\min}+O(\delta^2)) P_{\tilZ}(z)^{1/(1+\delta)}\right]
\\&\le-\frac{1+2\delta}{1+\delta}(1-\delta\log W_{\min}+O(\delta^2)) P_{\tilZ}(z)^{1/(1+\delta)}\log P_{\tilZ}(z)\label{eq:PtilZ1}
\end{align}
where in \eqref{eq:PtilZ1} we have simply dropped terms greater than $1$ inside the log. Here we need a technical result. For any $p\in[0,1]$, let $g_p(\delta)=-p^{1/(1+\delta)}\log p$. We claim that for all $\delta\ge 0$,
\be\label{eq:gq_claim}
g_p(\delta)\le -p\log p+4e^{-2}\delta.
\ee
Since $g_p(0)=-p\log p$, it is enough to show that $g'_p(\delta)\le 4e^{-2}$ for all $\delta$. The first and second derivatives of $g_p$ are
\begin{align}
g'_p(\delta)&=\frac{p^{1/(1+\delta)}\log^2 p}{(1+\delta)^2},
\\g''_p(\delta)&=p^{1/(1+\delta)}\log^2 p\left(\frac{-2}{(1+\delta)^3}-\frac{\log p}{(1+\delta)^4}\right).
\end{align}
Note that $g''_p(\delta)\le 0$ iff
\be
-2(1+\delta)-\log p\le 0.
\ee
That is, $g'_p(\delta)$ is maximized at $\delta=\frac{-\log p}{2}-1$. Thus
\be
g'_p(\delta)\le \frac{p^{\frac{2}{-\log p}}\log^2 p}{\left(\frac{-\log p}{2}\right)^2}
=4p^{\frac{2}{-\log p}}=4 \exp\left\{\log p \frac{2}{-\log p}\right\}=4e^{-2}.
\ee
This proves the claim in \eqref{eq:gq_claim}. Applying this result to \eqref{eq:PtilZ1} gives
\begin{align}
-P_Z(z)\log P_Z(z)
&\le \frac{1+2\delta}{1+\delta}(1-\delta\log W_{\min}+O(\delta^2)) \left[-P_{\tilZ}(z)\log P_{\tilZ}(z)+ 4e^{-2}\delta\right]
\\&\le -P_{\tilZ}(z)\log P_{\tilZ}(z)+\left[(1-\log W_{\min})e^{-1}+4e^{-2}\right]\delta+O(\delta^2)\label{eq:PtilZ2}
\end{align}
where in \eqref{eq:PtilZ2} we have used the fact that $-p\log p\le e^{-1}$. Therefore
\begin{align}
H(Z)-H(\tilZ)
&\le 8\min\{|\calX|,|\calY|\}\delta
+\sum_{z:P_{\tilZ}(z)<1/4}
\left(\left[(1-\log W_{\min})e^{-1}+4e^{-2}\right]\delta+O(\delta^2)\right)
\\&\le
\left[8\min\{|\calX|,|\calY|\}
+|\calZ|\left((1-\log W_{\min})e^{-1}+4e^{-2}\right)\right]\delta+O(\delta^2)\label{eq:HZ_bound}
\end{align}
Combining \eqref{eq:HZ_bound} with the bound on conditional entropy in \eqref{eq:conditional_bound} proves \eqref{eq:MI_DMC_bound}.

To prove the bound on $I(X;Z|Y)$ in \eqref{eq:MIX_DMC_bound}, we need to bound $H(Z|Y)$, or equivalently $H(Y,Z)$, since $H(Y)=H(\tilY)$. We may almost the same argument as above, but with the joint distribution $P_{YZ}$ in place of $P_Z$. In particular, if $P_{\tilY\tilZ}(y,z)\ge 1/4$, then
\be
-P_{YZ}(y,z)\log P_{YZ}(y,z)\le -P_{\tilY\tilZ}(y,z)\log P_{\tilY\tilZ}(y,z)+2\min\{|\calX|,|\calY|\}\delta.
\ee
To deal with $P_{\tilY\tilZ}(y,z)< 1/4$, let $r_{z|y}=\sum_x W(z|x,y)$. If $r_{z|y}=0$, then $P_{YZ}(y,z)=P_{\tilY\tilZ}(y,z)=0$, so this letter pair can be discarded. Otherwise, $r_{z|y}\ge W_{\min}$, so
\begin{align}
P_{YZ}(y,z)&=\sum_x P_{XY}(x,y)W(z|x,y)
\\&\le \sum_x (1+2\delta) (P_X(x)P_Y(y))^{1/(1+\delta)}W(z|x,y)
\\&\le (1+2\delta)r_{z|y}^{-\delta/(1+\delta)} P_{\tilY\tilZ}(y,z)^{1/(1+\delta)}
\\&\le (1+2\delta)W_{\min}^{-\delta/(1+\delta)}P_{\tilY\tilZ}(y,z)^{1/(1+\delta)}.
\end{align}
The remainder of the proof is essentially identical, and so we find
\be
H(Z|Y)\le H(\tilZ|\tilY)+\left[8\min\{|\calX|,|\calY|\}
+|\calY|\cdot|\calZ|\left((1-\log W_{\min})e^{-1}+4e^{-2}\right)\right]\delta+O(\delta^2).
\ee
Combining with the bound on the entropy conditioned on $X,Y$ in \eqref{eq:conditional_bound} proves \eqref{eq:MIX_DMC_bound}. The bound on $I(Y;Z|X)$ in \eqref{eq:MIY_DMC_bound} is proved by the same argument.

\section{Proof of Prop.~\ref{prop:bamac}}\label{appendix:bamac}

If $\delta\ge \frac{1-\log_2(1+2^{-\alpha})}{1+\log_2(1+2^{-\alpha})}$, then we may simply ignore the constraint on the wringing dependence, so
\be
C_{1,\alpha}(\delta)\le \max_{P_{XY}} \bar\gamma\left[\alpha H(X+Y)+(1-\alpha)H(X|Y)\right]
=\bar\gamma\left[\log(1+2^{-\alpha})+\alpha\log 2\right].
\ee
Now consider $\delta< \frac{1-\log_2(1+2^{-\alpha})}{1+\log_2(1+2^{-\alpha})}$. We define for convenience $r_{z}=\bbP(X+Y=z)$ for $z=0,1,2$. Note that
\be
\alpha H(X+Y)+(1-\alpha)H(X|Y)
\le \alpha H(X+Y)+(1-\alpha)H(X\oplus Y)
\\=\alpha H(r_0,r_1,r_2)+(1-\alpha)H_b(r_0+r_2)
\ee
where $\oplus$ is modulo 2 addition, and we have used the fact that $X\oplus Y=0$ iff $X+Y\in\{0,2\}$. Since $\Delta(X;Y)\le\delta$, using the properties of the wringing dependence in Thm.~\ref{thm:props}, there exist $Q_X,Q_Y\in\calP(\{0,1\})$ such that
\begin{align}
r_0&=P_{XY}(0,0)
\le (Q_X(0)Q_Y(0))^{1/(1+\delta)}.
\end{align}
Similarly $r_2\le (Q_X(1)Q_Y(1))^{1/(1+\delta)}$. Thus
\begin{align}
\sqrt{r_{0}}+\sqrt{r_2}
&\le (Q_X(0)Q_Y(0))^{1/(2(1+\delta))}+(Q_X(1)Q_Y(1))^{1/(2(1+\delta))}\label{eq:r0r2_constraint0}
\\&\le 2^{1-1/(1+\delta)}\label{eq:r0r2_constraint}
\end{align}
where \eqref{eq:r0r2_constraint} holds because $(pq)^{\rho}$ is concave in $(p,q)$ for $0\le\rho\le 1$, and so the quantity in \eqref{eq:r0r2_constraint0} is maximized with $Q_X(0)=Q_Y(0)=1/2$. We may rewrite the constraint in \eqref{eq:r0r2_constraint} as
\be
4r_0r_2\le (2^{1-1/(1+\delta)}-r_0-r_2)^2.\label{eq:quadratic_constraint}
\ee
Thus
\begin{align}
&\alpha H(r_0,r_1,r_2)+(1-\alpha) H_b(r_0+r_2)
\\&\le \max_{\substack{r_0,r_2\in[0,1]:\\ r_0+r_2\le 1,\\4r_0r_2\le (2^{1-1/(1+\delta)}-r_0-r_2)^2}}
\big[-(1-r_0-r_2)\log(1-r_0-r_2)+\alpha(-r_0\log r_0-r_2\log r_2)-(1-\alpha)(r_0+r_2)\log(r_0+r_2)\big]
\\&\le \min_{\lambda\ge 0}\ \max_{\substack{r_0,r_2\in[0,1]:\\ r_0+r_2\le 1}}\ 
\big[-(1-r_0-r_2)\log(1-r_0-r_2)+\alpha(-r_0\log r_0-r_2\log r_2)-(1-\alpha)(r_0+r_2)\log(r_0+r_2)\nonumber
\\&\qquad+\lambda((2^{1-1/(1+\delta)}-r_0-r_2)^2-4r_0r_2)\big].\label{eq:r0r2_function}
\end{align}
Let $f(r_0,r_2;\lambda)$ be the function in \eqref{eq:r0r2_function}. We claim that for any $\lambda\le \alpha$, $f(r_0,r_2;\lambda)$ is concave in $(r_0,r_2)$. The Hessian with respect to $(r_0,r_2)$ is given by
\be
\nabla^2 f(r_0,r_2;\lambda)=\left[\begin{array}{cc} 
-\frac{r_0+r_2(1-r_0-r_2)\alpha}{r_0(1-r_0-r_2)(r_0+r_2)}+\lambda
 & -\frac{1-(1-r_0-r_2)\alpha}{(1-r_0-r_2)(r_0+r_2)}-\lambda\\
-\frac{1-(1-r_0-r_2)\alpha}{(1-r_0-r_2)(r_0+r_2)}-\lambda
 & -\frac{r_2+r_0(1-r_0-r_2)\alpha}{r_2(1-r_0-r_2)(r_0+r_2)}+\lambda
\end{array}\right].
\ee
We need to show that $\nabla^2 f(r_0,r_2;\lambda)$ is negative semi-definite; this requires that the upper left element is non-positive, and the determinant is non-negative. The upper left element is given by
\begin{align}
-\frac{r_0+r_2(1-r_0-r_2)\alpha}{r_0(1-r_0-r_2)(r_0+r_2)}+\lambda
&\le-\frac{1}{(1-r_0-r_2)(r_0+r_2)}+\lambda\label{eq:upper_left1}
\\&\le -4+\lambda\label{eq:upper_left2}
\\&\le -3\label{eq:upper_left3}
\end{align}
where \eqref{eq:upper_left1} holds because $\alpha\ge 0$, \eqref{eq:upper_left2} holds because $x(1-x)\le 1/4$, and \eqref{eq:upper_left3} holds by the assumption that $\lambda\le\alpha\le 1$. The determinant of the Hessian is given by
\begin{align}
|\nabla^2 f(r_0,r_2;\lambda)|&=\frac{(r_0+r_2)\alpha-(4r_0r_2+(1-r_0-r_2)(r_0-r_2)^2\alpha)\lambda}{r_0r_2(1-r_0-r_2)(r_0+r_2)}
\\&\ge \frac{\alpha\left[r_0+r_2-4r_0r_2-(1-r_0-r_2)(r_0-r_2)^2\alpha\right]}{r_0r_2(1-r_0-r_2)(r_0+r_2)}\label{eq:hessian0}
\\&\ge \frac{\alpha\left[r_0+r_2-4r_0r_2-(1-r_0-r_2)(r_0-r_2)^2\right]}{r_0r_2(1-r_0-r_2)(r_0+r_2)}\label{eq:hessian1}
\\&=\frac{\alpha\left[1-r_0(1-r_0)-r_2(1-r_2)-2r_0r_2\right]}{r_0r_2(1-r_0-r_2)}\label{eq:hessian5}
\\&\ge 0\label{eq:hessian6}
\end{align}
where \eqref{eq:hessian0} holds by the assumption that $\lambda\le\alpha$, \eqref{eq:hessian1} holds since $\alpha\le 1$, and \eqref{eq:hessian6} holds again since $x(1-x)\le 1/4$ and since $r_0+r_2\le 1$.
We may upper bound \eqref{eq:r0r2_function} by choosing any $\lambda\ge 0$. With some hindsight, we choose
\be
\lambda=2^{-2+1/(1+\delta)} \left[\log\left(2^{-1+2/(1+\delta)}-1\right)+\alpha\log 2\right].
\ee
Note that $\lambda\ge 0$ if 
\be
1\le 2^{\alpha}\left(2^{-1+2/(1+\delta)}-1\right).
\ee
This indeed holds by the assumption that $\delta< \frac{1-\log_2(1+2^{-\alpha})}{1+\log_2(1+2^{-\alpha})}$. In addition, noting that $\lambda$ is decreasing in $\delta$,
\be
\lambda\le 2^{-1} \left[\log (2^{1}-1)+\alpha\log 2\right]=\frac{\alpha\log 2}{2}<\alpha.
\ee
Thus, by the above claim, for this value of $\lambda$, $f(r_0,r_2;\lambda)$ is concave. Since the function is also symmetric between $r_0$ and $r_2$, it is maximized at $r_0=r_2=r$. Differentiating this function, the maximizing value of $r$ is found at
\be
0=\frac{d}{dr} f(r,r;\lambda)=
2\log(1-2r)-2\log r-(1-\alpha)2\log 2-4\cdot 2^{1-1/(1+\delta)}\lambda
\ee
This is solved at $r=2^{-2/(1+\delta)}$. At this value, the constraint in \eqref{eq:quadratic_constraint} holds with equality. Thus the upper bound from \eqref{eq:r0r2_function} becomes
\begin{align}
\alpha H(r_0,r_1,r_2)+(1-\alpha) H_b(r_0+r_2)
&\le H_b(2^{1-2/(1+\delta)})+\alpha 2^{1-2/(1+\delta)}\log 2.
\end{align}
This gives an upper bound on $C_{1,\alpha}(\delta)$ that exactly matches the lower bound in \eqref{eq:BAMAC_lb}.

\section{Proof of Thm.~\ref{thm:gaussian}}\label{appendix:gaussian}

\subsection{Bounding \texorpdfstring{$C'_{\alpha_1,\alpha_2}(0)$}{Calpha1alpha2'(0)}}

Let $(\alpha_1,\alpha_2)=(1,\alpha)$ for $\alpha\in[0,1]$. Recall that
\be\label{eq:gaussian_Cdelta}
C_{1,\alpha}(\delta)=\sup_{\substack{X,Y,U:\Delta(X;Y|U=u)\le\delta\ \forall u,\\ \bbE [X^2]\le S_1,\\ \bbE [Y^2]\le S_2}} \big[\alpha I(X,Y;Z|U)+(1-\alpha) I(X;Z|Y,U)\big].
\ee
Note that 
\be
C_{1,\alpha}(0)=\alpha \frac{1}{2}\log (1+S_1+S_2)+(1-\alpha)\frac{1}{2}\log (1+S_1).
\ee
Since $C_{1,\alpha}(\delta)$ is convex in $\alpha$,
\be
C_{1,\alpha}(\delta)\le \alpha C_{1,1}(\delta)+(1-\alpha)C_{1,0}(\delta).
\ee
We may easily bound the second term:
\begin{align}
C_{1,0}(\delta)&=\sup_{\substack{X,Y,U:\Delta(X;Y|U=u)\le\delta\ \forall u,\\ \bbE [X^2]\le S_1,\\ \bbE [Y^2]\le S_2}} I(X;Z|Y,U)
\\&\le \sup_{X,Y:\bbE [X^2]\le S_1,\bbE [Y^2]\le S_2} h(X+N)-h(N)
\\&\le \frac{1}{2}\log(1+S_1)
\\&=C_{1,0}(0)
\end{align}
where $h(\cdot)$ denotes the differential entropy. This implies that $C'_{1,0}(0)=0$. Thus, to uniformly bound $C'_{1,\alpha}(\delta)$ for all $\alpha$, it is enough to prove that $C'_{1,1}(0)<\infty$. Let $X,Y,U$ be any set variables satisfying the constraints in the infimum in \eqref{eq:gaussian_Cdelta}.  Note that
\begin{align}
I(X,Y;Z|U)
&\le  h(Z|U)-h(N)
\\&=  h(Z|U)-\frac{1}{2}\log 2\pi e.
\end{align}
Now it is enough to show $h(Z|U)\le \frac{1}{2}\log 2\pi e (1+S_1+S_2)+O(\delta)$. For each $u$, let $S_{1u}=\bbE [X^2|U=u],S_{2u}=\bbE[Y^2|U=u]$. Thus $\sum_u P_U(u) S_{1u}\le S_1$, $\sum_u P_U(u) S_{2u}\le S_2$. Our goal is to show that, for each $u$
\be\label{eq:gaussian_goal}
h(Z|U=u)\le \frac{1}{2}\log 2\pi e(1+S_{1u}+S_{2u})+O(\delta)
\ee
which implies
\be
h(Z|U)=\sum_u P_U(u)h(Z|U=u)\le \frac{1}{2}\log 2\pi e(1+S_1+S_2)+O(\delta)
\ee
where we have used the concavity of the log. For convenience, for the remainder of the proof we drop the conditioning on $u$. Throughout this proof, we are careful to use $O(\cdot)$ notation only when the implied constant is universal, and in particular does not depend on $S_1,S_2$. 

We may assume without loss of generality that $X$ and $Y$ have zero mean, since if they do not, shifting their means to zero does not change $h(Z)$, and only reduces $\bbE [X^2],\bbE [Y^2]$. For convenience define $S=1+S_1+S_2$. Since our goal to is to prove \eqref{eq:gaussian_goal}, we may  assume
\be
h(Z)\ge  \frac{1}{2}\log(2\pi eS)\label{eq:hz_simple_bd}
\ee
because otherwise we have nothing to prove. Let $\sigma_Z^2=\bbE [Z^2]$. Since $\Delta(X;Y)\le\delta$, from Lemma~\ref{lemma:max_corr}, $\rho_m(X;Y)\le O(\delta\log\delta^{-1})$. This implies that $\bbE [XY]\le \sqrt{S_1S_2}\, O(\delta\log \delta^{-1})$. Thus,
\begin{align}
\sigma_Z^2&=\bbE [(X + Y+N)^2]
\\&=S+2\,\bbE [XY]\label{eq:sigmaZ_bd1}
\\&\le S+2\sqrt{S_1S_2}\,O(\delta\log\delta^{-1})
\\&\le S+S\,O(\delta\log\delta^{-1})\label{eq:sigmaZ_bd}
\end{align}
where in \eqref{eq:sigmaZ_bd1} we have used the fact that $N$ is independent from $(X,Y)$, and \eqref{eq:sigmaZ_bd} follows because $2\sqrt{S_1S_2}\le S_1+S_2\le S$. Let $\tilZ\sim\calN(0,S)$, so
\begin{align}
h(Z)&=\frac{1}{2}\log 2\pi S+\frac{\sigma_Z^2}{2S} -D(P_Z\|P_{\tilZ})
\\&\le \frac{1}{2}\log 2\pi S+\frac{1}{2}+O(\delta\log\delta^{-1})-2 d_{TV}(P_Z\|P_{\tilZ})^2\label{eq:pinsker1}
\\&=\frac{1}{2}\log 2\pi e S+O(\delta\log\delta^{-1})-2 d_{TV}(P_Z\|P_{\tilZ})^2
\end{align}
where the \eqref{eq:pinsker1} follows from the bound on $\sigma_Z^2$ in \eqref{eq:sigmaZ_bd} and from Pinsker's inequality. Applying the lower bound on $h(Z)$ from \eqref{eq:hz_simple_bd} gives
\be\label{eq:tv_bound}
d_{TV}(P_Z\|P_{\tilZ})\le O(\sqrt{\delta \log\delta^{-1}}).
\ee
For any function $f:\bbR\to[0,f_{\max}]$,
\begin{align}
\left|\bbE [f(Z)]-\bbE [f(\tilZ)]\right|
&=\left|\int_0^{f_{\max}} [\bbP(f(Z)>a)-\bbP(f(\tilZ)>a)]da\right|
\\&\le \int_0^{f_{\max}} \left|\bbP(f(Z)>a)-\bbP(f(\tilZ)>a)\right|da
\\&\le f_{\max} d_{TV}(P_Z\|P_{\tilZ})\label{eq:f_tv_bound3}
\\&\le f_{\max}\, O(\sqrt{\delta \log\delta^{-1}}).\label{eq:f_tv_bound}
\end{align}
where \eqref{eq:f_tv_bound3} follows from the fact that for any $\setA\subset\bbR$, $|P_Z(\setA)-P_{\tilZ}(\setA)|\le d_{TV}(P_Z,P_{\tilZ})$.

The following definitions will be key to the remainder of the proof:
\begin{align}
\tau_X&=\frac{S_1}{\sqrt{S}}-\frac{\sqrt{S}}{8}\log\delta,\\
\tau_Y&=\frac{S_2}{\sqrt{S}}-\frac{\sqrt{S}}{8}\log\delta,\\
\tau_N&=\frac{1}{\sqrt{S}},\\
\tau_Z&=\tau_X+\tau_Y+\tau_N=
\sqrt{S}\left(1-\frac{1}{4}\log\delta\right),\\
m_X&=\bbE \left[e^{X/\sqrt{S}}1(X<\tau_X)\right],\\
m_Y&=\bbE \left[e^{Y/\sqrt{S}}1(Y<\tau_Y)\right].
\end{align}
Similarly to the proof of Lemma~\ref{lemma:max_corr}, the core of the proof involves upper and lower bounding 
\be\label{eq:EXY_difference}
\bbE [XY1(X>0,Y>0)]-\bbE [X1(X>0)]\, \bbE [Y1(Y>0)].
\ee
Since $\Delta(X;Y)\le\delta$, the same argument as in \eqref{eq:integral_expansion}--\eqref{eq:EXY_upper_bd} shows that the quantity \eqref{eq:EXY_difference} is upper bounded by 
\be
\sqrt{S_1S_2}\,O(\delta)\le S\,O(\delta).
\ee
To lower bound \eqref{eq:EXY_difference}, we cannot use precisely the same argument as in Lemma~\ref{lemma:max_corr}, since we need a bound that eliminates the $\log\delta^{-1}$ term. We first divide \eqref{eq:EXY_difference} into four terms:
\begin{align}
&\bbE [XY1(X>0,Y>0)]-\bbE [X1(X>0)]\, \bbE [Y1(Y>0)]\nonumber
\\&=\big(\bbE [XY1(0<X<\tau_X,0<Y<\tau_Y)]-\bbE [X1(0<X<\tau_X)]\, \bbE [Y1(0<Y<\tau_Y)]\big)\nonumber
\\&\qquad+\big(\bbE [XY1(X\ge\tau_X,0<Y<\tau_Y)]-\bbE [X1(X\ge \tau_X)]\, \bbE [Y1(0<Y<\tau_Y)]\big)\nonumber
\\&\qquad+\big(\bbE [XY1(0<X<\tau_X,Y\ge\tau_Y)]-\bbE [X1(0<X<\tau_X)]\, \bbE [Y1(Y\ge\tau_Y)]\big)\nonumber
\\&\qquad+\big(\bbE [XY1(X\ge \tau_X,Y\ge\tau_Y)]-\bbE [X1(X\ge\tau_X)]\, \bbE [Y1(Y\ge\tau_Y)]\big).\label{eq:four_terms}
\end{align}

In order to bound the first term in the RHS of \eqref{eq:four_terms}, we tighten the proof technique of Lemma~\ref{lemma:max_corr} by bounding $m_X,m_Y$. Since $m_X,m_Y$ are essentially values of the moment generating functions for $X$ and $Y$, bounding $m_X,m_Y$ allows us to apply Chernoff bounds to probabilities involving $X$ and $Y$. We exploit the fact that Chernoff bounds are stronger than the Chebyshev's bounds used in the proof of Lemma~\ref{lemma:max_corr} to prove a tighter bound in this context. We first relate $m_X,m_Y$ to a moment generating function for $Z$, by writing
\begin{align}
&\bbE \left[e^{Z/\sqrt{S}}1(Z<\tau_Z)\right]\label{eq:Z_moment0}
\\&=\bbE \left[e^{(X+ Y+N)/\sqrt{S}}1(X+Y+N< \tau_X+ \tau_Y+\tau_N)\right]\label{eq:Z_moment1}
\\&\ge \bbE \left[e^{(X+Y+N)/\sqrt{S}}1(X<\tau_X,Y<\tau_Y,N<\tau_N)\right]\label{eq:Z_moment2}
\\&=\bbE \left[e^{(X+Y)/\sqrt{S}}1(X<\tau_X,Y<\tau_Y)\right] \frac{1}{2}e^{1/(2S)}\label{eq:Z_moment3}
\\&\ge \frac{1}{2} \bigg(\bbE \left[e^{X/\sqrt{S}}1(X<\tau_X)\right]\, \bbE \left[e^{Y/\sqrt{S}}1(Y<\tau_Y)\right]\nonumber
\\&\qquad-O(\delta\log\delta^{-1})\sqrt{
\var \left(e^{X/\sqrt{S}}1(X<\tau_X)\right)\,\var \left(e^{Y/\sqrt{S}}1(Y<\tau_Y)\right)
}\bigg)\label{eq:Z_moment4}
\\&\ge \frac{1}{2}\bigg(\bbE \left[e^{X/\sqrt{S}}1(X<\tau_X)\right]\, \bbE \left[e^{Y/\sqrt{S}}1(Y<\tau_Y)\right]\nonumber
\\&\qquad-O(\delta\log\delta^{-1})\sqrt{
\bbE \left[e^{2X/\sqrt{S}}1(X<\tau_X)\right]\,\bbE \left[e^{2Y/\sqrt{S}}1(Y<\tau_Y)\right]
}\bigg) 
\\&\ge \frac{1}{2} \left[m_Xm_Y-O(\delta\log \delta^{-1})\exp\left\{\frac{\tau_X+\tau_Y}{\sqrt{S}}\right\}\right]\label{eq:Z_moment5}
\\&=\frac{1}{2} \left[m_Xm_Y-O(\delta\log \delta^{-1})\exp\left\{\frac{S_1+S_2}{S}-\frac{1}{4}\log\delta\right\}\right]
\\&\ge \frac{1}{2} \left[m_Xm_Y-O(\delta^{3/4}\log \delta^{-1})\right]\label{eq:Z_moment7}
\end{align}
where \eqref{eq:Z_moment2} holds because the random quantity in \eqref{eq:Z_moment1} is non-negative and since $X<\tau_X,Y<\tau_Y,N<\tau_N$ implies $Z<\tau_Z$, \eqref{eq:Z_moment3} holds since $N$ is a standard Gaussian independent of $(X,Y)$, \eqref{eq:Z_moment4} holds by the bound on $\rho_m(X;Y)$ from Lemma~\ref{lemma:max_corr}, \eqref{eq:Z_moment5} holds from the simple upper bound on $\bbE \left[e^{2X/\sqrt{S}}1(X<\tau_X)\right]$ found by plugging in $X=\tau_X$, and \eqref{eq:Z_moment7} holds since $S_1+S_2\le S$. We now apply the total variational bound in \eqref{eq:f_tv_bound} to upper bound the quantity in \eqref{eq:Z_moment0}. Specifically, since $e^{z/\sqrt{S}}1(z<\tau_Z)\le e^{\tau_Z/\sqrt{S}}$,
\begin{align}
\bbE \left[e^{Z/\sqrt{S}}1(Z<\tau_Z)\right]
&\le \bbE \left[e^{\tilZ/\sqrt{S}}1(Z<\tau_Z)\right]+e^{\tau_Z/\sqrt{S}}O(\sqrt{\delta\log\delta^{-1}})
\\&\le e^{1/2}+e\, \delta^{-1/4}O(\sqrt{\delta\log\delta^{-1}})\label{eq:Z_moment8a}
\\&=e^{1/2}+O(\delta^{1/4}\sqrt{\log\delta^{-1}})\label{eq:Z_moment8}
\end{align}
where in \eqref{eq:Z_moment8a} we have used the fact that $\tilZ\sim\calN(0,S)$. Combining the bounds in \eqref{eq:Z_moment7} and \eqref{eq:Z_moment8} yields
\be\label{eq:mxmy_bound}
m_Xm_Y\le 2e^{1/2}+O(\delta^{1/4}\sqrt{\log \delta^{-1}}).
\ee
Since $2e^{1/2}<4$, and recalling that the implied constant in the $O(\cdot)$ term in \eqref{eq:mxmy_bound} is universal, we may assume that $\delta$ is sufficiently small that $m_Xm_Y\le 4$.

We now lower bound the first term in \eqref{eq:four_terms}, or equivalently upper bound the negative of this term. As in the proof of Lemma~\ref{lemma:max_corr}, we will use the function $k_\delta$,  defined in \eqref{eq:k_delta_def}. By an identical argument as in \eqref{eq:PXY_lower_bd0}--\eqref{eq:PXY_lower_bd}, 
\begin{multline}
\bbP(x<X<\tau_X)\bbP(y<Y<\tau_Y)-\bbP(x<X<\tau_X,y<Y<\tau_Y)
 \le k_\delta\left(\min\{\bbP(x<X<\tau_X),\,\bbP(y<Y<\tau_Y)\}\right).
\end{multline}
Thus
\begin{align}
&\bbE[ X1(0<X<\tau_X)]\,\bbE [Y1(0<Y<\tau_Y)]-\bbE [XY1(0<X<\tau_X,0<Y<\tau_Y)]\label{eq:first_of_four}
\\&=\int_0^{\tau_X}dx \int_0^{\tau_Y} dy \left[\bbP(x<X<\tau_X)\bbP(y<Y<\tau_Y)-\bbP(x<X<\tau_X,y<Y<\tau_Y)\right]
\\&\le \int_0^{\tau_X}dx \int_0^{\tau_Y} dy\, k_\delta\left(\min\{\bbP(x<X<\tau_X),\, \bbP(y<Y<\tau_Y)\}\right).
\end{align}
For any $x\le \tau_X$, a Chernoff-type bound gives
\be
\bbP(x<X<\tau_X)\le e^{-x/\sqrt{S}} \bbE \left[e^{X/\sqrt{S}}1(X<\tau_X)\right]=e^{-x/\sqrt{S}}m_X
\ee
and similarly $\bbP(y<Y<\tau_X)\le e^{-y/\sqrt{S}}m_Y$, so the difference in \eqref{eq:first_of_four} is at most
\begin{align}
&\int_0^{\tau_X}dx \int_0^{\tau_Y} dy\, k_\delta\left(\min\{e^{-x/\sqrt{S}}m_X,\, e^{-y/\sqrt{S}}m_Y\}\right)
\\&\le \int_0^\infty dx \int_0^\infty dy\, k_\delta\left(e^{-(x+y)/(2\sqrt{S})}\sqrt{m_Xm_Y}\right)\label{eq:chernoff2}
\\&\le \int_0^\infty dx \int_0^\infty dy\, k_\delta\left(2e^{-(x+y)/(2\sqrt{S})}\right)\label{eq:chernoff3}
\\&= 4S \int_0^\infty z\,k_\delta\left(2 e^{-z}\right)dz\label{eq:chernoff4}
\\&= 4S\left[\int_0^{\log 2} 2\delta z dz+\int_{\log 2}^\infty z\left((1+2\delta)(2e^{-z})^{1/(1+\delta)}-2e^{-z}\right)dz\right]\label{eq:chernoff5}
\\&=4S\left[(\log^2 2)\delta+(1+2\delta)(1+\delta)(1+\delta+\log 2)-(1+\log 2)\right]
\\&=S\, O(\delta)
\end{align}
where \eqref{eq:chernoff2} follows since the integrand is non-negative, so the upper limits of the integral may be extended to $\infty$, as well as because $\min\{a,b\}\le \sqrt{ab}$ and $k_\delta$ is non-decreasing; \eqref{eq:chernoff3} holds by the above conclusion that $m_Xm_Y\le 4$; \eqref{eq:chernoff4} holds by the change of variables $z=\frac{x+y}{2\sqrt{S}}$; and \eqref{eq:chernoff5} follows from the definition of $k_\delta$. This proves that the first term in \eqref{eq:four_terms} is lower bounded by $-S\,O(\delta)$.

We now consider the second term in \eqref{eq:four_terms}. Applying again the bound on $\rho_m(X;Y)$ from Lemma~\ref{lemma:max_corr} gives
\begin{align}
&\bbE [XY1(X\ge \tau_X,0<Y< \tau_X)]-\bbE [X1(X\ge \tau_X)]\,\bbE [Y1(0<Y<\tau_Y)]
\\&\ge -O(\delta\log\delta^{-1})\sqrt{\bbE [X^21(X\ge \tau_X)]\,\bbE [Y^21(0<Y< \tau_X)]}
\\&\ge -O(\delta\log\delta^{-1})\sqrt{\bbE [X^21(X\ge \tau_X)]\,S}\label{eq:ax_goal}
\end{align}
where the second inequality holds since $\bbE [Y^21(0<Y< \tau_X)]\le \bbE [Y^2]\le S_2\le S$. We now need to upper bound $\bbE [X^21(X\ge \tau_X)]$. Define
\begin{align}
p_X&=\bbP(X\ge \tau_X),\\
a_X&=\bbE [X^21(X\ge \tau_X)].
\end{align}
Intuitively, if $X\ge \tau_X$, then we expect $Z$ also to be large, and so we expect $p_X$ to be small. This intuition can be formalized by writing
\begin{align}
\bbP(Z\ge \tau_X-2\sqrt{S_2})
&=\bbP(X+Y+N\ge \tau_X-2\sqrt{S_2})
\\&\ge \bbP(X\ge \tau_X,Y\ge -2\sqrt{S_2},N\ge 0)
\\&=\frac{1}{2} \bbP(X\ge \tau_X,Y\ge -2\sqrt{S_2})\label{eq:pX1}
\\&\ge \frac{1}{2} \bbP(X\ge \tau_X)\bbP(Y\ge-2\sqrt{S_2})-\delta\label{eq:pX2}
\\&\ge \frac{3}{8} p_X-\delta\label{eq:pX3}
\end{align}
where \eqref{eq:pX1} holds because $N$ is Gaussian and independent of $X,Y$, \eqref{eq:pX2} holds by the consequence of $\Delta(X;Y)\le\delta$ in \eqref{eq:gdep_abs_bd}, and \eqref{eq:pX3} holds by Chebyshev's inequality on $Y$. Thus
\begin{align}
p_X&\le \frac{8}{3}\bbP(Z\ge \tau_X-2\sqrt{S_2})+O(\delta)
\\&\le \frac{8}{3}\bbP(\tilZ\ge \tau_X-2\sqrt{S_2})+O(\sqrt{\delta\log\delta^{-1}})\label{eq:px_bd2}
\\&=\frac{8}{3}P\left(\tilZ\ge \frac{S_1}{\sqrt{S}}-\frac{\sqrt{S}}{8}\log\delta-2\sqrt{S_2} \right)+O(\sqrt{\delta\log\delta^{-1}})
\\&\le \frac{8}{3} P\left(\tilZ\ge \sqrt{S}\left(-\frac{1}{8}\log \delta-2\right)\right)+O(\sqrt{\delta\log\delta^{-1}})\label{eq:px_bd4}
\\&\le \frac{8}{3} \exp\left\{-\frac{1}{2}\left(-\frac{1}{8}\log \delta-2\right)^2\right\}+O(\sqrt{\delta\log\delta^{-1}})\label{eq:px_bd5}
\\&=O(\sqrt{\delta\log\delta^{-1}})\label{eq:px_bd6}
\end{align}
where \eqref{eq:px_bd2} holds by the bound on total variational distance in \eqref{eq:tv_bound}, \eqref{eq:px_bd4}  holds since $S_2\le S$, \eqref{eq:px_bd5} holds since $\tilZ\sim\calN(0,S)$ and by the Chernoff bound on the Gaussian CDF, and \eqref{eq:px_bd6} holds since $\exp\{-O(\log^2\delta)\}$ vanishes faster than $O(\sqrt{\delta\log\delta^{-1}})$. In order to bound $a_X$, we bound the mean-squared of $Z$ conditioned on either $X<\tau_X$ or $X\ge\tau_X$. In particular, 
\begin{align}
\bbE [Z^2 1(X<\tau_X)]&=\bbE[(X+Y+N)^21(X<\tau_X)]
\\&=1+\bbE [X^21(X<\tau_X)]+\bbE [Y^2 1(X<\tau_X)]+2\,\bbE [XY1(X<\tau_X)]
\\&\le 1+S_1-a_X+S_2+O(\delta\log\delta^{-1})\,\sqrt{\bbE [X^2 1(X<\tau_X)]\,\bbE [Y^2]}\label{eq:z2_bd1}
\\&\le S-a_X+S\, O(\delta\log\delta^{-1})\label{eq:z2_bd2}
\end{align}
where \eqref{eq:z2_bd1} again uses the maximal correlation bound from Lemma~\ref{lemma:max_corr}, and \eqref{eq:z2_bd2} follows from the mean squared bounds on $X$ and $Y$. Thus
\be\label{eq:z2_conditional1}
\bbE[Z^2|X<\tau_X]\le \frac{S-a_X+S\,O(\delta\log\delta^{-1})}{1-p_X}.
\ee
Moreover
\be\label{eq:z2_conditional2}
\bbE [Z^2|X\ge \tau_X]\le \frac{\sigma_Z^2}{p_X}\le \frac{S+S\,O(\delta\log\delta^{-1})}{p_X}.
\ee
We now apply these two bounds to upper bound the differential entropy of $Z$. In particular, if we let $F=1(X\ge \tau_X)$, then
\begin{align}
h(Z)&\le H(F)+h(Z|F)
\\&=H_b(p_X)+(1-p_X)h(Z|X<\tau_X)+p_X h(Z|X\ge \tau_X)
\\&\le H_b(p_X)+(1-p_X)\frac{1}{2}\log 2\pi e\frac{S-a_X+S\,O(\delta\log\delta^{-1})}{1-p_X}
+p_X \frac{1}{2}\log 2\pi e \frac{S+S\,O(\delta\log\delta^{-1})}{p_X}\label{eq:z2_bd3}
\\&=\frac{3}{2} H_b(p_X)+(1-p_X)\frac{1}{2}\log 2\pi e(S-a_X+S\,O(\delta\log\delta^{-1}))
+p_X\frac{1}{2}\log 2\pi e(S+S\,O(\delta\log\delta^{-1}))
\end{align}
where \eqref{eq:z2_bd3} follows from the fact that differential entropy is upper bounded by that of a Gaussian with the same variance and the bounds in \eqref{eq:z2_conditional1}--\eqref{eq:z2_conditional2}. Recalling the assumption that $h(Z)\ge \frac{1}{2}\log 2\pi eS$, we have
\begin{align}
0&\le \frac{3}{2} H_b(p_X)+(1-p_X)\frac{1}{2}\log\left(1+\frac{-a_X+S\,O(\delta\log\delta^{-1})}{S}\right)
+p_X \frac{1}{2} \log\left(1+ O(\delta\log\delta^{-1})\right)
\\&\le \frac{3}{2} H_b(p_X)+(1-p_X)\frac{-a_X+S\,O(\delta\log\delta^{-1})}{2S}+p_X O(\delta\log\delta^{-1})
\\&=\frac{3}{2} H_b(p_X)-\frac{(1-p_X)a_X}{2S}+O(\delta\log\delta^{-1}).
\end{align}
Rearranging gives
\begin{align}
a_X&\le \frac{S}{1-p_X}\left[3H_b(p_X)+O(\delta\log\delta^{-1})\right]
\\&\le S (1+O(\sqrt{\delta\log\delta^{-1}}))\left[O(\delta^{1/2}(\log\delta^{-1})^{3/2})+O(\delta\log\delta^{-1})\right]\label{eq:ax_bd1}
\\&=S\, O(\delta^{1/2}(\log\delta^{-1})^{3/2})
\end{align}
where in \eqref{eq:ax_bd1} we have applied the bound on $p_X$ from \eqref{eq:px_bd6}, as well as the fact that for small $p$, $H_b(p)=O(p\log p^{-1})$. Plugging this bound back into \eqref{eq:ax_goal}, we find
\be\label{eq:term2_bd}
\bbE [XY1(X\ge \tau_X,0<Y< \tau_X)]-\bbE [X1(X\ge \tau_X)]\,\bbE [Y1(0<Y<\tau_Y)]
\ge -S\,O(\delta^{5/4}(\log \delta^{-1})^{7/4}).
\ee
By the same argument as the above bound on $a_X$, we may similarly find
\be
\bbE [Y^2 1(Y\ge \tau_Y)]\le S\,O(\delta^{1/2}(\log\delta^{-1})^{3/2}).
\ee
This implies that the third term in \eqref{eq:four_terms} is lower bounded by
\be\label{eq:term3_bd}
\bbE [XY1(X<\tau_X,Y\ge \tau_Y)]-\bbE [X1(X<\tau_X)]\bbE [Y1(Y\ge \tau_Y)]
\ge -S\,O(\delta^{5/4}(\log \delta^{-1})^{7/4})
\ee
and the fourth term in \eqref{eq:four_terms} is lower bounded by
\be\label{eq:term4_bd}
\bbE [XY1(X\ge \tau_X,Y\ge \tau_Y)]-\bbE [X1(X\ge \tau_X)]\bbE [Y1(Y< \tau_Y)]
\ge -S\,O(\delta^{3/2}(\log\delta^{-1})^{5/2}).
\ee
Note that for each of the bounds in \eqref{eq:term2_bd}, \eqref{eq:term3_bd}, and \eqref{eq:term4_bd}, the function of $\delta$ grows smaller than $O(\delta)$. Putting everything together, we now have
\be
|\bbE [XY1(X>0,Y>0)]-\bbE [X1(X>0)]\,\bbE [Y1(Y>0)]|\le S\, O(\delta).
\ee
Applying this bound by swapping $X$ with $-X$ and/or $Y$ with $-Y$ gives
\be
\bbE [XY]\le S\,O(\delta).
\ee
Therefore
\be
h(Z)\le \frac{1}{2}\log 2\pi eS(1+O(\delta))=\frac{1}{2}\log 2\pi e S+O(\delta).
\ee
This proves \eqref{eq:gaussian_goal}.

\subsection{Bounding \texorpdfstring{$V_{\max}$}{Vmax}}

Recall that
\be
V_{\max}=\sup_{P_{UXY}:\bbE [X^2]\le S_1,\bbE [Y^2]\le S_2} 
\max\{V(W\|P_{Z|U}|P_{UXY}),\,
V(W\|P_{Z|YU}|P_{UXY}),\,
V(W\|P_{Z|XU}|P_{UXY})\}.
\ee
Each of the terms in the maximum can be shown to be finite by showing that the equivalent point-to-point quantity is finite:
\be
\sup_{P_{UX}:\bbE [X^2]\le S} V(W'\|P_{Z|U}|P_{UX})
\ee
where $W'\in\calP(\bbR\to \bbR)$ is the point-to-point channel where $Z=X+N$, $N\sim\calN(0,1)$. Consider any $P_{UX}$ where $\bbE [X^2]\le S$. Fix $u$, and let $S_u=\bbE[X^2|U=u]$. To simplify notation, we again drop the conditioning on $U=u$. Define the information density
\be
\imath(x;z)=\log \frac{dW'_{x}}{dP_Z}(z).
\ee
Note that
\begin{align}
V(W'\|P_Z|P_{X})&=\bbE\left[\var(\imath(X;Z)|X)\right]
\\&\le \bbE[ \imath(X;Z)^2]
\\&=\bbE[\imath(X;Z)^2 1(\imath(X;Z)\le 0)]+\bbE[\imath(X;Z)^2 1(\imath(X;Z)\ge 0)]\label{eq:two_density_terms}
\end{align}
where $(X,Z)$ are distributed according to $P_{X}W'$. To lower bound the information density, we may upper bound the Radon-Nikodym derivative
\begin{align}
\frac{dP_Z}{dW'_{x}}(z)&=\int dP_{X}(x') \frac{dW'_{x'}}{dW'_{x}}(z)
\\&=\int dP_{X}(x') \exp\left\{-\frac{(z-x')^2}{2}+\frac{(z-x)^2}{2}\right\}
\\&\le \exp\left\{\frac{(z-x)^2}{2}\right\}.
\end{align}
Thus
\be
\imath(x;z)\ge -\frac{(z-x)^2}{2}.
\ee
Thus the first term in \eqref{eq:two_density_terms} may now be upper bounded by
\begin{align}
\bbE[\imath(X;Z)^2 1(\imath(X;Z)\le 0)]
&\le \bbE \left[\left(\frac{(Z-X)^2}{2}\right)^2 1(\imath(X;Z)\le 0)\right]
\\&\le \bbE \left[\frac{(Z-X)^4}{4}\right]
\\&=\frac{3}{4}
\end{align}
where we have used the fact that $Z-X=N$ is a standard Gaussian. 

We now upper bound the second term in \eqref{eq:two_density_terms}. For any integer $k$, let $\setA_k=[k,k+1)$. Let $p_k=\bbP(X\in \setA_k)$. Also let $\mu_k=\bbE[X|X\in \setA_k]$ and $\sigma_k^2=\var(X|X\in \setA_k)$. Since $\setA_k$ is an interval of length 1, $\sigma_k^2\le 1/4$. Then for any integer $k$, the PDF of $P_Z$ is lower bounded by
\begin{align}
f_Z(z)&=\int dP_{X}(x) \frac{1}{\sqrt{2\pi}} \exp\left\{-\frac{(z-x)^2}{2}\right\}
\\&\ge \int_{x\in \setA_k} dP_{X}(x) \frac{1}{\sqrt{2\pi}} \exp\left\{-\frac{(z-x)^2}{2}\right\}
\\&\ge p_k \frac{1}{\sqrt{2\pi}} \exp\left\{\bbE\left[-\frac{(z-X)^2}{2}\bigg|X\in \setA_k\right]\right\}\label{eq:PDF_bd3}
\\&=p_k  \frac{1}{\sqrt{2\pi}} \exp\left\{-\frac{(z-\mu_k)^2}{2}-\frac{\sigma_k^2}{2}\right\}\label{eq:PDF_bd4}
\\&\ge p_k  \frac{1}{\sqrt{2\pi}} \exp\left\{-\frac{(z-\mu_k)^2}{2}-\frac{1}{8}\right\}\label{eq:PDF_bd5}
\end{align}
where \eqref{eq:PDF_bd3} holds by the convexity of the exponential, \eqref{eq:PDF_bd4} holds by the definitions of $\mu_k$ and $\sigma_k$, and \eqref{eq:PDF_bd5} holds since $\sigma_k^2\le 1/4$. Thus, for any $k$ the information density can be upper bounded by
\be
\imath(x;z)\le \frac{-(z-x)^2+(z-\mu_k)^2}{2}+\frac{1}{8}-\log p_k
\ee
Applying this bound to the second term in \eqref{eq:two_density_terms} gives
\begin{align}
&\bbE[\imath(X;Z)^2 1(\imath(X;Z)\ge 0)]
\\&\le \sum_{k=-\infty}^\infty  \int_{x\in \setA_k} dP_{X}(x) \bbE\left[\left(\frac{-(Z-x)^2+(Z-\mu_k)^2}{2}+\frac{1}{8}-\log p_k\right)^2\bigg|X=x\right]
\\&=\sum_k \int_{x\in \setA_k} dP_{X}(x) \left[(x-\mu_k)^2+\left(\frac{(x-\mu_k)^2}{2}+\frac{1}{8}-\log p_k\right)^2\right]
\\&\le \sum_k p_k \left[1+\left(\frac{5}{8}-\log p_k\right)^2\right]\label{eq:density_bd3}
\\&\le 2+\sum_k \left[-2 p_k\log p_k+p_k\log^2 p_k\right]\label{eq:density_bd4}
\end{align}
where \eqref{eq:density_bd3} holds since $|x-\mu_k|\le 1$ for $x\in \setA_k$, because $\mu_k\in \setA_k$ and $\setA_k$ has length $1$, and in \eqref{eq:density_bd4} we have upper bounded $5/8$ by $1$ to simplify the expression. By Chebyshev's inequality, for $k>0$
\be
p_k=\bbP(X\in \setA_k)\le \bbP(X\ge k)\le \frac{S_u}{k^2}.
\ee
Note that for $p\in[0,1]$, $-p\log p\le 1/e$, and this function is increasing for $p\le 1/e$. Thus, if we consider the sum of $-p_k\log p_k$ for $k\ge 0$, we have
\begin{align}
\sum_{k=0}^\infty -p_k\log p_k
&\le\sum_{k=0}^{\ceil{\sqrt{eS_u}}} \frac{1}{e}+\sum_{k=\ceil{\sqrt{e S_u}}+1}^\infty -\frac{S_u}{k^2}\log \frac{S_u}{k^2}
\\&\le \frac{1}{e}(\sqrt{eS_u}+2)+\int_{\sqrt{eS_u}}^\infty -\frac{S_u}{r^2}\log \frac{S_u}{r^2} dr
\\&=\frac{\sqrt{S_u}}{\sqrt{e}}+\frac{2}{e}+\frac{3\sqrt{S_u}}{\sqrt{e}}
\\&=\frac{4\sqrt{S_u}}{\sqrt{e}}+\frac{2}{e}.\label{eq:pklogpk_bd}
\end{align}
By an identical calculation, $\sum_{k=-\infty}^{-1}-p_k\log p_k\le \frac{4\sqrt{S_u}}{\sqrt{e}}+\frac{2}{e}$. Similarly, note that $p\log^2 p\le 4/e^2$, and this function is increasing for $p\le 1/e^2$. Thus
\begin{align}
\sum_{k=0}^\infty p_k\log^2 p_k
&\le \sum_{k=0}^{\ceil{e\sqrt{S_u}}} \frac{4}{e^2}+\sum_{\ceil{e\sqrt{S_u}}+1}^\infty \frac{S_u}{k^2}\log^2 \frac{S_u}{k^2}
\\&\le \frac{4}{e^2}(e\sqrt{S_u}+2)+\int_{e\sqrt{S_u}}^\infty \frac{S_u}{r^2}\log^2 \frac{S_u}{r^2}dr
\\&=\frac{4}{e^2}(e\sqrt{S_u}+2)+\frac{20\sqrt{S_u}}{e}
\\&=\frac{24\sqrt{S_u}}{e}+\frac{2}{e^2}.\label{eq:pklog2pk_bd}
\end{align}
Again the same holds for the summation over $k<0$. Applying the bounds in \eqref{eq:pklogpk_bd} and \eqref{eq:pklog2pk_bd} to \eqref{eq:density_bd4} gives
\be
\bbE[\imath(X;Z)^2 1(\imath(X;Z)\le 0)]
\le 2+\frac{8\sqrt{S_u}}{\sqrt{e}}+\frac{4}{e}+\frac{48\sqrt{S_u}}{e}+\frac{4}{e^2}.
\ee
Now combining the bounds on each of the terms in \eqref{eq:two_density_terms} gives
\begin{align}
V(W'\|P_{Z|U}|P_{UX})&\le \sum_u P_U(u)\left[
 \frac{11}{4}+\frac{4}{e}+\frac{4}{e^2}+\left(\frac{8}{\sqrt{e}}+\frac{48}{e}\right)\sqrt{S_u}\right]
 \\&\le \frac{11}{4}+\frac{4}{e}+\frac{4}{e^2}+\left(\frac{8}{\sqrt{e}}+\frac{48}{e}\right)\sqrt{S}.
\end{align}

\bibliographystyle{IEEETran}
\bibliography{KosutMAC}

\end{document}